\documentclass{article}[12pt,a4paper] 
 \pdfoutput=1

\usepackage{amssymb} 
\usepackage{amsmath}
\usepackage{amsthm}
\usepackage{amsfonts}

\numberwithin{equation}{section}

\usepackage{times}

\usepackage[matrix,arrow]{xy}
\usepackage{tikz}
\usetikzlibrary{shapes, arrows, shadows}

\def\be{\begin{equation}}
\def\ee{\end{equation}}
\def\bea{\begin{eqnarray}}
\def\eea{\end{eqnarray}}

\def\thefootnote{\fnsymbol{footnote}}

\usepackage{bbold}
\usepackage{simplewick}
\usepackage{array}
\usepackage{xcolor}
\usepackage{caption}
\usepackage{subcaption}

\def\eq{\begin{equation}}
\def\en{\end{equation}}
\def\eqa{\begin{eqnarray}}
\def\ena{\end{eqnarray}}
\def\expval#1{\langle \, #1 \,\rangle}
\def\expvalc#1{\expval{#1}_{c}}
\newcommand{\OL}[1]{\mathcal{O}\!\ \left(\lambda^{#1}\right)}

\newcommand{\cC}{\mathcal{C}}

\newcommand{\cO}{\mathcal{O}}

\newcommand{\cU}{\mathcal{U}}

\newcommand{\oA}{A}
\newcommand{\oB}{B}
\newcommand{\oC}{C}
\newcommand{\oD}{D}

\newcommand{\fF}[3]{f_{#1 #2}{}^{ #3}}
\newcommand{\fFb}[3]{\bar{f}_{\bar{#1}\bar{#2}}{}^{\bar{#3}}}

\newcommand{\FF}[3]{f_{#1 #2 #3}}
\newcommand{\FFb}[3]{\bar{f}_{\bar{#1}\bar{#2}\bar{#3}}}

\usepackage{hyperref}
\hypersetup{colorlinks=true, breaklinks=true, linkcolor=black, menucolor=black, urlcolor=black,citecolor=green!20!black}

\begin{document}

\begin{titlepage}
\today         \hfill

\begin{center}
% \hfill hep-th/yymmnnn  \\

\vskip .5in
\renewcommand{\thefootnote}{\fnsymbol{footnote}}
{\Large \bf Renormalization and redundancy in 2d quantum field theories}\\
 
\vskip .50in

\vskip .5in
{\large Nicolas Behr}\footnote{email address: N.Behr@hw.ac.uk} and {\large Anatoly Konechny}\footnote{email address: anatolyk@ma.hw.ac.uk}

\vskip 0.5cm
{\large \em Department of Mathematics,\\
Heriot-Watt University,\\
Riccarton, Edinburgh, EH14 4AS, UK \\
and\\
Maxwell Institute for Mathematical Sciences\\
Edinburgh, UK}
\end{center}

\vskip .5in

\begin{abstract} \large

\end{abstract}
We analyze renormalization group (RG) flows in two-dimensional quantum field theories in the presence of redundant directions. We use the operator picture  in which redundant operators are total derivatives. 
Our analysis has  three levels of generality. We introduce a redundancy anomaly equation which is analyzed 
together with the  RG anomaly equation previously considered by H.~Osborn \cite{Osborn} and D.~Friedan and A.~Konechny  \cite{FK_grad}. 
The Wess-Zumino consistency conditions between these anomalies yield a number of general relations which should hold to all orders in perturbation theory. We further  use conformal perturbation theory to study field theories in the vicinity of a fixed point when some of the symmetries of the fixed point are broken by the perturbation. We relate various anomaly coefficients to OPE coefficients at the fixed point and analyze  which operators become redundant and how they participate in the RG flow. Finally, we illustrate our findings by three explicit models constructed as current-current perturbations of ${\rm SU}(2)_{k}$ WZW model. At each generality level we discuss the geometric picture behind redundancy and how one can reduce the number of couplings by taking a quotient with respect to the redundant directions. We point to the special role of polar representations for the redundancy groups. 

\end{titlepage}
\large

\newpage
\renewcommand{\thepage}{\arabic{page}}
\setcounter{page}{1}
\setcounter{footnote}{0}
\renewcommand{\thefootnote}{\arabic{footnote}}
%%%%%%%%%%%
\tableofcontents
\large

\section{Introduction}
\renewcommand{\theequation}{\arabic{section}.\arabic{equation}}
The subject of two-dimensional quantum field theories (2d QFTs) has provided us with the richness of 
nonperturbative techniques such as the ones related to integrability and conformal symmetry, 
as well as with a number of powerful general results. Among such results are the $c$-theorem
\cite{Zam} and the $g$-theorem \cite{AL}, \cite{FK_g} describing certain general properties of renormalization group (RG) 
flows in 2d QFTs. The $c$- and $g$-theorems proved to be very useful in establishing the 
phase diagrams and patterns of RG flows for various 2d systems with and without a boundary, see e.g. \cite{Zam2},
\cite{RRS}, \cite{FGSC}. 

The $c$-theorem explicitly constructs  a special function of the coupling constants, called the $c$-function, that  decreases monotonically along the RG flow 
and that is equal to the Virasoro central charge at fixed points of the flow. 
The $c$-theorem was proved in \cite{Zam} by deriving the relation  
\be\label{Zam_f}
\mu \frac{\partial c}{\partial  \mu} = - \beta^{i}g_{ij}\beta^{j} \, . 
\ee
Here $\mu$ is the RG scale\footnote{We define $\mu$ such that renormalized correlation functions with insertions at $x_i$ depend, up to the classical dimension, on the dimensionless combinations $\mu x_i$. Thus, although $\mu$ has dimensions of momentum, $\sum_i (2+x_i\tfrac{\partial}{\partial x_i})=\mu\tfrac{\partial}{\partial \mu}$. In this convention, the far infrared corresponds to $\mu\to\infty$.}, $c$ is the $c$-function, $\beta^{i}$ are the components of the beta function vector field, and $g_{ij}$ is 
the Zamolodchikov metric on the theory space, which is  positive definite. An even richer geometric structure is uncovered by a 
gradient formula for the beta function. A gradient formula relates the beta function vector field to the gradient of some potential function. In \cite{FK_grad}, a gradient formula for the beta function of 2d QFTs was proved under fairly general assumptions.
The formula has the form
\be\label{eq:gradientFormula}
  \partial_{i} c =- (g_{ij} + \Delta g_{ij})\beta^{j} - b_{ij}\beta^{j} 
  \ee
   where $c$, $g_{ij}$ and $\beta^{j}$ are the same as in (\ref{Zam_f}), $b_{ij}$ is the Osborn antisymmetric tensor \cite{Osborn} on the theory space, and 
   $\Delta g_{ij}$ is a certain correction to the Zamolodchikov metric. We review this formula in more detail in section \ref{sec:grad}.
    The objects $\beta^{i}, c, g_{ij}, \Delta g_{ij}, b_{ij}$ are the basic geometric data associated to the RG flows of 2d QFTs. 
   
   The gradient formula (\ref{eq:gradientFormula}) in particular applies to two-dimensional sigma models. In string theory 2d sigma models describe the space-time 
   background on which the strings propagate. Conformal sigma models, i.e. sigma models with vanishing beta functions, correspond to solutions 
   to classical equations of motion for the string. In this context the gradient formula has a special significance -- it provides a string action principle. 
   
   For sigma models with vanishing target space torsion (antisymmetric 2-form), the RG flow in the one loop approximation reduces to the Ricci flow for the target space metric. 
   The RG gradient formula involving metric and dilation couplings has interesting connections with the work of G. Perelman \cite{Perelman}, \cite{TseytlinonPerelman}.
   
   Geometric structures often provide us with useful tools to study the topology of the underlying spaces. For the spaces of quantum field theories, very little is currently known about their topology (a recent discussion can be found in \cite{Douglas}). There have been attempts to use Zamolodchikov's theorem and Morse theory 
   to obtain some information about the topology of the spaces related to perturbed minimal models \cite{Vafa}, \cite{Das_etal}, but the results are sparse and are still at the level  of conjectures. A better understanding of the geometry related to RG flows may help to advance our understanding of the topology of spaces of 2d QFTs. 
   
   In the current paper, we study aspects of the geometry of 2d QFTs and of the gradient formula (\ref{eq:gradientFormula}) related to redundant operators. We study the spaces of CFTs 
   abstractly in terms of correlation functions of local operators. In this context redundant operators are total derivative operators. If the set $\{\phi_{i}\} $ forms a basis of spin zero operators, then for any current $J^{\mu}(x)$ we have an expansion 
   \be \label{tot_der0}
   \partial_{\mu}J^{\mu}(x) = r^{i}\phi_{i}(x)  
   \ee
    which describes how total derivatives are embedded into the set of spin zero operators.    
   In particular there may be total derivative combinations of those operators $\phi_{i}$  which couple to the coupling constants parameterizing our QFTs. 
   As any operator equation, formula (\ref{tot_der0}) in general holds up to contact terms. Shifting the couplings so that we move along a redundant 
   direction amounts to a local redefinition of the local fields. Such redefinitions are stored in the contact terms related to (\ref{tot_der0}). More intuitively, one 
   can imagine inserting a total derivative into a correlation function  in which divergences are regulated by cutting out small circles around the insertions. 
   Integrating the total derivative will result in having contour integrals around each insertion. Shrinking the contours and subtracting divergences will result in 
   a local redefinition of the inserted operators. Such a picture and the related broken Ward identities were considered in \cite{Moore} (see sec. 9 in particular).  
   
   In the Lagrangian formulation of QFT, a  coupling is called redundant if  the change in the action when this coupling is varied vanishes on the equations of motion 
   (this definition is given e.g. in sec. 7.7 of \cite{Weinberg}). The local operator that couples to such a coupling equals a total derivative up to the terms proportional to 
   equations of motion, which are pure contact terms. To make this more explicit consider the following elementary example: a scalar field theory with action
   \be
   S= \int\!\! d^2x\, \frac{1}{2}Z ( \partial_{\mu} \phi \partial^{\mu} \phi + m^2 \phi^2) \, .
   \ee
   This action depends on 2 couplings: $m$ and $Z$. The coupling $Z$ couples to the local operator 
   \bea
    \phi_{Z}(x)= \frac{1}{2}( \partial_{\mu} \phi \partial^{\mu} \phi (x)+ m^2 \phi^{2}(x)) &=& \frac{1}{2}\partial_{\mu}(\phi\partial^{\mu} \phi)(x) + \frac{1}{2}\phi[m^2 \phi - \partial_{\mu}\partial^{\mu} \phi](x)  \nonumber \\
  & = &  \frac{1}{2}\partial_{\mu}(\phi\partial^{\mu} \phi)(x) + \frac{1}{2}Z^{-1}\phi \frac{\delta S}{\delta \phi} \, . 
        \eea
   Here the first term on the right hand side is a total derivative, while the second term is proportional to the equations of motion and is thus a pure contact term. 
   The coupling $Z$ is therefore redundant -- changing it can be 
   compensated by rescaling the field $\phi(x)$ by a constant factor.    
   
   In the context of exact RG equations redundant couplings were discussed in \cite{Wegner} and recently in \cite{Morris}. In \cite{Fortin} an RG anomaly equation was 
   analyzed in connection with an example in which the RG trajectory has a cycle along redundant directions.
   
   The $S$-matrix and thermodynamic quantities are independent of field redefinitions and thus are independent of the redundant couplings. Moreover, at the level of local correlation functions, moving along the redundant directions 
   only reparameterizes the local observables so that all essential physical information is stored in correlators evaluated on a slice of the coupling space transverse to the redundant directions. One can imagine reducing the 
   number of couplings by  eliminating  the redundant couplings (i.e.\ taking a quotient). Since redundant operators get admixed to other operators  when we change the scale (see sec. \ref{sec:red} for a detailed discussion),  
   it is not immediately clear how such an elimination  can be performed in an RG covariant way. For Lagrangian field theories, such an elimination was discussed in  \cite{Arzt}, \cite{Einhorn} (see also \cite{Morris}).   
   
   In this paper we first discuss the redundant operators in very general terms. We write out the most general form for the contact terms in (\ref{tot_der0}) which holds perturbatively. We analyze the compatibility of 
    redundancy equations with the renormalization group equations via the Wess-Zumino consistency conditions on  the contact terms (the anomaly). This yields a number of relations between contact terms in 
    the RG equations (the Weyl anomaly) and contact terms in the redundancy equations. These relations, derived in section \ref{sec:red},  allow us to show the existence of reduced beta functions. 
    
    Besides being able to reduce the beta functions, we are interested in showing that other geometric data associated with the gradient formula (\ref{eq:gradientFormula}) can be reduced onto the quotient space. 
    In search for a general procedure we 
    made calculations in conformal perturbation theory for RG flows near fixed points with symmetries perturbed by marginally relevant operators breaking (some of) the  symmetries.   
    These calculations are presented in section \ref{general_cp}.
    In particular, in sections \ref{sec:anomalyFP} and \ref{sec:metricCorrection} we relate the leading order  anomaly coefficients (in the RG equation and in the redundancy equation) to certain 
    OPE coefficients calculated at the fixed point. In section \ref{sec:red1} we calculate the redundancy equations in a point-splitting scheme up to the quadratic order in the couplings.
      Among other results we   have also obtained  a general formula for the two loop beta function of marginal operators expressed in terms of an integrated four-point function (\ref{beta3}). 
      
      For illustration purposes we apply the findings of section \ref{general_cp} to three particular models constructed as current-current perturbations of the ${\rm SU}(2)_{k}$ WZW model. 
      In section \ref{sec:models} we present explicit calculations related to these models and discuss the geometric structure of redundancy as well as the reduction procedure. We show that 
      a consistent reduction is possible up to two loops for any model in which the (fixed point) representation of the redundancy group is polar. 
      In section  \ref{discussion} we try for a general geometric picture of redundancy and RG flows that emerges from our studies and point out some loose ends and future directions. 
      The appendices contain some more technical details of the calculations.
       
\section{Gradient formula} \label{sec:grad}
\setcounter{equation}{0}
In this section we introduce some notations, explain the basic principles and 
 formulate the gradient formula of \cite{FK_grad}.

We consider two-dimensional Euclidean quantum field theories equipped 
with a conserved stress-energy tensor $T_{\mu \nu}(x)$. In response to 
a metric variation $g_{\mu \nu}(x)=\delta_{\mu \nu} + \delta g_{\mu \nu}(x)$, 
the partition function $Z[g_{\mu \nu}]$ changes as 
\be
\delta \ln Z = \frac{1}{2}\iint d^{2}x \langle \delta g_{\mu \nu}(x) T^{\mu \nu}(x)\rangle. 
\ee
For a conformally flat metric $g_{\mu \nu}(x) = \mu^{2}(x)\delta_{\mu \nu}$ the function 
$\mu(x)$ sets the local scale. Changing that local scale gives 
\be \label{mu}
\mu(x) \frac{ \delta \ln Z}{\delta \mu(x)}= \langle \Theta(x) \rangle
\ee
where $ \Theta(x)=g^{\mu \nu}T_{\mu \nu}(x)$ is the trace of the stress-energy tensor. For correlation functions on ${\mathbb R}^{2}$ with constant $\mu(x)=\mu={\rm Const}$, the change of 
scale in correlation functions is given by integrating $\Theta(x)$: 
\be \label{1}
\mu \frac{\partial}{\partial \mu} \langle {\cal O}_{1}(x_{1}) \dots {\cal O}_{n}(x_{n})\rangle_{c}
= \!\! \int\!\! d^{2}x\, \langle \Theta(x) {\cal O}_{1}(x_{1}) \dots {\cal O}_{n}(x_{n})\rangle_{c}
\ee
where ${\cal O}_{i}$ are local operators and $\langle \dots \rangle_{c}$ stands for a connected 
correlator. 

We further assume that we have a family of  quantum field theories parameterized by renormalized 
coupling constants $\lambda^{i}$, $i=1, \dots , N$. 
Each coupling $\lambda^{i}$ couples to a 
local operator $\phi_{i}(x)$ in such a way that the action principle is satisfied \cite{actionpr}. 
This means that  changing $\lambda^{i}$  
in any local correlation function  is given by integrating an insertion of $\phi_{i}(x)$: 
\be \label{2}
\frac{\partial }{\partial \lambda^{i}} \langle {\cal O}_{1}(x_{1}) \dots {\cal O}_{n}(x_{n})\rangle_{c} 
= \int\!\! d^{2}x \, \langle \phi_{i}(x) {\cal O}_{1}(x_{1}) \dots {\cal O}_{n}(x_{n})\rangle_{c} \,  .
\ee
The renormalized correlation functions in (\ref{1}) and (\ref{2}) are distributions, so they 
are always locally integrable, but the existence of integrals over the entire ${\mathbb R}^{2}$ assumes a suitable 
infrared behavior.  Note that we allow for any scalar operator $\phi_{i}$, in particular 
among the $\phi_{i}$ there can be total derivative operators. 
 
 We further assume that the coupling constants $\lambda^{i}$ can be promoted to local sources $\lambda^{i}(x)$. 
 The partition function $Z[g_{\mu, \nu}]$ generalizes to a generating functional that depends on the local 
 scale factor $\mu(x)$ and the sources $\lambda^{i}(x)$  so that in addition to (\ref{mu}) we have 
 \be\label{lambda}
  \frac{\delta\ln  Z}{\delta \lambda^{i}(x)} = \langle \phi_{i}(x) \rangle \, . 
 \ee
 Correlation functions on flat space involving the fields $\phi_{i}(x)$ and the trace of the stress-energy tensor can be obtained by taking a number of variational derivatives with respect to the sources and the scale factor, followed by setting the scale factor and sources to constant values. 
 
 In a renormalizable QFT, a change of scale $\mu$ can be compensated by changing the coupling constants according to the beta function vector field $\beta^{i}(\lambda)$.
 It follows from the action principle (\ref{2}) that 
 \be \label{theta}
 \Theta(x)= \beta^{i} \phi_{i}(x) 
 \ee  holds as an operator equation. As we remarked above there can be total derivatives 
 among the operators $\phi_{i}$. Strictly speaking, the coefficients $\beta^{i}$ standing at total derivatives are not called beta functions, but for the sake of uniformity we 
 will use the same notation for them, and -- by a slight abuse of terminology -- will refer to all $\beta^{i}$'s as  beta functions\footnote{In some papers, the authors use the notation $B^i$ for the expansion coefficients in~\eqref{theta} which contain total derivatives, reserving the notation $\beta^i$ for the usual beta functions.}.  Equation (\ref{theta}) holds inside correlation functions 
 up to constant terms (i.e.\ up to distributions supported on a set of measure zero). Using the sources and non-constant scale factor we can store the contact terms in 
 derivatives of $\lambda^{i}(x)$ and $\mu(x)$. To this end we expand the difference $\Theta(x) -  \beta^{i}(\lambda(x))\phi_{i}(x)$ in such derivatives.  
 The form of this expansion is constrained by 2d covariance and locality. One can write 
 \be \label{RGop}
 \Theta(x) - \beta^{i}\phi_{i}(x) = \frac{1}{2}\mu^{2}R_{2}(x)C(\lambda) + \partial_{\mu} \lambda^{i}J_{i}^{\mu}(x) +
 \partial^{\mu}[W_{i}(\lambda)\partial_{\mu} \lambda^{i}] + \frac{1}{2}\partial^{\mu} \lambda^{i}\partial_{\mu}\lambda^{j} G_{ij}(\lambda) + \dots 
 \ee
 where 
 $$
 \mu^2 R_{2}(x) = -2\partial_{\mu}\partial^{\mu} \ln \mu(x) \, .
 $$
 In (\ref{RGop}) we wrote explicitly all possible terms containing one and two derivatives of $\mu$ and $\lambda^{i}$. In the vicinity of a fixed point QFT where perturbation theory applies there can be nothing else. As in \cite{FK_grad}, we say that in this situation a strict power counting applies. In such a case the coefficients $C(\lambda)$, $W_{i}(\lambda)$, 
 $G_{ij}(\lambda)$ are functions and $J_{i}^{\mu}$ is a quantum field of spin 1. This restriction can be relaxed to a loose power counting in which the coefficients 
 $C$, $W_{i}$, $G_{ij}$ are allowed to have a non-trivial operator content. The loose power counting applies when one considers perturbation theory for nonlinear sigma models. 
 More generally, when perturbation theory does not apply,  one can allow for arbitrary order derivatives to appear in (\ref{RGop}). Equation (\ref{RGop}) generalizes the equation for the conformal anomaly in curved space. In a sense one can call it an equation for the renormalization anomaly. 
 
In this paper we will use perturbation theory around a 2d CFT so that the strict power counting applies, such that the full anomaly is given by the terms explicitly written in (\ref{RGop}). 
 In this case one derives  Callan-Symanzik equations by applying  (\ref{RGop}) to the generating functional $\ln Z$ and taking additional variational derivatives that give  insertions of $\Theta$'s  and $\phi_{i}$'s: 
\bea \label{CS}
&&  \Bigl( \mu \frac{\partial}{\partial \mu} - \beta^{i}\frac{\partial}{\partial \lambda^{i}}  \Bigr) \langle \phi_{i_{1}}(x_1) \phi_{i_{2}}(x_2) \dots \Theta(y_{1}) \dots \rangle_{c} = \nonumber \\
 && \langle \Gamma \phi_{i_{1}}(x_1)  \phi_{i_{2}}(x_2) \dots \Theta(y_{1}) \dots \rangle_{c} + \langle  \phi_{i_{1}}(x_1)  \Gamma \phi_{i_{2}}(x_2) \dots \Theta(y_{1}) \dots \rangle_{c}
 + \dots
  \eea
 where 
 \be\label{GammaCS}
 \Gamma \phi_{i_{1}}(x_1)= \partial_{i_{1}}\beta^{i}\phi_{i}(x_1) - \partial_{\mu} J_{i_{1}}^{\mu}(x_1) \, . 
 \ee
We see that the  operators  $\Gamma$ that give mixings of operators under RG include the standard part $\partial_{i} \beta^{j}$ which comes from the beta functions and 
additional admixtures of total derivatives that come from the anomaly (\ref{RGop}).   We can rewrite (\ref{CS}) as 
\bea
 \label{CS2}
&&  \Bigl( \mu  \frac{\partial}{\partial \mu} - {\cal L}_{\beta}  \Bigr) \langle \phi_{i_{1}}(x_1) \phi_{i_{2}}(x_2) \dots \Theta(y_{1}) \dots \rangle_{c} = \nonumber \\
 && -\langle   \partial_{\mu} J_{i_{1}}^{\mu}(x_1) \phi_{i_{2}}(x_2) \dots \Theta(y_{1}) \dots \rangle_{c} - \langle  \phi_{i_{1}}(x_1)  \partial_{\mu} J_{i_{2}}^{\mu}(x_2)  \dots \Theta(y_{1}) \dots \rangle_{c}
 + \dots
  \eea
   where ${\cal L}_{\beta}$ denotes the Lie derivative with respect to the beta function vector field. The last equation shows that the currents $J_{i}$ from the anomaly are responsible 
   for the noncovariant behavior of the correlators under the change of scale.   
 
 Besides the above considerations, the terms in (\ref{RGop}) are subject to Wess-Zumino consistency conditions. We can write using  (\ref{mu}) and (\ref{lambda}) both sides of 
 (\ref{RGop}) as functional differential operators acting on functionals of sources and the scale factor: 
 \be
\mu(x) \frac{\delta}{\delta \mu(x)} - \beta^{i}(\lambda(x))\frac{\delta}{\delta \lambda^{i}(x)} = {\cal D}(x) \, 
 \ee
 where $ {\cal D}(x) $ is a differential operator\footnote{To write the differential operator representing the vector field $J_{i}^{\mu}(x)$ one needs sources for vector fields. Such sources and additional terms in 
 the anomaly related to them are introduced in the next section. For the purpose of deriving   the gradient formula, the vector field sources can be largely ignored, so we do not explicitly use them in this section. } representing the right hand side of (\ref{RGop}). 
 The Wess-Zumino consistency conditions are then the zero commutator equations for these operators,
 \be\label{WZ}
 [\mu(x) \frac{\delta}{\delta \mu(x)} - \beta^{i}(\lambda(x))\frac{\delta}{\delta \lambda^{i}(x)} - {\cal D}(x), \mu(y) \frac{\delta}{\delta \mu(y)} - \beta^{i}(\lambda(y))\frac{\delta}{\delta \lambda^{i}(y)} -{\cal D}(y)]=0\, .
   \ee
   These equations lead to various relations between the anomaly terms in (\ref{RGop}). When strict power counting applies, one of the consequences is the operator 
   equation\footnote{More generally, when strict power counting does not apply, e.g. for nonlinear sigma 
   models,  equation (\ref{WZop}) is replaced by $ \beta^{i}  J^{\mu}_{i} (x) =\partial^{\mu} C(x)$ for a scalar operator $C(x)$.
   The combinations $\partial_{\mu} \partial_{\nu} C - \delta_{\mu \nu} \partial^{2} C$ are the improvement currents that  get admixed to the stress-energy tensor under the RG flow, see \cite{FK_grad}. 
   In the context of nonlinear sigma models $C(x)$ is the dilation beta functions operator and the generalization of (\ref{WZop}) is called Curci-Paffuti relation \cite{CP}. \label{ft}}
   \be\label{WZop}
   \beta^{i}  J^{\mu}_{i} (x) = 0 \, . 
   \ee  
   This equation implies that while $\Theta(x) = \beta^{i}\phi_{i}$ and each of the $\phi_{i}$ fields may get an anomalous admixture of a total derivative under the RG flow, 
    $\Theta(x)$  does not get such an admixture (cf. equation \eqref{CS}).  
   
  The consequences of  equations (\ref{WZ}) were systematically explored  in \cite{Osborn}. Under certain assumptions a gradient formula for the beta function was derived in \cite{Osborn} as a consequence 
  of equations (\ref{WZ}). 
  In \cite{FK_grad}, the same method was used to derive under a more general set of assumptions a gradient formula of the form 
  \be\label{grad_f}
  \partial_{i} c + (g_{ij} + \Delta g_{ij})\beta^{j} + b_{ij}\beta^{j} = 0\,.
  \ee
Here, $c$ is the Zamolodchikov c-function, $g_{ij}$ is the Zamolodchikov metric \cite{Zam}, $b_{ij}$ is the Osborn antisymmetric tensor \cite{Osborn}, and 
   $\Delta g_{ij}$ is a certain correction to the Zamolodchikov metric. Explicitly we have 
   \be 
c = 4\pi^{2}\left (
 x^{\mu}x^{\nu} x^{\alpha}x^{\beta}
- x^{2}g^{\mu\nu}  x^{\alpha}x^{\beta}
- \frac{1}{2}x^{2} x^{\mu}g^{\nu\alpha}x^{\beta}
\right )
{\expvalc{T_{\mu\nu}(x)\,T_{\alpha\beta}(0)}}_{\big /\Lambda|x|=1}
\label{eq:corig}
\ee
\be
g_{ij} = 6\pi^{2}  \Lambda^{-4} \,{\expvalc{\phi_{i}(x)\,\phi_{j}(0)}}_{\big / \Lambda 
|x|=1}
\label{eq:gijorig}
\ee
where $\Lambda^{-1}$ is a fixed arbitrary 2d distance.
The tensor $b_{ij}$ is an antisymmetric 2-form that can be expressed as 
\begin{equation}\label{eq:bijorig}
b_{ij}=\partial_{i}w_{j} - \partial_{j}w_{i}\, , \quad w_{i} = 3\pi \int\!\! d^{2}x\,x^{2}\theta(1-\Lambda|x|)
\langle \phi_{i}(x)\Theta(0)\rangle_{c} 
\end{equation}
where $\Lambda$ is the same mass scale used in the definition of $c$ and $g_{ij}$.      The metric correction $\Delta g_{ij}$ is constructed using the 
anomaly currents $J_{i}^{\mu}(x)$: 
\be  \label{delta_g}
\Delta g_{ij}
= \lim\limits_{L\to \infty}
3\pi \int_{|x|<L} d^{2}x \, x^2 \theta(\Lambda |x| - 1) \, [
\expvalc{
\phi_{i}(x)\,\partial_{\mu}J_{j}^{\mu}(0)}
+
\expvalc{ \phi_{j}(x)\, \partial_{\mu}J_{i}^{\mu}(0)
}] \, . 
\ee 
  where subtractions may be needed to take the limit $L\to \infty$ (see \cite{FK_grad}  for details). 
   %%%% assumptions in the proof 

   The gradient formula (\ref{grad_f}) was proven under a number of assumptions of a rather general nature: stress-energy tensor conservation, locality, the validity of the action 
   principle (\ref{2}) and the absence of spontaneous breaking of global conformal symmetry. The last assumption  means that for any 
   vector field $J_{\mu}(x)$ we have an infrared condition
   \be
   \lim\limits_{|x|\to \infty} |x|^3 \langle J_{\mu}(x) T_{\alpha \beta}(0)\rangle_{c} = 0 \, . 
   \ee
   
   Contracting the gradient formula with the beta function one obtains 
   \be 
   \beta^{i} \partial_{i} c - \beta^{i}\Delta g_{ij} \beta^{j} = - \beta^{i} g_{ij} \beta^{j} \, . 
   \ee
   One can show that the left hand side of the above formula gives the scale derivative of the $c$-function  \cite{FK_grad} (the second term on the left hand side accounts for the anomalous admixtures of 
   improvement currents to the stress-energy tensor; it vanishes when strict power counting applies). So  one obtains the celebrated Zamolodchikov formula
   \be 
   \mu \frac{\partial c}{\partial \mu} = - \beta^{i} g_{ij} \beta^{j} \, . 
      \ee 
We also note that the extension of the analysis of the Wess-Zumino consistency conditions~\eqref{WZ} to higher-dimensional theories was done in~\cite{jack_analogs_1990,Osborn,nakayama_consistency_2013,grinstein_consequences_2013}.
   %%%%%%%%%%%%%%%%%%%%%%%%%%%%%%%%%%%%%%%%%%%%%%%%%%%%%%%%%%%%%%%%%%%%%%%%%%%%%%%%%%%%%%%%%%%%%%%%%%
   
\section{Redundant operators} \label{sec:red}
\setcounter{equation}{0}

 Redundant operators arise in the RG anomaly (\ref{RGop}) and subsequently enter the gradient formula via the metric correction (\ref{delta_g}). They are also responsible for the noncovariance of the RG transformation of the correlators (\ref{CS2}) and, as a consequence, for  the noncovariance of the metric $g_{ij}$ and of the antisymmetric tensor $b_{ij}$. 
 On the other hand, it is clear that if among operators $\phi_{i}$ there are total derivatives, those directions are physically redundant and there must be a way to reduce the number 
 of couplings by taking a quotient via projecting out such directions. One of the main motivations for this paper was to investigate how such a reduction can be implemented systematically 
 and how all geometric objects in the gradient formula reduce.
 In this section we discuss the general theory of redundancy in the operator formalism. 
 
 To account for total derivatives  among scalar fields, one introduces a basis of vector fields $J_{\mu}^{a}(x)$ so that if $\phi_{i}(x)$ form a complete basis of scalar fields one has 
 \be \label{red0}
 \partial_{\mu} J^{\mu}_{a}(x) = r^{ i}_{a}(\lambda ) \phi_{i}(x)
 \ee
  where $r_{a}^{ i}(\lambda)$ are some coefficients giving the embedding of total derivatives into the set of scalar operators. Equation (\ref{red0}) is an operator equation that holds 
  inside correlation functions up to contact terms.
  As in the case of the renormalization equation $\Theta(x) = \beta^{i}\phi_{i}$ we can store such contact terms in an expansion similar to (\ref{RGop}). Since we have local 
  vector fields involved, we should introduce sources $\lambda^{a}_{ \mu}(x)$ for them so that 
  \be \label{vect_s}
  \frac{\delta \ln Z}{\delta \lambda^{a}_{\mu}(x)} = \langle J_{a}^{\mu}(x) \rangle 
  \ee
  where $Z$ now stands for a generating functional of correlators involving $\Theta$, $\phi_{i}$, and $J_{a}^{\mu}$ (see also \cite{Nakayama} for a recent discussion of such sources). 
  Note that to get a correlator involving $J_{a}^{\mu}$, we vary with respect to $\lambda^{a}_{\mu}$ as in (\ref{vect_s}) and then, after all variational derivatives are taken, we set 
  $\lambda^{i}$ to constants and $\lambda^{a}_{\mu}$ to zero.  In addition to the derivatives of $\lambda^{i}$ and $\mu(x)$, the vector sources themselves can be present both in 
  the expansion in (\ref{RGop}) and in (\ref{red0}).

  Assuming the currents $J_{a}^{\mu} $, derivatives $\partial_{\nu}$ and the sources $\lambda^{a}_{\mu}$ have engineering scaling dimension one, we can write out all possible "anomaly"  terms in  (\ref{red0}) 
  up to the second order in this dimension:
  \be \label{red_op1}
  \partial_{\mu} J^{\mu}_{a}(x) - r^{ i}_{a}(\lambda ) \phi_{i}(x) = -{ R}_{a}^{(0)} (x) - { R}^{(1)}_{a}(x) \, 
  \ee
  where 
  \bea \label{red_op2}
   { R}_{a}^{(0)}(x) &=&   r^{b}_{ai}(\lambda) \partial_{\mu}\lambda^{i} J^{\mu}_{b}(x) + \Gamma_{ba}^{c}(\lambda) \lambda^{b}_{\mu}J^{\mu}_{c}(x) \, , \nonumber \\
   { R}_{a}^{(1)}(x) &=& k_{a}\mu^{2}R_{2}(x)  + \partial^{\mu}[ k_{ai} \partial_{\mu} \lambda^{i}(x) + k_{ab} \lambda_{\mu}^{b}(x) ]  \nonumber \\
  &&  + \mu^2 g^{\mu \nu}\Bigl[ \frac{k_{abc} }{2}\lambda_{\mu}^{b} \lambda_{\nu}^{c}(x) + k_{aib}\partial_{\mu}\lambda^{i}\lambda_{\nu}^{b}(x) + \frac{k_{aij} }{2}\partial_{\mu}\lambda^{i} 
  \partial_{\nu}\lambda^{j}(x) \Bigr] \, . 
    \eea
  % Here we grouped the terms in the "redundancy anomaly" in two separate groups    arranged according to explicit powers of $\mu^2$ involved. 
  In the vicinity of a fixed point QFT, the engineering dimension is preserved perturbatively to all orders, and if we only study perturbation theory, the expansion formulas (\ref{red_op1}) and (\ref{red_op2}) are exact.
  In this case the terms $k_{a}$, $k_{ai}$, $k_{ab}$, $k_{abc}$, $k_{aib}$, $k_{aij}$ are all functions of $\lambda^{i}$ proportional to the identity operators. (In the sigma model 
  context loose power counting applies and these terms will have a nontrivial operator content.) Thus the terms in  ${ R}_{a}^{(1)}(x)$ are all proportional to the identity operator, while the 
  terms in   ${ R}_{a}^{(0)}(x)$ have a nontrivial operator content.  
      
  For conserved currents the coefficients $r_{a}^{i}$ vanish and the terms on the right hand side of (\ref{red_op1}) measure various  anomalies in the conservation law. We will express some of these terms 
 in terms of the OPE coefficients in a current algebra  in section \ref{sec:anomalyFP}. The parallel between (\ref{RGop}) and  (\ref{red_op1})  becomes even closer if we notice that $\Theta(x)$ is the divergence of the dilation current. 
  The beta functions $\beta^{i}$ then play a similar role to the coefficients $r_{a}^{i}$.

  The operator expressions $R_{a}^{(0)}(x)$ and $R_{a}^{(1)}(x)$ give rise to functional differential operators ${\cal R}^{(0,1)}(x)$. One can use them to calculate various contact terms in  correlation 
  functions by commuting  them with variational derivatives. For illustration and for later reference we  calculate  
  \bea \label{red_cor1}
   &&r^{j}_{a}  \langle \phi_{j}(x) J^{\nu}_{b}(y) \phi_{i}(z)\rangle_{c} = \langle \partial_{\mu} J^{\mu}_{a}(x) J^{\nu}_{b}(y) \phi_{i}(z)\rangle_{c}   + r_{ai}^{c} \partial_{\mu}\delta(x-z) \langle J^{\mu}_{c}(x) J^{\nu}_{b}(y)\rangle
  \nonumber \\ 
  && - \partial_{i} r^{j}_{a} \delta(x-z) \langle \phi_{j}(x) J^{\nu}_{b}(y)\rangle + \Gamma_{ba}^{c} \delta(x-y) \langle J_{c}^{\nu}(x) \phi_{i}(z)\rangle
  \nonumber \\ 
  && + k_{aib} \partial^{\nu}\delta(x-z) \delta(x-y)  + \partial_{i}k_{ab} \delta(x-z)\partial^{\nu}\delta(x-y) \, , 
  \eea
  \bea\label{redcor2}
 && r^{k}_{a}  \langle \phi_{k}(x) \phi_{i}(y) \phi_{j}(z)\rangle_{c} = \langle \partial_{\mu} J^{\mu}_{a}(x) \phi_{i}(y) \phi_{j}(z)\rangle_{c}   + \partial_{\mu}\delta(x-y) r_{ai}^{b}
  \langle J_{b}^{\mu}(y) \phi_{j}(z)\rangle  \nonumber \\
   && + \partial_{\mu}\delta(x-z) r_{aj}^{b}
  \langle \phi_{i}(y) J_{b}^{\mu}(z) \rangle - \partial_{i}r_{a}^{k} \delta(x-y)\langle \phi_{k}(x)\phi_{j}(z) \rangle_{c} - \partial_{j}r_{a}^{k} \delta(x-z)\langle \phi_{i}(y)\phi_{k}(z) \rangle_{c} \nonumber \\
  &&     + \Delta  \delta(x-y)  \delta(x-z) \partial_{j}k_{ai}   + \Delta \delta(x-z)  \delta(x-y) \partial_{i}k_{aj}   + k_{aij}\partial_{\mu}\delta(x-y)\partial^{\mu}\delta(x-z) \, , 
   \eea
   \bea \label{redcor3}
   &&r^{j}_{a}  \langle \phi_{j}(x) J^{\mu}_{b}(y) J^{\nu}_{c}(z)\rangle_{c}= \langle \partial_{\alpha} J^{\alpha}_{a}(x) J^{\mu}_{b}(y) J^{\nu}_{c}(z)\rangle_{c}    + 
   \Gamma^{d}_{ba}\delta(x-y) \langle J^{\mu}_{d}(x)J^{\nu}_{c}(z)\rangle \nonumber \\
   && +   \Gamma^{d}_{ca}\delta(x-z) \langle J^{\mu}_{b}(y)J^{\nu}_{d}(z)\rangle + \delta(x-y)\delta(x-z)g^{\mu \nu} k_{abc}
   \eea
 where $\Delta = \partial_{\mu}\partial^{\mu}$. 
 
 Integrating equation (\ref{red_cor1}) over $x$, we obtain the following identity 
 \bea
  r^{j}_{a}\partial_{j} \langle J_{b}^{\nu}(y)\phi_{i}(z)\rangle& =& -\partial_{i} r_{a}^{j} \langle J^{\nu}(y) \phi_{j}(z) \rangle - r_{ai}^{c}\langle J^{\nu}_{b}(y)\partial_{\mu}J^{\mu}_{c}(z)\rangle \nonumber \\
&& +\Gamma^{c}_{ba} \langle J_{c}^{\nu}(y)\phi_{i}(z)\rangle + (\partial_{i}k_{ab} - k_{aib})\partial^{\nu}\delta(z-y) \, . 
 \eea
  For finite separation $|y-z|>0$, using (\ref{red0}) we can rewrite the last equation as 
  \be \label{red_connection}
    r^{j}_{a}\partial_{j} \langle J_{b}^{\nu}(y)\phi_{i}(z)\rangle =  \langle J^{\nu}_{b}(y) \, \Gamma_{ia}^{j} \phi_{j}(z) \rangle  +\Gamma^{c}_{ba} \langle J_{c}^{\nu}(y)\phi_{i}(z)\rangle 
  \ee
  where we defined 
  \be\label{Gamma_a}
  \Gamma^{j}_{ia} =  -\partial_{i} r_{a}^{j} -r_{ai}^{c}r_{c}^{j} \, . 
    \ee
    Equation (\ref{red_connection}) easily generalizes to any multi-point  correlator of the fundamental scalar and vector fields inserted at finite separations (so that various contact terms drop out). 
    This means that differentiating a correlator along a redundant direction merely results in field redefinitions given by connection coefficients $\Gamma_{ab}^{c}$ and $\Gamma_{ia}^{j}$. 
    
    As for the renormalization anomaly, we can represent the anomaly equation (\ref{red_op1}) in terms of  functional derivative operators:    
    \be
   {\cal R}_{a}(x) \equiv  \partial^{\mu} \frac{\delta}{\delta \lambda_{\mu}^{a}(x)} - r^{ i}_{a}(\lambda ) \frac{\delta}{\delta \lambda^{i}(x)}   + {\cal  R}_{a}^{(0)} (x) + { \cal R}^{(1)}_{a}(x) \, . 
        \ee
    We can then write out the Wess-Zumino consistency conditions as 
   \be\label{WZred}
   [{\cal R}_{a}(x), {\cal R}_{b}(y)] = 0 \, . 
   \ee 
    This results in a number of equations on the redundancy anomaly coefficients which can be interpreted in geometrical terms. In particular these equations include the zero curvature conditions on the 
    connection defined in (\ref{red_connection}). In this paper we are not going to explore these equations. Their detailed analysis will appear in \cite{FK_redundancy}.   
    
  The renormalization anomaly (\ref{RGop}) similarly generalizes to 
 \be\label{newD1}
 \Theta(x) - \beta^{i}\phi_{i}(x) = { D}^{(0)}(x) + { D}^{(1)}(x)
 \ee 
  where\footnote{Note that in (3.14) as well as in (3.4) we are assuming \emph{parity conservation}. This excludes terms in the anomaly containing the anti-symmetric 2-tensor $\epsilon_{\mu\nu}$ (we would like to thank H. Osborn for pointing this out to us). Since such terms would enter only into the contributions $R_a^{(1)}$ and $D^{(1)}$, the equations (3.18) -- (3.21) hold also for parity violating theories.} 
  \bea \label{newD2}
  { D}^{(0)}(x) &=& \partial_{\mu} \lambda^{i} v_{i}^{a}(\lambda)J_{a}^{\mu}(x) + \lambda_{\mu}^{a}\gamma_{a}^{b}(\lambda) J_{b}^{\mu}(x) \, , \nonumber \\
  { D}^{(1)}(x) & = & \frac{C}{2}\mu^{2}R_{2}(x) +
 \partial_{\mu}[\mu^2g^{\mu\nu}(  W_{i}\partial_{\nu} \lambda^{i}(x)    + w_{a}\lambda^{a}_{\nu}(x) ] \nonumber \\
&& + \mu^2 g^{\mu\nu}  \Bigl[ \frac{1}{2}G_{ij} \partial_{\mu} \lambda^{i}\partial_{\nu}\lambda^{j}(x)  + g_{aj}\lambda_{\mu}^{a}\partial_{\nu}\lambda^{j}(x) + 
g_{ab} \frac{1}{2} \lambda_{\mu}^{a}\lambda^{b}_{\nu}(x)  \Bigr]  
 \eea
  Here we introduced coefficients $v_{i}^{a}$ so that $J_{i}^{\mu}(x)= v_{i}^{a}J_{a}^{\mu}(x)$. When strict power counting applies, the terms $v_{i}^{a}$, $\gamma_{a}^{b}$, 
  $C$, $W_{i}$, $w_{a}$, $G_{ij}$, $g_{aj}$, $g_{ab}$ are all functions of $\lambda$, while in the sigma model situation they can have a nontrivial operator content. 
  
  The Callan-Symanzik equation for correlators (at finite separation) involving the fundamental scalar and vector fields has the form
  \begin{gather}
  \begin{aligned}
   \label{CS3}
&  \Bigl( \mu \frac{\partial}{\partial \mu} - \beta^{i}\frac{\partial}{\partial \lambda^{i}}  \Bigr) \langle \phi_{i_{1}}(x_1) \dots J_{a_{1}}^{\mu_{1}}(y_1) \dots \Theta(z_{1}) \dots \rangle_{c} =  \\
 & \langle \Gamma \phi_{i_{1}}(x_1)  \dots  J_{a_{1}}^{\mu_{1}}(y_1) \dots \Theta(y_{1}) \dots \rangle_{c} + 
 \langle  \phi_{i_{1}}(x_1)  \dots \gamma_{a_{1}}^{b}J_{b}^{\mu_{1}}(y_1) \dots \Theta(y_{1}) \dots \rangle_{c} + \dots 
 \end{aligned}
 \end{gather}%
 where $\Gamma$ is defined in (\ref{GammaCS}) and $\gamma_{a}^{b}$ is the anomalous dimension matrix for vector fields. (It coincides   with the matrix  $\gamma_{a}^{b}$ appearing in $D^{(0)}(x)$.)
 
   In addition to the Wess-Zumino consistency conditions (\ref{WZ}) and (\ref{WZred}), there are Wess-Zumino conditions involving the commutators of the renormalization anomaly with the 
   redundancy anomaly:
   \be\label{RD}
   [ \mu(x) \frac{\delta}{\delta \mu(x)} - \beta^{i}(\lambda(x))\frac{\delta}{\delta \lambda^{i}(x)} - {\cal D}(x)   ,{\cal R}_{a}(y)] = 0 
   \ee 
   where ${\cal D}(x)= {\cal D}^{(0)}(x) + {\cal  D}^{(1)}(x)$  are the functional differential operators corresponding to 
    (\ref{newD1}) and (\ref{newD2}).  By direct inspection we find that  the terms in (\ref{RD}) containing ${\cal R}^{(0)}_{a}$ and ${\cal D}^{(0)}$ give rise to separate equations. We find 
    \bea
  && [ \mu \frac{\delta}{\delta \mu(x)} - \beta^{i}\frac{\delta}{\delta \lambda^{i}(x)} - {\cal D}^{(0)}(x)   ,   \partial^{\mu} \frac{\delta}{\delta \lambda_{\mu}^{a}(y)} - r^{ j}_{a}(\lambda ) \frac{\delta}{\delta \lambda^{j}(y)}   + {\cal  R}_{a}^{(0)} (y)  ] 
  \nonumber \\ 
  && = \delta(x-y) \Bigl[  (\beta^{j}\partial_{j}r^{i}_{a} - r^{j}_{a}\partial_{j}\beta^{i})\frac{\delta}{\delta \lambda^{i}(x)} + \Bigl(  (- \beta^{j} \partial_{j}r_{ai}^{b} - r_{a}^{j}\partial_{j}v^{b}_{i}   - \Gamma_{ca}^{b}v_{i}^{c}  -
r_{ai}^{c}\gamma_{c}^{b} )\partial_{\mu}\lambda^{i}
    \nonumber \\
  &&    
  - (r_{a}^{i}\partial_{i}\gamma_{c}^{b} + \beta^{i}\partial_{i}\Gamma_{ca}^{b} + \Gamma_{da}^{b}\gamma^{d}_{c} - \Gamma_{ca}^{d}\gamma_{d}^{b}) \lambda_{\mu}^{c}  \Bigr)\frac{\delta}{\delta \lambda^{b}_{\mu}(x)}     \Bigr]   
  \nonumber \\
  && + \Delta \delta(x-y)\Bigl[ \gamma_{a}^{b} - \beta^{i}r_{ai}^{b} + r^{i}_{a}v_{i}^{b} \Bigr]\frac{\delta}{\delta \lambda^{b}_{\mu}(x)} \, . 
     \eea
Setting this expression to zero gives rise to four separate equations: 
   \be \label{WZ1}
 \gamma_{a}^{b}=-r_{a}^{i}v_{i}^{b} + r_{ai}^{b}\beta^{i}\, , 
 \ee  
 \be\label{WZ2}
 \beta^{j}\partial_{j}r^{i}_{a} - r_{a}^{j}\partial_{j}\beta^{i} = -\beta^{j}r_{aj}^{b}r^{i}_{b}\, , 
 \ee
   \be\label{WZ3}
   \beta^{j}(r_{aj}^{c}r_{ci}^{b} + \partial_{j}r_{ai}^{b} - \partial_{i} r_{aj}^{b}  ) + r_{a}^{j}\partial_{j} v_{i}^{b} +v_{j}^{b}\partial_{i}r^{j}_{a} + \Gamma_{ca}^{b}v_{i}^{c} - r_{ai}^{c}\gamma_{c}^{b} =0\, , 
   \ee
   \be\label{WZ4}
   \beta^{j}( r_{aj}^{c}\Gamma_{dc}^{b} + \partial_{j}\Gamma^{b}_{da}) + \gamma_{d}^{c}\Gamma_{ca}^{b} - \gamma_{c}^{b}\Gamma_{da}^{c} + r^{i}_{a}\partial_{i}\gamma_{d}^{b}= 0 \, . 
   \ee
   Here, to separate the equations, we used the redundancy equation (\ref{red_op1}) again. 
   
   The meaning of the first two of the above equations is quite transparent. Equation (\ref{WZ1}) expresses the anomalous dimensions of the currents through the terms in the anomaly related to the scalar field. 
   This relation stems from the fact that the divergence of a current, which has the same anomalous dimension, is expressible according to (\ref{red0}) via scalar operators. Equation (\ref{WZ2}) can be rewritten in terms of a
   commutator of vector fields acting on the space of couplings,
   \be \label{Rbeta_comm}
   [ \hat \beta , \hat R_{a} ] = - \beta^{j} r_{aj}^{b}  \hat R_{b}
   \ee   
   where 
   \be
   \hat \beta = \beta^{i}\partial_{i} \, , \qquad \hat R_{a} = r_{a}^{j}\partial_{j} \, . 
   \ee
    Equation (\ref{Rbeta_comm}) shows that the commutator of the beta function vector field $\hat \beta$ with the redundancy vector fields $\hat R_{a}$ closes again on the redundancy vector fields. 
   This condition is crucial for the reduction of the RG  flow onto the quotient space in which we identify points on the orbits  generated by the redundancy vector fields. 
   In the present paper we are not going to explore the meaning of equations (\ref{WZ3}) and (\ref{WZ4}) nor any of the other equations following from (\ref{RD}). Equations (\ref{WZ1}) and (\ref{WZ2}) 
   will be checked by explicit calculations in conformal perturbation theory in  sections \ref{sec:red_beta} and \ref{sec:anom_curr}. 

By taking two variational derivatives with respect to the scale $\mu(x)$ we obtain from (\ref{red_op2}) 
\be \label{TTd}
\langle \Theta(x) \Theta(y) \partial_{\mu} J_{a}^{\mu}(z)\rangle_{c} = \langle \Theta(x) \Theta(y) r^{i}_{a}\phi_{i}(z)\rangle_{c}
\ee
where both sides are distributions. The only contact term between $\Theta$ and the redundancy operation for $J_{a}^{\mu}$ is in the term proportional to $\kappa_{a}$ in (\ref{red_op2}), 
which goes away when we consider a 3-point connected correlator in  (\ref{TTd}). Integrating both sides of (\ref{TTd}) over $z$ we obtain 
\be \label{red_c0}
r_{a}^{i}\partial_{i} \langle \Theta(x) \Theta(y)\rangle = 0 \, , 
\ee
which holds at the level of distributions. The Zamolodchikov $c$-function (\ref{eq:corig}) can also be  written as 
\be \label{c2}
c = - 3\pi \int\!\!d^2 x\,  x^2 \theta(1- \Lambda |x|) \langle\Theta(x) \Theta(0)\rangle_{c}
\ee
where one integrates a distributional 2-point function. Thus (\ref{red_c0}) implies\footnote{This holds when strict power counting applies. With loose power counting, the term $\kappa_{a}$ in 
(\ref{red_op2}) may have nontrivial operator content, and moving along a redundant direction may result in shifting $\Theta(x)$ by a Laplacian of $\kappa_{a}(x)$ - an improvement current.} 
\be \label{red_c}
r_{a}^{i}\partial_{i} c = 0 \, , 
\ee
i.e.\ the $c$-function is  independent of the redundant couplings.

\section{General conformal perturbation analysis}\label{general_cp}
\setcounter{equation}{0}

We will analyze a 2d Euclidean, unitary conformal field theory  with current symmetry algebras, perturbed by dimension $2$ spin $0$ operators $\phi_i$. The 
Euclidean action perturbation is 
\begin{equation} \label{deltaS}
\delta S=\sum_i\int d^2z \lambda^i\phi_i(z)\, . 
\end{equation} 
Here $z=x + iy$ is the complex coordinate on ${\mathbb R}^{2}$, and $d^{2}z = dxdy$ is the standard volume element. The fixed point theory does not have to come from a 
particular Euclidean action. The perturbation given by (\ref{deltaS}) merely says that the correlation functions in the perturbed theory are calculated according to the following formal 
perturbation theory expansion
 \bea
&&  \left\langle  [\cO_{a_1}] (z_1)\ldots [\cO_{a_n}](z_k)  \right\rangle_{\lambda} = \sum_{n=0}^{\infty} \frac{1}{n !}\sum_{i_1,\ldots,i_{n}}\lambda^{i_1}\ldots \lambda^{i_{n}} \nonumber \\ 
 && \times
 \int\!\! d^2x_1\ldots  \int\!\! d^2x_{n} \left\langle  \phi_{i_1}(x_1)\ldots\phi_{i_{n}}(x_{n})\cO_{a_1}(z_1)\ldots\cO_{a_n}(z_k) \right\rangle_{0}\,.
  \eea
Here $\langle ... \rangle_{\lambda}$ denotes the correlator in the perturbed theory, while $\langle ... \rangle_{0}$ stands for the correlators at the $\lambda^{i}=0$ fixed point.
By default  correlators are assumed to be connected, though sometimes to emphasize this we will use the explicit notations $\langle ...\rangle_{0;c}\, $ and $\langle ... \rangle_{\lambda; 0}$. 
The operators $\cO_{a}(z)$ are local operators at the fixed point. The integrals on the right hand 
side are divergent, so some regularization and renormalization is assumed. The divergences coming from several $\phi_{i}$ insertions colliding away from the points $z_{i}$ in general result in 
nontrivial beta functions for the couplings $\lambda^i$,  
while collisions with the points $z_i$ are dealt with by counter terms that renormalize the operators  $\cO_{a}(z)$. On the left hand side, we denote by $[\cO_{a}](z)$ the renormalized operators of the deformed theory.
As standard in conformal perturbation theory, we label this renormalized operators by the unperturbed (bare) operators $\cO_{a}(z)$.  In explicit calculations below we will usually omit the square 
brackets as the role of the operators will be clear from the context. 

In terms of  concrete realizations of the perturbations considered in this section we have a large class of current-current perturbations of WZW models. Another, more general class is obtained 
by considering tensor products of WZW theories. Primary fields in each copy have rational conformal dimensions. We can consider perturbations by tensor products of such primaries that have 
total dimension 2 -- for example, take a WZW model with symmetry ${\rm SU}(2)_{3}\times {\rm SU(2)}_{1}$  perturbed by $\psi_{s=3/2}\otimes \psi_{s=1/2}$. 
In this paper, we present in section 5 three concrete current-current models for illustration of the general results developed in this section. 

\subsection{At the fixed point}

\subsubsection{OPE algebra}

Next we discuss the OPE algebra at the fixed point. The fixed point CFT we consider has  a  symmetry algebra generated by currents $J^a$ and  $\bar{J}^{\bar{a}}$ with levels $k_{L}$ and $k_{R}$. The currents have  the OPEs 
\begin{align}
\begin{split}\label{eq:JaJbOPE}
J_a(z_a)J_b(z_b)&= \frac{k_L\delta_{ab}}{(z_{ab})^2}+\frac{i\fF{a}{b}{c}J_c(z_b)}{(z_{ab})}+r.p.\\
\bar{J}_{\bar{a}}(\bar{z}_a)\bar{J}_{\bar{b}}(\bar{z}_b)
&= \frac{k_R\delta_{ab}}{(\bar{z}_{ab})^2}
+\frac{i\fFb{a}{b}{c}\bar{J}_{\bar{c}}(\bar{z}_b)}{(\bar{z}_{ab})}+r.p.
\end{split}
\end{align}
where $r.p.$ stands for the regular part of the OPEs, and where the structure constants 
\[
f_{abc}=\fF{a}{b}{d}\eta_{cd}\,,\quad \bar{f}_{\bar{a}\bar{b}\bar{c}}=\fFb{a}{b}{d}\bar{\eta}_{\bar{c}\bar{d}}
\]
are real and totally antisymmetric. We employ the Einstein summation convention throughout, using contractions with the metrics
\begin{equation}
\eta_{ab}=k_L\delta_{ab}\,,\quad \bar{\eta}_{\bar{a}\bar{b}}=k_R\delta_{\bar{a}\bar{b}}
\end{equation}
to raise and lower indices where necessary. In a generic theory, the holomorphic and anti-holomorphic chiral algebras could be of a different type, so in particular the levels $k_L$ and $k_R$ could be different. 
The perturbing operators $\phi_i$, which have dimension $2$ and spin $0$, possess the OPEs 
{\allowdisplaybreaks
\begin{align}
\begin{split}\label{eq:PhiIPhiJOPEs}
\phi_i(z_i)\phi_j(z_j)&= 
\frac{\delta_{ij}}{|z_{ij}|^4}
+\frac{i\oA^a_{ij}J_a(z_j)}{(z_{ij})(\bar{z}_{ij})^2}
+\frac{i\bar{\oA}^{\bar{a}}_{ij}\bar{J}_{\bar{a}}(z_j)}{(z_{ij})^2(\bar{z}_{ij})} \\
&\hphantom{\sim}+\frac{1}{|z_{ij}|^2} \oC_{ij}{}^k\phi_k(z_j)+\ldots+r.p.\;
\end{split}
\end{align}}%
where the ellipsis stands for  other singular terms. We assume that no  relevant spin zero fields  appear in  (\ref{eq:PhiIPhiJOPEs}) so that 
the omitted singular terms contain irrelevant scalar fields, fields of spin 1 with dimension 
larger than 1 and fields of spin 2 and 3. The precise form of the omitted singular terms  will not be important to us. 
The OPE structure tensors $\oA^a_{ij}$ and $\bar{\oA}^{\bar{a}}_{ij}$ are  antisymmetric under the exchange of $i$ and $j$, while the $\oC_{ijk}$ are totally symmetric in all indices. 
Note that the metric for the scalar operator indices is trivial: $\eta_{ij}=\delta_{ij}$.

The OPEs of currents $J^a$ and $\bar{J}^{\bar{a}}$ with perturbing operators $\phi_i$ in the unperturbed theory have the form
\begin{subequations}\label{eq:JphiOPEs}
\begin{align}
\begin{split}\label{eq:holJphiOPE}
J_a(z_a)\phi_i(z_i)&=
\frac{1}{(z_{ai})^2}\oB_{ai}{}^{\bar{b}}\bar{J}_{\bar{b}}(\bar{z}_i)
+\frac{i}{(z_{ai})}\oA_{ai}{}^{j}\phi_j(z_i)
+\frac{i}{(z_{ai})}\oA_{ai}{}^{\tilde{j}}\chi_{\tilde{j}}(z_i)+r.p.
\end{split}\\
\begin{split}\label{eq:anti-holJphiOPE}
\bar{J}_{\bar{a}}(\bar{z}_a)\phi_i(z_i)&=
\frac{1}{(\bar{z}_{ai})^2}\bar{\oB}_{\bar{a}i}{}^{b}J_b(z_i)
+\frac{i}{(\bar{z}_{ai})}\bar{\oA}_{\bar{a}i}{}^{j}\phi_j(z_i)
+\frac{i}{(\bar{z}_{ai})}\bar{\oA}_{\bar{a}i}{}^{\tilde{j}}\chi_{\tilde{j}}(z_i)+r.p.\,.
\end{split}
\end{align}
\end{subequations}%
Here, the operators $\chi_{\tilde{j}}$ together with the perturbing operators $\phi_i$ are assumed to form a complete orthonormal basis for the space of dimension $2$ spin zero operators. For later convenience, we introduce the notation
\begin{equation}\label{eq:fullBasis}
\{\Phi_{I}\}=\{\phi_i\}\cup\{\chi_{\tilde{j}}\}
\end{equation}
for the full basis of dimension $2$ spin $0$ operators. 
The OPEs (\ref{eq:PhiIPhiJOPEs}) and (\ref {eq:JphiOPEs}) are then extended to include operators $\chi_{\tilde{j}}$ with the OPE coefficients denoted the same way but with the tilted indices.

Since the leading order $\beta$-functions of the perturbed theory are proportional to the OPE coefficients,
\[
\beta^i= \pi \oC^{i}{}_{jk}\lambda^j\lambda^k+\OL{3}\,,
\]
renormalizability of the perturbed theory demands OPE closure of the set of perturbing fields $\phi_i$:
\begin{equation}
\oC_{\tilde{i}jk}=0\quad \forall \; \tilde{i},j,k\,.
\end{equation}

The OPE coefficients in (\ref{eq:PhiIPhiJOPEs}) and (\ref {eq:JphiOPEs}) satisfy some identities stemming from the Ward identities for correlators. We denote the charges 
corresponding to the currents $J_{a}$ and $\bar J_{\bar b}$ as 
%\footnote{The signs in the definition of the charges may be motivated from the complex Stokes theorem:
%\[
%\int_{\cU} d^2z\; \left\lbrace (\bar{\partial}J_a)(z)+(\partial\bar{J}_{\bar{a}})(z)\right\rbrace\ldots
%=\frac{1}{2i}\oint_{\partial\cU}dz J_a(z)\ldots-\frac{1}{2i}\oint_{\partial\cU}d\bar{z} \bar{J}_{\bar{a}}(\bar{z})\ldots\,.
%\]
%The additional factor of $\tfrac{1}{\pi}$ is conventional.}
\begin{equation}
Q_a=\frac{1}{2\pi i}\oint dx\; J_a(x)\,,\quad \bar{Q}_{\bar{b}}=-\frac{1}{2\pi i}\oint d\bar{x}\; \bar{J}_{\bar{b}}(\bar{x})\, . 
\end{equation}
The action of the charges $Q_{a}$ on a local operator $\Phi_{I}$ reads
\be
Q_a \Phi_{I}(y) =\frac{1}{2\pi i}\oint_{\cC_y} dx\; J_a(x)\Phi_{I}(y) = 
i A_{a I}^{J} \Phi_{J}(y)  \,,  
\ee 
and analogously 
\be 
\bar{Q}_{\bar{b}}\Phi_{I}(y) =i \bar A_{\bar b I}^{J} \Phi_{J}(y) \, .
\ee 
The  Ward identities for the $n$-point functions of operators  $\Phi_I$ read 
\begin{equation}\label{eq:chiralWardIdentity}
Q_a(\langle \Phi_{I_1}(z_1)\ldots \Phi_{I_n}(z_n)\rangle_0)=
\sum_{\ell=1}^n i\oA_{aI_{\ell}}{}^R \langle \Phi_{I_1}(z_1)\ldots \Phi_{R}(z_{\ell})\ldots \Phi_{I_n}(z_n)\rangle_0=0 \, . 
\end{equation}
Specializing this identity to 3-point functions we obtain a relation
\begin{align}\label{eq:ACeqn}
%Q_a(\langle \phi_i(z_i)\phi_j(z_j)\rangle_0)&\propto \left(\oA^a_{ij}+\oA^a_{ji}\right)=0\\
\oA_{aI}{}^R\oC_{RJK}+\oA_{aJ}{}^R\oC_{RKI}+\oA_{aK}{}^R\oC_{RIJ}=0\; ,
\end{align} 
which means that the structure constants $C_{IJK}$ form an invariant tensor under the symmetry algebra. 
Since the holomorphic and anti-holomorphic current algebras commute, so do the corresponding charges, hence we have 
\begin{equation}
\oA_{aI}{}^{R}\bar{\oA}_{\bar{b}RJ}=\oA_{aJ}{}^R\bar{\oA}_{\bar{b}RI}\,.
\end{equation}
A relation of a different kind is obtained by calculating 
\be 
\langle  J_a \bar{J}_{\bar{b}} (\infty)\Phi_I(x)\Phi_J(y)\rangle_0
=-\frac{\oA_{aI}{}^{R}\bar{\oA}_{\bar{b}RJ}}{|x-y|^2}= \frac{\oC_{IJ}{}^R\bar{\oB}_{\bar{b}R}^{c}\eta_{ac}}{|x-y|^2}
\ee
where the first equality is obtained by using a Ward identity, while the second equality is obtained by taking the OPE of $\Phi_{I}$ with $\Phi_{J}$. 
Thus we have an identity
\begin{equation}\label{eq:AABCrels}
\oA_{aI}{}^{R}\bar{\oA}_{\bar{b}RJ}=\oA_{aJ}{}^R\bar{\oA}_{\bar{b}RI}
=-\oC_{IJ}{}^R\oB_{aR\bar{b}}
=-\oC_{IJ}{}^R\bar{\oB}_{\bar{b}Ra}\,.
\end{equation}

\subsubsection{Anomaly terms for conserved currents}\label{sec:anomalyFP}
  Here we  explore the anomaly terms in   (\ref{red_op1}) at the fixed point. In this case  we have conserved currents and equations (\ref{red_cor1}), (\ref{redcor2}), (\ref{redcor3}) read 
   \bea\label{r1}
   -\langle \partial_{\mu} J^{\mu}_{a}(x) J^{\nu}_{b}(y) \phi_{i}(z)\rangle_{c}   &= &r_{ai}^{c} \partial_{\mu}\delta(x-z) \langle J^{\mu}_{c}(x) J^{\nu}_{b}(y)\rangle   + k_{aib} \partial^{\nu}\delta(x-z) \delta(x-y) 
   \nonumber \\
 && + \partial_{i}k_{ab} \delta(x-z)\partial^{\nu}\delta(x-y) \, ,   
  \eea
  \bea \label{r2}
     \langle \partial_{\mu} J^{\mu}_{a}(x) \phi_{i}(y) \phi_{j}(z)\rangle_{c}  & =&   \partial_{i}r_{a}^{k} \delta(x-y)\langle \phi_{k}(x)\phi_{j}(z) \rangle_{c} + \partial_{j}r_{a}^{k} \delta(x-z)\langle \phi_{i}(y)\phi_{k}(z) \rangle_{c} \nonumber \\
  &&     - \Delta  \delta(x-y)  \delta(x-z) \partial_{j}k_{ai}   - \Delta \delta(x-z)  \delta(x-y) \partial_{i}k_{aj}   \nonumber \\
  && - k_{aij}\partial_{\mu}\delta(x-y)\partial^{\mu}\delta(x-z) \, ,
  \eea
  \bea\label{r3}
    -\langle \partial_{\alpha} J^{\alpha}_{a}(x) J^{\mu}_{b}(y) J^{\nu}_{c}(z)\rangle_{c}    &=&
   \Gamma^{d}_{ba}\delta(x-y) \langle J^{\mu}_{d}(x)J^{\nu}_{c}(z)\rangle +   \Gamma^{d}_{ca}\delta(x-z) \langle J^{\mu}_{b}(y)J^{\nu}_{d}(z)\rangle + 
    \nonumber \\
   &&  \delta(x-y)\delta(x-z)g^{\mu \nu} k_{abc}\, .
   \eea  
  Here,  we did set $r_{a}^{i}|_{\lambda^{i}=0}=0$ but we kept  $\partial_{i}r_{a}^{j}=\partial_{i}r_{a}^{j}|_{\lambda^i=0}$ which give the charge matrices of the fields $\phi_{i}$.
  As we are considering here a current algebra in  conformal field theory, it is convenient to use complex coordinates. The currents $J^{\mu}_{a}$ are then replaced by the $(1,0)$ and 
  $(0,1)$ conformal fields $J_{a}(z)$ and $\bar J_{\bar b}(\bar z)$. We thus switch to using the homomorphic and antiholomorphic  labels $a, \bar b$. 
  
  The contact terms in (\ref{r1})-(\ref{r3}) depend on the regularization scheme chosen. If the left and right current algebras are isomorphic, one can choose a gauge invariant regularization. 
  More generally, any local prescription of contact terms can be chosen. 
  Thus the coefficients of the double contact terms in (\ref{r1})-(\ref{r3}),  $\partial_{i}r_{a}^{j}$, $r_{ai}^c$ and $\Gamma_{ab}^{c}$, can be obtained (prescribed) by taking  distributional derivatives of the OPE's 
  (\ref{eq:JaJbOPE}) and (\ref{eq:holJphiOPE}). 
  For example, using 
  $$
  \partial_{\bar z} \frac{1}{z-w} = \pi \delta(z-w) 
  $$  
  and differentiating 
  $$
  J_{a}(z) \phi_{i}(w,\bar w) \sim \frac{1}{(z-w)^2} B_{ai}{}^{\bar b} \bar J_{\bar b}(\bar w) + \dots\,,
  $$
  we obtain
  $$
   \partial_{\bar z}  J_{a}(z) \phi_{i}(w,\bar w) \sim -2\pi \partial_{z} \delta(z-w)B_{ai}{}^{\bar b} \bar J^{\bar b}(\bar w) + \dots \, ,
  $$
  hence in  this prescription we have
  \be \label{rabj}
  r_{ai}{}^{\bar b} = \pi B_{ai}{}^{\bar b} \, , \quad  r_{\bar b i}{}^{ a} = \pi B_{\bar b i}{}^{a} \, , 
  \quad r_{a i}{}^{b}=r_{\bar a i}{}^{\bar b}= 0\, .  
      \ee
      Similarly, we find 
\be
      \partial_{i}r_{a}{}^{k}=i\pi A_{ai}{}^{k}\, , \qquad \partial_{i}r_{\bar b}{}^{k}= i\pi A_{\bar bi}{}^{k}\, , 
\ee
\be
\kappa_{ab} = \pi \eta_{ab} = \pi k_{L}\delta_{ab} \, , \quad \kappa_{\bar a\bar b} = \pi \bar \eta_{\bar a\bar b} = \pi k_{R}\delta_{\bar a\bar b} \, , \quad 
\kappa_{a\bar b} = \kappa_{\bar b a} = 0\, , 
\ee
\be
\Gamma_{ab}{}^c = i \pi f_{ab}{}^{c} \, , \qquad \Gamma_{\bar a\bar b}{}^{\bar c} = i \pi f_{\bar a\bar b}{}^{\bar c}\, , 
\ee            
   and all components of $\Gamma$ which contain both holomorphic and antiholomorphic indices vanish.    
   
   We also note that the coefficients $k_{a}$ in  (\ref{red_op1}) give  background charges  (mixed anomalies).

%%%%%%%%%%%%%%%%%%%%%%%%%%%%%%%%%%%%%%%%%%%%%%%%%%%%%%%%%%%%%%%%%%%%%%%%%%%%%%%%%%%%%%%
\subsection{Away from the fixed point}
\subsubsection{Beta functions}\label{sec:betaFunctionGeneral}
The perturbative beta functions for the couplings $\lambda^{i}$ in (\ref{deltaS}) have the form 
\begin{equation}\label{eq:beta:components}
\beta^{i}(\lambda)=\sum_{\ell>1}\beta^{i}_{(\ell)}(\lambda)\,,\quad 
\beta^{i}_{(\ell)}(\lambda)=\beta^i_{j_1\ldots j_{\ell}}\lambda^{j_1}\ldots\lambda^{j_{\ell}}
\end{equation}
where the leading terms are well known: 
\begin{equation}\label{betaLO}
\beta^i_{jk}=\pi\oC_{jk}{}^{i}\; .
\end{equation}
These terms are scheme independent. 
To calculate the two loop contributions $\beta^{i}_{(3)}$, we will follow the method of ~\cite{Gaberdiel:2008fn} which is reviewed in Appendix A. The approach of ~\cite{Gaberdiel:2008fn}
uses a sharp position space cutoff (point splitting) and gives a recursion formula for calculating the beta function coefficients. We specialize this method to the case of 
perturbing operators having dimension 2. This allows one to use conformal invariance to obtain an especially compact formula for the two loop coefficients as a single integral of 
four-point functions over the conformal cross-ratio. 
 We also pay particular attention to regularization in this integral, which is subtle due to the conditionally convergent terms coming from  the currents $J_{a}$, $\bar J_{\bar b}$.

Relegating all details  to appendix~\ref{app:betaNLOcomputations},  here we state the result 
\bea  
\beta^i_{jk\ell}=\frac{\pi}{3!}\lim\limits_{\frac{\epsilon}{L}\to0}
\biggr\{& 2\int_{U_{I}\cup U_{II} \cup U_{III}}d^2\eta\;
\langle \phi_i(\infty)\phi_j(0)\phi_k(1)\phi_{\ell}(\eta)\rangle_{0;c} \nonumber \\
&-2\pi\ln(L/\epsilon)\left(
\sum_{\text{perm}(j,k,\ell)}\oC_{i j}{}^{m}\oC_{mk\ell}\right) \biggr\}   \label{NLOgen1}
\eea
where the symbol $\sum_{\text{perm}(j,k,\ell)}$ stands for the sum over all permutations of the index set $\{j,k,\ell\}$. The regions of integration 
$U_{I}$, $U_{II}$, $U_{III}$ are as depicted in figure~\ref{fig:integrationRegionsAndDetails} below. 

\begin{figure}[h!]
        \centering
        \begin{subfigure}[b]{0.45\textwidth}
                \centering
		\includegraphics[width=\textwidth]{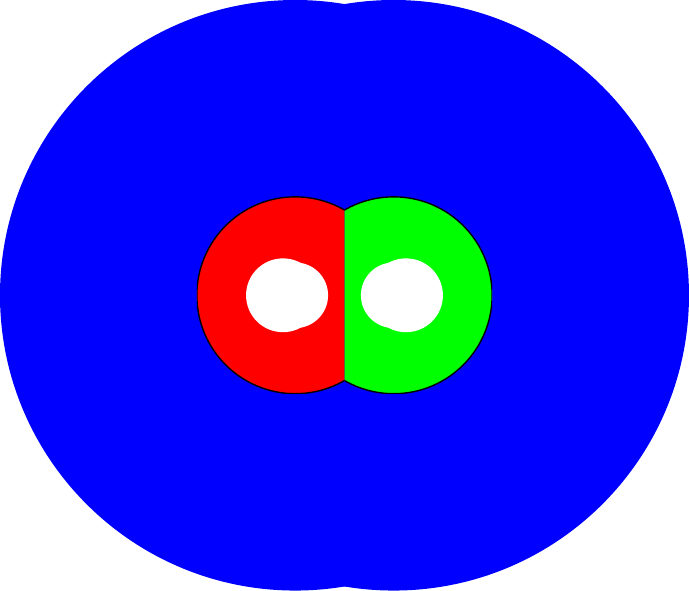}
                 \caption{Integration regions}
                \label{fig:integrationRegions}
        \end{subfigure}
	\qquad
	\begin{subfigure}[b]{0.45\textwidth}
              \centering
		\includegraphics[width=\textwidth]{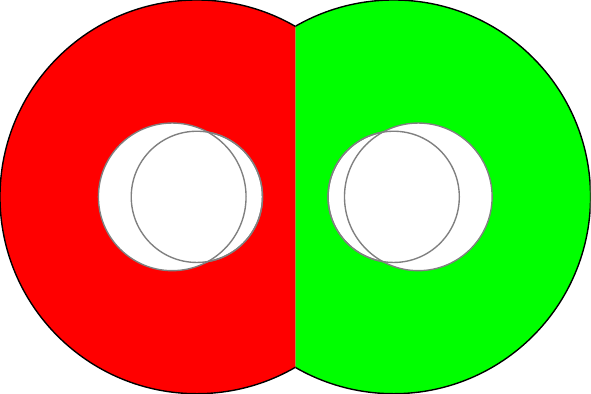}
		\caption{Details}
		\label{fig:integrationRegionDetails}
        \end{subfigure}%        
        \caption{The integration regions $U_I$ (blue), $U_{II}$ (green) and $U_{III}$ (red). See appendix~\ref{app:integrationRegionDescr} for the precise formulae defining the regions.}
        \label{fig:integrationRegionsAndDetails}
\end{figure}

The white regions around $\eta=0$ and $\eta=1$ that are zoomed in in part (b) of the picture look like deformed circles of the size of the order $\epsilon/L$. 
More precisely, they are constructed out of two arcs of slightly offset circles, see formulas~\eqref{eq:regionUII} and~\eqref{eq:regionUIII} in appendix~\ref{app:integrationRegionDescr}. These regions are cut out around the singularities of the four-point function. Analogously, the boundary of the blue region 
  is given by arcs of slightly offset circles with radii of the order $L/\epsilon$ and provides an infrared regulator near $\eta=\infty$. Fields of nonzero spin present  in the OPEs 
  of operators $\phi_{i}$, including dimension 1 currents,  may result in singularities which render  the integral to be only conditionally convergent around $\eta=0,1$ 
  and $\infty$. 
  Therefore even though in the limit $\epsilon/L\to 0$ the cut off regions look approximately like small (large at infinity) circles, the precise shape may be important in dealing with these 
  singularities.  We will argue shortly that this is not the case and for practical purposes one can use the circular regularization around $\eta=0,1$ and $\infty$. 
  The precise cutoff shapes however are instrumental  in establishing the general properties of the coefficients $\beta^{i}_{jk\ell}$.
  The three regions $U_{I}$, $U_{II}$, $U_{III}$ as well as the cut off regions (unlike  circles centered at the singularities) have the special property that they 
  are  mapped to each other by global conformal transformations permuting $\eta=0,1$ and $\infty$. Combining these mappings applied to the four point functions with an appropriate  change of the
  integration variable, we can rewrite formula (\ref{NLOgen1}) in terms of an integral over just one out of the three regions,
  \bea \label{beta2}
\beta^i_{jk\ell}=\frac{\pi}{3!}\lim\limits_{\frac{\epsilon}{L}\to0}\biggr\{& 
\int_{U_I} d^2\eta\; \sum_{\text{perm}(j,k,\ell)} 
\left\langle \phi_i(\infty)\phi_{j}(0)\phi_{k}(1)\phi_{\ell}(\eta)\right\rangle_{0;c} \nonumber \\
& -2\pi\ln(L/\epsilon)\left(
\sum_{\text{perm}(j,k,\ell)}\oC_{ij}{}^m\oC_{mk\ell}\right)
\biggr\}\:.
\eea  
The last equation proves efficient in explicit calculations since the integration region $U_I$ is comparatively easy to parameterize and one can use  Stokes theorem to calculate the integral. 

The resulting two loop coefficients $\beta^i_{jk\ell}$ are totally symmetric under the exchange of all four indices. While the symmetry under the permutation of the indices $j$, $k$ and $\ell$ is manifest in~\eqref{beta2}, 
there exist combinations of global conformal and coordinate transformations (cf.~appendix~\ref{app:permRules}) that permute the insertion point of $\phi_r(\infty)$ with any of the other insertion points 
 $0,1$ and $\eta$. 

Using the permutation  symmetry  we can argue for an alternative form of the regularization prescription -- cutting out circles around the singularities. The regions cut out around $\eta=0,1$ differ from round circles of radius $\epsilon/L$ by regions whose area is of the order of $(\epsilon/L)^3$, so that the only conditionally convergent singularities which are sensitive to the difference 
are those that come from dimension one currents. The OPE coefficients for those fields are antisymmetric and thus drop out from the gradient formula. At large values of $\eta$ the leading asymptotics 
of the order $1/\eta$, $1/\bar \eta$ comes again from dimension one currents. Only these terms are sensitive to the details of the infrared cutoff, but they also drop out under symmetrization. Thus we can substitute the infrared regulator by a round circle of radius $L/\epsilon$ centered at the origin. This gives us the following alternative 
representation  
 \bea \label{beta3}
 \beta^{i}_{jk\ell}= \frac{\pi}{4!}\lim\limits_{\epsilon \to 0}\sum_{\text{perm}(i, j,k,\ell)}
 \biggr\{& \frac{1}{3} \int\limits_{{\mathbb R}^{2}}d^2\eta\; \theta_{\epsilon}(\eta)
\langle \phi_i(\infty)\phi_j(0)\phi_k(1)\phi_{\ell}(\eta)\rangle_{0;c} \nonumber \\
&+ \left(
\oC_{i j}{}^{m}\oC_{mk\ell}\right)2\pi\ln\! \epsilon  \biggr\}
 \eea
where 
\be
\theta_{\epsilon}(\eta) = \theta(|\eta |-\epsilon)\theta(|1-\eta |-\epsilon)\theta(1-|\eta-1/2|/\epsilon) \, .
\ee

Formula (\ref{NLOgen1}), or its other representations  (\ref{beta2}), (\ref{beta3}),  give the two loop beta functions in the sharp cutoff followed by the minimal subtraction scheme. Any other renormalization scheme will result in a coupling 
constants redefinition. Under a redefinition of the form
\begin{equation}
\lambda^{\ell}\mapsto \tilde{\lambda}^{\ell}=\lambda^{\ell}+\sum_{i,j}c^{\ell}_{ij}\lambda^i\lambda^j+\sum_{i,j,k}f^{\ell}_{ijk}\lambda^i\lambda^j\lambda^k+\OL{4}\,,
\end{equation}
where w.l.o.g.\ the coefficient tensors $c^{\ell}_{ij}$ and $f^{\ell}_{ijk}$ are symmetric under the exchange of lower indices, the beta function transforms according to 
\begin{align}
\begin{split}\label{eq:betaRedef}
\tilde{\beta}^{\ell}&=\sum_{i,j}\beta^{\ell}_{ij}\tilde{\lambda}^i\tilde{\lambda}^j+\sum_{i,j,k}\beta^{\ell}_{ijk}\tilde{\lambda}^i\tilde{\lambda}^j\tilde{\lambda}^k\\
&+\frac{1}{3}\sum_{i,j,k}\tilde{\lambda}^i\tilde{\lambda}^j\tilde{\lambda}^k\sum_{\text{perm}(i,j,k)}
\bigg(
c^{\ell}_{mi}\beta^{m}_{jk}-c^{m}_{ij}\beta^{\ell}_{mk}
\bigg)+\OL{4}\,.
\end{split}
\end{align}%
We see that while the leading order coefficients are universal, the next-to-leading order coefficients generically get modified. 
However, it follows from (\ref{eq:betaRedef}) that  if we have two schemes such that in each one the coefficients $\beta^{\ell}_{ijk}$ are symmetric in all four indices\footnote{Recall that we normalize the fields $\phi_{i}$ so that the natural metric at fixed point is trivial and we can raise and lower indices trivially.}, then these coefficients are the same. 
In other words, we have a class of renormalization schemes within which the next-to-leading order coefficients are universal.

\subsubsection{Redundant operators  and redundancy vector fields}\label{sec:red1}
Since the operators $\phi_{i}$ that appear in (\ref{deltaS}) are in general charged under the current algebra, we 
expect broken  symmetries in the deformed theory. The corresponding currents are no longer conserved 
and we get a number of redundant operators. Recalling that the operators $\Phi_{I}$ introduced in \eqref{eq:fullBasis} by assumption form a complete basis of spin 0 dimension 2 operators, we have 
\be \label{redd}
 (\bar{\partial}J_a)_{\lambda}(x) = r_a{}^{I}(\lambda )\Phi_{I}(x) \, , \qquad  
 (\partial\bar{J}_{\bar{a}})_{\lambda}(x) =
\bar{r}_{\bar{a}}{}^{I}(\lambda ) \Phi_{I}(x)\, . 
 \ee
 The coefficients $r_a{}^{I}(\lambda )$, $\bar{r}_{\bar{a}}{}^{I}(\lambda )$ can be expanded as  
\begin{align*}
r_{a}{}^{I}(\lambda)=\sum_{\ell=1}^{\infty} r^{(\ell)}_{ai_1\ldots i_{\ell}}{}^{I}\lambda^{i_1}\ldots\lambda^{i_{\ell}}\,,\quad \bar{r}_{\bar{a}}{}^{I}(\lambda)=
\sum_{\ell=1}^{\infty} \bar{r}^{(\ell)}_{\bar{a}i_1\ldots i_{\ell}}{}^{I}\lambda^{i_1}\ldots\lambda^{i_{\ell}}\,.
\end{align*}
In this section we will calculate the coefficients $r^{(1)}$, $\bar r^{(1)}$, $r^{(2)}$ and $\bar r^{(2)}$ in terms of the OPE coefficients of the fixed point theory.
  The redundancy equations  (\ref{redd}) hold up to contact terms, 
 which at the leading order were calculated in section \ref{sec:anomalyFP}.  
 
The coefficients in expansion (\ref{redd})  can be computed from perturbed correlation functions using
\begin{align}
\begin{split}\label{eq:genDivergenceA}
\langle (\bar{\partial}J_a)(x)\Phi_{I}(y)\rangle_{\lambda}&=r_{a}{}^{J}(\lambda)\langle \Phi_{J}(x)\Phi_{I}(y)\rangle_{\lambda}\\
\langle (\partial\bar{J}_{\bar{a}})(x)\Phi_{I}(y)\rangle_{\lambda}&=\bar{r}_{\bar{a}}{}^{J}(\lambda)\langle \Phi_{J}(x)\Phi_{I}(y)\rangle_{\lambda}\,.
\end{split}
\end{align}%
Differentiating both sides of equations  (\ref{eq:genDivergenceA}) at $\lambda^{i}=0$ and using the action principle\footnote{This method is quite similar to the method of 
\cite{GuidaMagnoli} for calculation of deformed OPE coefficients.} (\ref{2}), we obtain 
the following equations for the leading and next-to-leading order coefficients $r^{(l)}$:
\begin{equation*}
 \bar \partial_{\bar x} \int\!\! d^{2} z \langle J_a(x)\Phi_{I}(y) \phi_{i}(z)\rangle_{0}=r^{(1)}_{ai}{}^{J} \langle \Phi_{J}(x)\Phi_{I}(y)\rangle_{0}\, , 
 \end{equation*}
 \begin{eqnarray*}
&&  \frac{1}{2!} \bar \partial_{\bar x}\int\!\! dz_{1}\!\! \int \!\! dz_2 \langle J_a(x)\Phi_{I}(y)\phi_{i}(z_1) \phi_{j}(z_2) \rangle_{0}=
r^{(1)}_{a i}{}^{J} \int\!\! dz \langle \Phi_{J}(x)\Phi_{I}(y) \phi_{j}(z)\rangle_{0} \nonumber \\
&& + r^{(1)}_{aj}{}^{J} \int\!\! dz \langle \Phi_{J}(x)\Phi_{I}(y) \phi_{i}(z)\rangle_{0} 
 +r^{(2)}_{aij}{}^{J}\langle \Phi_{J}(x)\Phi_{I}(y)\rangle_{0}\,.
\end{eqnarray*}
Since 
\[
\langle \Phi_{I}(x)\Phi_{J}(y)\rangle_0=\frac{\delta_{IJ}}{|x-y|^4}\,,
\]
we obtain 
\be
r^{(1)}_{ai}{}^{I}=    |x-y|^4\bigg\{\bar \partial_{\bar x} \int\!\! d^{2} z \langle J_a(x)\Phi_{I}(y) \phi_{i}(z)\rangle_{0}\bigg\}  \, , 
\ee
\bea
r^{(2)}_{aij}{}^{I}&=& |x-y|^4\bigg\{\frac{1}{2!} \bar \partial_{\bar x}\int\!\! dz_{1}\!\! \int \!\! dz_2 \langle J_a(x)\Phi_{J}(y)\phi_{i}(z_1) \phi_{j}(z_2) \rangle_{0}
 \nonumber \\
&&- r^{(1)}_{aj}{}^{J} \int\!\! dz \langle \Phi_{J}(x)\Phi_{I}(y) \phi_{i}(z)\rangle_{0} -r^{(1)}_{a i}{}^{J} \int\!\! dz \langle \Phi_{J}(x)\Phi_{I}(y) \phi_{j}(z)\rangle_{0}\bigg\} \, . 
\eea
Similar expressions are also obtained for  $\bar{r}_{\bar{a}}{}^{I}(\lambda)$. Relegating the details  to appendix~\ref{app:redundancyLOandNLOv2},
after taking the integrals we arrive at the expressions
\begin{gather}
\begin{aligned}\label{rr}
r_{a}{}^{I}(\lambda)&=i\pi\lambda^i\oA_{ai}{}^{I}-i\pi^2\oB^{\bar{a}}_{ai}\bar{\oA}_{\bar{a}j}{}^{I}\lambda^{i}\lambda^{j}+\OL{3}\, , \\
\bar{r}_{\bar{a}}{}^{I}(\lambda)&=
i\pi\lambda^i\bar{\oA}_{\bar{a}i}{}^{I}-i\pi^2\bar{\oB}^{a}_{\bar{a}i}\oA_{aj}{}^{I}\lambda^{i}\lambda^{j}+\OL{3}\,.
\end{aligned}
\end{gather}

%%%%%%%%%%%%%%%%%%%%%%%%%%%%%  KAPPA COMBINATIONS 

 Formulas (\ref{rr}) apply to all broken symmetry currents. For the purposes of reducing the number of couplings we need to identify those linear combinations of operators $\phi_{i}$  present 
 in our perturbation that are total derivatives. To identify all such total derivatives  we would like to find
   a basis of linear combinations of currents
\begin{equation} \label{redK}
K_{\alpha}=\kappa_{\alpha}{}^{a}(\lambda)J_a\,,\quad \bar{K}_{\alpha}=\kappa_{\alpha}{}^{\bar{a}}(\lambda)\bar{J}_{\bar{a}}\, , 
\end{equation}
which may now contain both holomorphic and anti-holomorphic components,
such that   stronger equations  than (\ref{redd}) are satisfied:
\begin{equation}\label{eq:redV1}
(\bar{\partial}K_{\alpha})_{\lambda}(x)+(\partial\bar{K}_{\alpha})_{\lambda}(x) =r_{\alpha}{}^{i}(\lambda )\phi_i(x) \, . 
\end{equation}
Such combinations give redundant operators and identify redundant combinations of couplings. Associated to such combinations are {\it redundancy vector fields }
$$
\hat R_{\alpha}= r_{\alpha}{}^{i}(\lambda) \frac{\partial}{\partial \lambda^{i}} \, . 
$$

The coefficients $\kappa_{\alpha}{}^{a}(\lambda)$, $\kappa_{\alpha}{}^{\bar b}(\lambda)$ in (\ref{redK}) can be analyzed perturbatively. Using (\ref{rr}) we find that at the leading order ${\cal O}(\lambda^{0})$ the coefficients $\kappa^{(0)}_{\alpha}{}^{a}=\kappa_{\alpha}{}^{a}(0)$, 
$\kappa^{(0)}_{\alpha}{}^{\bar a}=\kappa_{\alpha}{}^{\bar a}(0)$ must satisfy
%Returning to the redundancy equation~\eqref{eq:redV1}, at leading order in $\lambda$'s, the defining equation for the coefficient tensors 
%$\kappa^{(0)}_{\alpha}{}^{a}$ and $\kappa^{(0)}_{\alpha}{}^{\bar{a}}$ read (for the index $J$ labeling ``orthogonal'' operators $\chi_{\tilde{j}}$)
\begin{align}\label{eq:DefKappaLO}
\kappa^{(0)}_{\alpha}{}^{a}\oA_{ai}{}^{\tilde{j}}
+\kappa^{(0)}_{\alpha}{}^{\bar{a}}\bar{\oA}_{\bar{a}i}{}^{\tilde{j}}=0\quad \forall i,{\tilde{j}}\,.
\end{align}%
Let us assume that  $\kappa^{(0)}_{\alpha}{}^{a}$,  $\kappa^{(0)}_{\bar b}{}^{a}$ form a complete orthonormal basis for the solutions of this equation (labelled by the index $\alpha$), normalized so  that 
\be
\kappa^{(0)}_{\alpha}{}^{a}\kappa^{(0)}_{\beta}{}^b\eta_{ab}=\delta_{\alpha\beta}\,,\quad
\kappa^{(0)}_{\alpha}{}^{\bar{a}}\kappa^{(0)}_{\beta}{}^{\bar{b}}\eta_{\bar{a}\bar{b}}=\delta_{\alpha\beta}\,.
\ee
%Inserting the results for the leading order coefficients $M_{ai}{}^{\rho}$ and $\bar{M}_{\bar{a}i}{}^{\rho}$ into~\eqref{eq:redV1}, we obtain
The leading order redundancy vector fields have the form 
\begin{equation}\label{eq:redLO}
\hat R^{(0)}_{\alpha} =
\pi \left( Q_{\alpha}\right)_r{}^{i}\lambda^r\partial_i \, , \qquad (Q_{\alpha})_i{}^j = i\left(
\kappa^{(0)}_{\alpha}{}^{a}\oA_{ai}{}^j+\kappa^{(0)}_{\alpha}{}^{\bar{a}}\bar{\oA}_{\bar{a}i}{}^j\right)\, 
\end{equation} 
where $(Q_{\alpha})_{i}{}^{j}$ are the charge matrices for the currents $(K_{\alpha}, \bar K_{\alpha})$.
The vector fields $\hat R^{(0)}_{\alpha}$ satisfy the commutation relations of a Lie algebra 
\be
[\hat R_{\alpha}^{(0)}, \hat R_{\beta}^{(0)}] = i f_{\alpha\beta}{}^{\gamma} \hat R_{\gamma}^{(0)} \, 
\ee
where 
\be
f_{\alpha\beta\gamma}=
\pi \kappa^{(0)a}_{\alpha}\kappa^{(0)b}_{\beta}\kappa^{(0)c}_{\gamma}\fF{a}{b}{d}\eta_{cd}
=\pi \kappa^{(0)\bar{a}}_{\alpha}\kappa^{(0)\bar{b}}_{\beta}\kappa^{(0)\bar{c}}_{\gamma}\fFb{a}{b}{d}\eta_{\bar{c}\bar{d}}\, .
\ee
At next-to-leading order in $\lambda$'s we have 
\be
\kappa_{\alpha}{}^{a} =  \kappa^{(0)a}_{\alpha} + \kappa_{\alpha i}^{(1)a} \lambda^{i} + \dots \, , \quad 
\kappa_{\alpha}{}^{\bar a} =  \kappa^{(0)\bar a}_{\alpha} + \kappa_{\alpha i}^{(1)\bar a} \lambda^{i} + \dots
\ee
Substituting (\ref{rr}) and (\ref{redK}) into (\ref{eq:redV1}) we obtain 
\begin{align}\label{eq:DefKappaNLO}
\kappa_{\alpha r}^{(1)a}\oA_{as}{}^{\tilde{j}}
-\pi \kappa_{\alpha }^{(0)a}\oB_{ar}{}^{\bar{a}}\bar{\oA}_{\bar{a}s}{}^{\tilde{j}}
+\kappa_{\alpha  r}^{(1)\bar{a}}\bar{\oA}_{\bar{a}s}{}^{\tilde{j}}
-\pi \kappa_{\alpha }^{(0)\bar{a}}\bar{\oB}_{\bar{a}r}{}^{a}\oA_{as}{}^{\tilde{j}}=0\quad \forall \tilde{j},r,s\,.
\end{align}%
Since we assume the leading order coefficients $\kappa^{(0)}_{\alpha}{}^{a}$ and $\kappa^{(0)}_{\alpha}{}^{\bar{a}}$ to form a complete basis of all solutions to~\eqref{eq:DefKappaLO}, 
 a solution to~\eqref{eq:DefKappaNLO} exists and must satisfy
\begin{align}
\left\lbrace\begin{array}{rcl}
\kappa_{\alpha  r}^{(1)a}\oA_{as}{}^{\tilde{j}}
+\pi \kappa_{\alpha}^{(0)\bar{a}}\bar{\oB}_{\bar{a}r}{}^{a}\oA_{as}{}^{\tilde{j}}
&=&\eta_{\alpha r}{}^{\beta}\kappa^{(0)a}_{\beta}\oA_{as}{}^{\tilde{j}}\\
\kappa_{\alpha r}^{(1)\bar{a}}\bar{\oA}_{\bar{a}s}{}^{\tilde{j}}
+\pi \kappa_{\alpha}^{(0)a}\oB_{ar}{}^{\bar{a}}\bar{\oA}_{\bar{a}s}{}^{\tilde{j}}
&=&\eta_{\alpha r}{}^{\beta}\kappa^{(0)\bar{a}}_{\beta}\bar{\oA}_{\bar{a}s}{}^{\tilde{j}}\;
\end{array}\right. \label{rcl}
\end{align}%
for some coefficients $\eta_{\alpha r}{}^{\beta}$. 
Since the OPE coefficients $\oA_a$ and $\bar{\oA}_{\bar{a}}$ are linearly independent with respect to the indices $a$ and $\bar{a}$, respectively, we thus obtain the relations
\begin{align}
\begin{split}\label{eq:kappaOne}
\kappa_{\alpha r}^{(1)a}
&=\eta_{\alpha r}{}^{\beta}\kappa^{(0)a}_{\beta}-\pi \kappa_{\alpha}^{(0)\bar{a}}\bar{\oB}_{\bar{a}r}{}^a\\
\kappa_{\alpha r}^{(1)\bar{a}}
&=\eta_{\alpha r}{}^{\beta}\kappa^{(0)\bar{a}}_{\beta}-\pi \kappa_{\alpha}^{(0)a}\oB_{ar}{}^{\bar{a}}\,.
\end{split}
\end{align}
Therefore
\be \label{ri_exp}
r^{i}_{\alpha} =\pi  (Q_{\alpha})_{r}{}^{i}\lambda^{r} + \pi \lambda^{r}\eta_{\alpha r}{}^{\beta}\lambda^{s}(Q_{\beta})_{s}{}^{i}  + {\cal O}(\lambda^{3}) \, . 
\ee
The corresponding  redundancy vector field that contains the leading and the next-to-leading order terms is 
\begin{align}
\begin{split}\label{red:NLO}
\hat R^{(1)}_{\alpha}    &=\pi (\delta_{\alpha}{}^{\beta}+\lambda^r\eta_{\alpha r}{}^{\beta})\lambda^s\left(Q_{\beta}\right)_s{}^i\partial_{i}  = (\delta_{\alpha}{}^{\beta}+\lambda^r\eta_{\alpha r}{}^{\beta})\hat R_{\beta}^{(0)}  \,.
\end{split}
\end{align}%
This formula means that the $\hat R^{(1)}_{\alpha}  $ are linear combinations of the Lie algebra vector fields $\hat R^{(0)}_{\alpha}  $. This implies that the $\hat R^{(1)}_{\alpha} $ 
satisfy the Frobenius integrability condition 
(the commutators close on linear combinations). Moreover, we can  change the  basis of redundancy vector fields to 
\begin{equation}\label{sp_basis}
\widehat{\mathcal{R}}_{\alpha}=(\delta_{\alpha}^{\beta} -\lambda^r\eta_{\alpha r}{}^{\beta}) \hat R_{\beta}  \,,
\end{equation}
so that
\begin{equation}
\left[\widehat{\mathcal{R}}_{\alpha},\widehat{\mathcal{R}}_{\beta}\right]=i\fF{\alpha}{\beta}{\gamma}\widehat{\mathcal{R}}_{\gamma}  +  \OL{3} \,.
\end{equation}
This means that in the special basis (\ref{sp_basis})  the deformed redundancy vector fields still form a subalgebra of the fixed point Lie algebra up to the next-to-leading order in perturbation theory, 
which we call the {\it redundancy subalgebra}. 
In this basis, the connection coefficients $\Gamma_{ i \alpha }^{j}$ defined in (\ref{Gamma_a}) take an especially simple form 
\be 
\tilde { \Gamma}_{i\alpha}^{j} = -\pi (Q_{\alpha})_{i}{}^{j} + {\cal O}(\lambda^2) 
\ee
which means (see (\ref{red_connection})) that, to this order,  when moving along the redundant directions the operators $\phi_{i}$ are rotated by the corresponding fixed point  Lie algebra action.

\subsubsection{Redundancy and the beta function}\label{sec:red_beta}
In section 3 we derived a general relationship  for the commutator of the redundancy vector fields and the beta function, 
\be \label{check_Br}
   [ \hat \beta , \hat R_{a} ] = - \beta^{j} r_{aj}{}^{b}   \hat R_{b} \, .
\ee   
Here the redundancy vector fields $\hat R_{a}$ act on the enlarged space of couplings for all scalar dimension 2 operators. This relationship can be checked to hold through the quartic order in the couplings using formulas (\ref{rabj}), (\ref{rr}), (\ref{betaLO}), (\ref{NLOgen1}) and the Ward identities at the fixed point.

We would also like  to check whether this general relation  can be  specialized to the case of the redundancy vector fields $\widehat{R}_{\alpha}$ acting on the space 
of flowing couplings. We  find the following relation:
\begin{equation} \label{reducedrel}
[\widehat{\beta},\widehat{R}_{\alpha}]=-\beta^{j}\eta_{\alpha j}{}^{\beta}\widehat{R}_{\beta} + {\cal O}(\lambda^4)  \,.
\end{equation}

At the leading order, we calculate the commutator on the left hand side of (\ref{check_Br})  to be 
\begin{align*}
[\widehat{\beta},\widehat{R}_{\alpha}]&=
\left[\pi \left(Q_{\alpha}\right)_i{}^j\lambda^i\partial_j,\pi\oC_{rs}{}^t\lambda^r\lambda^s\partial_t\right]+\OL{3}\\
&=\pi^2\lambda^r\lambda^s\bigg(\left(Q_{\alpha}\right)_r{}^i\oC_{is}{}^t
+\left(Q_{\alpha}\right)_s{}^i\oC_{ir}{}^t-\oC_{rs}{}^i\left(Q_{\alpha}\right)_i{}^t\bigg)+\OL{3} =0\,,
\end{align*}%
which vanishes by virtue of \eqref{eq:ACeqnVariant}.

At next-to-leading order, denoting by $\widehat{R}^{(\ell)}_{\alpha}$ and $\widehat{\beta}^{(\ell)}$ the terms in the vector fields at a given order in $\lambda$'s, we have two contributions to the commutator:
\begin{equation}
[\widehat{\beta},\widehat{R}_{\alpha}]^{(3)}=[\widehat{\beta}^{(3)},\widehat{R}_{\alpha}^{(1)}]+[\widehat{\beta}^{(2)},\widehat{R}_{\alpha}^{(2)}]\,.
\end{equation}
Recalling from~\eqref{NLOgen1}  that the two loop coefficients $\beta^i_{rst}$ of the $\beta$ function are defined in terms of (the regular part of) 
an integral over a $4$-point function of $\phi$'s, and since the commutator with $\widehat{R}_{\alpha}^{(1)}$ is proportional to the action of the linear combinations of chiral currents
\[
K_{\alpha}=\kappa^{(0)}_{\alpha}{}^{a}J_a\,,\quad \bar{K}_{\alpha}=\kappa^{(0)}_{\alpha}{}^{\bar{a}}\bar{J}_{\bar{a}}
\]
on this $4$-point function, the contribution $[\widehat{\beta}^{(3)},\widehat{R}_{\alpha}^{(1)}]$ to the commutator vanishes due to the Ward identity for the 4-point functions:
 \begin{equation}\label{eq:ACeqnVariant}
Q_{\alpha}(\langle \phi_{i}(z_1)\phi_{j}(z_2) \phi_{k}(z_3) \phi_{l}(z_4)\rangle_{0;c})=0\,.
\end{equation}

The other contribution to the commutator -- $[\widehat{\beta}^{(2)},\widehat{R}_{\alpha}^{(2)}]$ -- yields the right hand side of 
(\ref{check_Br})  (again making use of equations of type~\eqref{eq:ACeqnVariant}):
\begin{align*}
[\widehat{\beta}^{(2)},\widehat{R}_{\alpha}^{(2)}]&=\pi^2\lambda^r\lambda^s\lambda^t\bigg(
\eta_{\alpha r}{}^{\beta}\left(Q_{\beta}\right)_s{}^i\left(\oC_{it}{}^j+\oC_{ti}{}^j\right)
-\oC_{rs}{}^i\left(
\eta_{\alpha i}{}^{\beta}\left(Q_{\beta}\right)_t{}^j+\eta_{\alpha t}{}^{\beta}\left(Q_{\beta}\right)_i{}^j
\right)
\bigg)\partial_j\\
&=-\pi^2\lambda^r\lambda^s\lambda^t\oC_{rs}{}^i\eta_{\alpha t}{}^{\beta}\left(Q_{\beta}\right)_i{}^j\partial_{j} =-\beta^{i}\eta_{\alpha i}{}^{\beta}\widehat{R}_{\beta}+\OL{4}\,.
\end{align*}%
We have thus verified that up to two loop order the commutator of the beta function vector field with the redundancy vector fields closes on the redundancy vector fields. 
The coefficients $r_{aj}{}^{b}$ defined in (\ref{red_op2}) transform under a change of basis of the vector fields 
\be
J^{\mu}_{a}(x) \mapsto K^{\mu}_{a}(x) = M^{b}_{a}(\lambda) J_{b}^{\mu}(x) 
\ee
as
\be
r_{aj}{}^{b} \mapsto \tilde r_{aj}{}^{b}= M_{a}^{c}\, r_{cj}{}^{d}(M^{-1})_{d}^{b} + (\partial_{j}M_{a}^{d})(M^{-1})_{d}^{b} \, .
\ee
Comparing this with formulas (\ref{rcl}) and taking into account (\ref{rabj}), we find that the coefficients $\eta_{\alpha r}{}^{\beta}$ coincide with the corresponding redundancy anomaly coefficients calculated in the basis introduced in equation (\ref{redK}). 

Note that one cannot argue on general grounds that a relation of the type (\ref{reducedrel}) must hold to all orders in perturbation theory. Taking a commutator with the beta function could produce new redundancy vector fields which are not 
expressed as linear combinations of perturbing fields. To analyze such situations it seems appropriate  to add couplings corresponding to the extra redundant fields to have a set closed under the action of the beta function. 
At the first two orders in perturbation theory, we took advantage of the fact that some (or all) connection coefficients $r_{\alpha i}{}^{\beta}$ can be made to vanish at the origin $\lambda=0$ by a choice of basis for our vector 
fields.   

%Since the $\widehat{R}_{\alpha}$ act as endomorphisms on the space of couplings $\lambda^i$, this will allow to consistently quotient out the redundancy. 

\subsubsection{$\Theta$ and  redundancy}

Up to contact terms, the trace of the stress-energy tensor is given by
\[
\Theta(x)=\beta^i(\lambda)\phi_i(x)\,. 
\]
Given that some combinations of $\phi_{i}$'s are redundant, we may ask  whether the trace $\Theta$ contains any of 
these total derivatives. In other words, we want to see if there are beta functions for the redundant directions. Direct calculations show 
that 
\begin{equation}\label{noredTheta}
\langle\Theta(x)\bar \partial J_{a}(y)\rangle_{\lambda}=\OL{5}\,, \qquad 
\langle\Theta(x) \partial \bar J_{\bar b}(y)\rangle_{\lambda}=\OL{5}\, . 
\end{equation}

We will explain how this result is obtained for the correlator involving $J_{a}$ currents as the calculations for the one involving $\bar J_{\bar b}$ 
go in parallel. Expanding the expression
\begin{align}
\begin{split}
\left\langle \Theta(x)\bar \partial  J_{a}(y)\right\rangle_{\lambda}
&=\beta^{i}(\lambda)r_{a}{}^J(\lambda)\left\langle \phi_i(x)\Phi_{J}(y)\right\rangle_{\lambda}
\end{split}
\end{align}%
at finite separation $|x-y|>0$,  we obtain at the leading order
\begin{align*}
\left\langle \Theta(x)\bar \partial  J_{a}(y)\right\rangle^{(3)}
&=\pi \lambda^r\lambda^s\lambda^t\oC_{rs}{}^i A_{at}{}^j\frac{\delta_{ij}}{|x-y|^4}=0
\end{align*}%
by virtue of equation (\ref{eq:ACeqn}) and 
\be \label{C3}
C_{ij\tilde k}=0\, , \enspace \forall\;  i,j,\tilde k \, .
\ee
At  next-to-leading order  in $\lambda$'s, we have three contributions to the correlator:
\begin{align*}
\left\langle \Theta(x)\bar \partial  J_{a}(y)\right\rangle^{(4)}
&=\beta^{(2)i}r_{a}^{(2)j}\langle\phi_i(x)\phi_j(y)\rangle_0
+\beta^{(3)i}r_{a}^{(1)j}\langle\phi_i(x)\phi_j(y)\rangle_0\\
&\qquad+\beta^{(2)i}r_{a}^{(1)j}\langle\phi_i(x)\phi_j(y)\rangle^{(1)}\\
&=\frac{\delta_{ij}}{|x-y|^4}\left(
\beta^{(2)i}r_{a}^{(2)j}+ \beta^{(3)i}r_{a}^{(1)j}\right) + \beta^{(2)i}r_{a}^{(1)j}\langle\phi_i(x)\phi_j(y)\rangle^{(1)}\,.
\label{6}
\end{align*}%
Here the  indices in round brackets for all quantities stand for the perturbative contributions of the corresponding order.
The term proportional to 
\be \label{4}
\beta^{(3)i}r_{a}^{(1)j}\delta_{ij} = i \sum_{j}\beta^{j}_{k,l,m}A_{an}{}^{j}\lambda^{k}\lambda^{l}\lambda^{m}\lambda^{n}  
\ee
 vanishes due to Ward identities for the 4-point functions. 
More precisely, denote
\be
G_{IJKL} = \langle \phi_I(\infty)\phi_J(0)\phi_K(1)\phi_{L}(\eta)\rangle_{0;c} \, . 
\ee
Consider  a Ward identity generated by $J_{a}$:  
\be
A_{a i}{}^{R} G_{Rjkl} + A_{aj}{}^{R}G_{iRkl} + A_{ak}{}^{R}G_{ijRl} + A_{al}{}^{R}G_{ijkR} = 0 \, . 
\ee
Symmetrizing this identity over the four indices $i,j,k,l$ we obtain 
\be \label{5}
\sum_{\text{perm}(i, j,k,l)} \Bigl[  A_{a i}{}^{R} G_{Rjkl} + A_{aj}{}^{R}G_{Rikl} + A_{ak}{}^{R}G_{Rijl} + A_{al}{}^{R}G_{Rijk}  \Bigr] =0 \, .
\ee
Since the beta function coefficients $\beta^{i}_{jkl}$ given by (\ref{NLOgen1}) are totally symmetric in all four indices, taking into account (\ref{C3}) and integrating (\ref{5}) over $\eta$ 
we obtain 
\be \label{6}
\sum_{\text{perm}(i, j,k,l)} \sum_{R} \Bigl[  A_{a i}{}^{R} \beta^{R}_{jkl} + A_{aj}{}^{R}\beta^{R}_{ikl} + A_{ak}{}^{R}\beta^{R}_{ijl} + A_{al}{}^{R}\beta^{R}_{ijk}  \Bigr] =0 \, .
\ee
Since by assumption of two loop renormalizability $\beta^{\tilde r}_{ijk}=0$, equation (\ref{6}) reduces to 
\be
\sum_{\text{perm}(i, j,k,l)} \sum_{m} \Bigl[  A_{a i}{}^{m} \beta^{m}_{jkl} + A_{aj}{}^{m}\beta^{m}_{ikl} + A_{ak}{}^{m}\beta^{m}_{ijl} + A_{al}{}^{m}\beta^{m}_{ijk}  \Bigr] =0 \, .
\ee
Comparing this to the right hand side of (\ref{4}) we conclude that $\beta^{(3)i}r_{a}^{(1)j}\delta_{ij} = 0$.

Furthermore, from (\ref{eq:ACeqn}) and (\ref{C3}) we find 
  \be
\beta^{(2)i}r_{a i}^{(2)}=\sum_{i} \pi \lambda^p\lambda^q\lambda^r\lambda^s \oC_{pq}{}^ir_{a r}{}^{\bar b}A_{\bar bs}{}^{i}   =0\, . 
\ee 
For the remaining  contribution on the right hand side of (6), we need the correction to the metric, which is proportional to the OPE coefficients $\oC$ (see section~\ref{sec:metricCorrection} for the details)
\begin{equation}
\langle\phi_i(x)\phi_j(y)\rangle_{(1)} \propto  \oC_{ijk}\lambda^k\
\end{equation}
and hence again drops out by  (\ref{eq:ACeqn}) and (\ref{C3}). This concludes the proof of (\ref{noredTheta}).

\subsubsection{Currents $J_{i}$ and corrections to the Zamolodchikov metric}\label{sec:metricCorrection} 
The renormalization anomaly (\ref{RGop}) contains terms 
$ \partial_{\mu}\lambda^{i}J_{i}^{\mu}$ where the currents $J_{i}$ are expanded in a basis $J_{a}^{\mu}$ as $J_{i}^{\mu}=v_{i}^{a}J_{a}^{\mu} $.
Using the basis associated with  holomorphic and antiholomorphic currents at the fixed point we have coefficients $v_{i}^{a}$ and $v_{i}^{\bar b}$.
At leading order these coefficients were calculated in Appendix A of \cite{FK_curv}:
\begin{align}
\begin{split}
J_i(z,\bar{z})&\equiv v_{i}{}^a(\lambda)J_a(z)
=i\pi\oA_{ij}^a\lambda^j J_a(z)+\OL{2}\\
\bar{J}_i(z,\bar{z})&\equiv \bar{v}_{i}{}^{\bar{a}}(\lambda)\bar{J}_{\bar{a}}(z)
=i\pi\bar{\oA}_{ij}^{\bar{a}}\lambda^j\bar{J}_{\bar{a}}(\bar{z})+\OL{2}\,.
\end{split}\label{Ji}
\end{align}%
This result follows from the term in the deformed OPE 
 \begin{equation}
\bar{T}(\bar{x})\phi_i(y)= \frac{ i\pi\oA_{ij}^a\lambda^j J_a(y)}{(\bar{x}-\bar{y})^3}+\ldots
\end{equation}
and a similar cubic term in the $T(x)\phi_i(y)$ OPE.

It follows from (\ref{Ji}) and (\ref{eq:ACeqn}) that the Wess-Zumino consistency conditions 
\be
\beta^{i} J_{i}^{\mu} = 0 
\ee
are satisfied at the leading order in perturbation. 

Combining (\ref{Ji}) with (\ref{rr}) we obtain 
\begin{align}
\begin{split}\label{eq:divJi}
\partial_{\mu}J^{\mu}_i(x)&=
\left(v_{i}{}^{a}(\lambda)r_a^{I}(\lambda)+\bar{v}_{i}{}^{\bar{a}}(\lambda)\bar{r}_{\bar{a}}{}^{I}(\lambda)\right)
\Phi_{I}(x)\\
&=-\pi^2\lambda^r\lambda^s\left(
\oA_{ir}^a\oA_{as}{}^I
+\bar{\oA}_{ir}{}^{\bar{a}}\bar{\oA}_{\bar{a}s}{}^I\right)\Phi_{I}(x)+\OL{3}\,.
\end{split}
\end{align}

Earlier we defined  a set of currents
\begin{equation}
K_{\alpha}=\kappa_{\alpha}^{(0)a}J_a\,,\quad \bar{K}_{\alpha}=\kappa_{\alpha}^{(0)\bar{a}}\bar J_{\bar a}\,,
\end{equation}
which together with an auxiliary set of currents 
\begin{equation}
K_{\widetilde{\alpha}}=\kappa_{\widetilde{\alpha}}^{(0)a}J_a\,,\quad \bar{K}_{\widetilde{\alpha}}=\kappa_{\widetilde{\alpha}}^{(0)\bar{a}}\bar J_{\bar a}
\end{equation}
form a complete alternative basis. 
Using this basis we can rewrite formula (\ref{eq:divJi}) as
\be
\label{eq:divJi2}
\partial_{\mu}J^{\mu}_i(x)=
\pi^2\lambda^r\lambda^s\left(
(Q^{\alpha})_{ir}(Q_{\alpha})_{ s}{}^{j}\phi_{j}(x)
+(Q^{\widetilde{\alpha} })_{ir} (Q_{{\widetilde{\alpha}}})_ {s}{}^I \Phi_{I}(x)   \right)+\OL{3}\,.
\ee
We see from this formula that if for some $i,r,s,\tilde j$ 
\be
(Q^{\widetilde{\alpha} })_{ir} (Q_{{\widetilde{\alpha}}})_ {s}{}^{\tilde j} + (Q^{\widetilde{\alpha} })_{is} (Q_{{\widetilde{\alpha}}})_ {r}{}^{\tilde j} \ne 0
% Q_{i(r}^{\widetilde{\alpha} }Q_{{{\widetilde{\alpha}}}s)}^{\tilde j}\ne 0
\ee
scale transformations will admix to fields $\phi_{i}$ new redundant fields for which there were no couplings. It is easy to engineer current-current 
perturbations of WZW theories for which this is the case at the leading order. However we could not find such example which would be also closed under the beta function at two loops, 
that is to say in the examples we tried at two loops one would need to include   counter terms for new fields and to introduce more flowing couplings. But in general this remains a 
possibility.  If this happens, it would be  natural in our opinion to enlarge the space  of couplings to include all redundant operators which appear in the Callan-Symanzik equations. 

The correction to  Zamolodchikov metric $\Delta g_{ij}$ is defined in equation (\ref{delta_g}). It is constructed by integrating  correlation functions
\be
\langle \phi_{i}(x) \partial_{\mu} J^{\mu}_{j}(0) \rangle \, . 
\ee 
The tensor $\Delta g_{ij}$ is defined up to symmetric matrices orthogonal to the beta function. The contraction of $\Delta g_{ij}$ with the beta function which enters the 
gradient formula is free from such ambiguities. When strict power counting applies, due to equation (\ref{WZop}) we have 
\be
\Delta g_{ij}\beta^{j} = \lim\limits_{L\to \infty}
3\pi \int_{|x|<L} d^{2}x \, x^2 \theta(\Lambda |x| - 1) \, \expvalc{  \partial_{\mu}J_{i}^{\mu}(x) \Theta(0)} \, .  
\ee
Using  (\ref{eq:divJi}) and (\ref{noredTheta}) we conclude that
\be
\Delta g_{ij}\beta^{j} = {\cal O}(\lambda^6) \, .
\ee

Next we discuss the first perturbative correction to the fixed point Zamolodchikov metric. The metric is defined as 
\[
g_{ij}(\lambda)=\frac{6\pi^2}{\Lambda^4}\langle \phi_i(x)\phi_j(y)\rangle_{\lambda}\bigg\vert_{\Lambda|x-y|=1}\;
\]
where $\Lambda$ is some arbitrary, but fixed scale. At the fixed point $g_{ij} = g_{ij}^{(0)}=6\pi^2\delta^{ij}$. Using the point splitting cutoff and 
minimal subtraction we obtain the first correction 
\be\label{Zmetric_corr}
g_{ij}^{(1)}=
%\bigg\lbrace6\pi^2\lambda^k\frac{\oC_{ijk}}{\Lambda^4|x-y|^2}\lim\limits_{\frac{\epsilon}{L}\to0}\int d^2v\; 
%\frac{1}{|x-v|^2|y-v|^2}\bigg\rbrace\bigg\vert_{\Lambda|x-y|=1}\\
%&=
-24\pi^3 \ln\left(\frac{\Lambda}{\mu} \right)\sum_k\oC_{ijk}\lambda^k \,.
\ee
where $\mu$ is the subtraction scale. 
Zamolodchikov's choice \cite{Zam} is $\Lambda=\mu$ which results in no first order correction (the minimal subtraction scheme gives coordinates in which the Christoffel symbols 
vanish at $\lambda^i=0$). More generally $\zeta= \Lambda/\mu$ is some arbitrary dimensionless parameter which we consider to be fixed\footnote{The reader should not be worried about an apparent loss of positivity 
in the sum $g^{(0)}_{ij} + g^{(1)}_{ij}$ as the leading logarithms sum up to power corrections corresponding to the anomalous dimensions of $\phi_{i}$'s. }. 

\subsection{The gradient formula}
We have discussed all  quantities  that enter the gradient formula (\ref{grad_f}) except for the $c$-function and the Osborn antisymmetric tensor $b_{ij}$ defined in (\ref{eq:bijorig}). 
At a fixed point the one-form $w_{i}$ can be read off the contact term 
\be
\langle \phi_{i}(x) \Theta(y) \rangle_{c} = \frac{1}{12\pi} w_{i} \partial_{\mu}\partial^{\mu} \delta(x-y) \, .   
\ee
 The same contact term can be obtained from the one-point function of $\phi_{i}$ on ${\mathbb R}^{2}$ with nontrivial metric. We have
 $$
 \langle \phi_{i}(x) \rangle = -\frac{w_i}{24 \pi} \mu^2 R_{2}(x) + \mbox{ nonlocal terms} \, . 
 $$ 
 This implies that $w_{i}$ is exact and thus at the fixed point $b_{ij}=0$.  The first correction to $w_{i}$ comes from the leading order beta function and is thus 
 of the form $w^{(2)}_{i} \sim C_{ijk} \lambda^{j}\lambda^{k}$ which is again a closed 1-form. We conclude that $b_{ij} = {\cal O}(\lambda^2)$.

Since we showed that $b_{ij}={\cal O}(\lambda^2)$ and $\Delta g_{ij}={\cal O}(\lambda^4)$, the gradient formula \eqref{grad_f} has the form
\begin{equation} \label{approx_grad}
\partial_ic=-g_{ij}\beta^j   + {\cal O}(\lambda^4)\,.
\end{equation}
With the results for the beta function up to two loops  and for the metric up to the leading order corrections (\ref{Zmetric_corr}), we obtain the following 
expression for the $c$-function:
\begin{equation}\label{eq:cCC}
c=c_0+2\pi^3\oC_{rst}\lambda^r\lambda^s\lambda^t
+\lambda^r\lambda^s\lambda^t\lambda^u\left(
\frac{3\pi^2}{2}\beta_{rstu}-6\pi^4\ln(\zeta)\oC_{rs}{}^m\oC_{mtu}
\right)\;
\end{equation}
where $c_{0}$ is the central charge of the UV fixed point.

%%%%%%%%%%%%%%  SOME GENERAL RELATIONS FROM THE GRADIENT FORMULA REDUCED TO REDUNDANT DIRECTIONS

\subsection{Anomalous dimensions of the currents}\label{sec:anom_curr}
The general relation (\ref{WZ1}) in the basis corresponding to fixed point holomorphic and antiholomorphic currents reads
\be
\gamma_{a}^{b} = - r_{a}^{i} v_{i}^{b} + r_{ai}{}^{b}\beta^{i} \, , \quad 
\gamma_{\bar a}^{\bar b} = - r_{\bar a}^{i} v_{i}^{\bar b} + r_{\bar ai}{}^{\bar b}\beta^{i} \, , 
\ee
\be
\gamma_{a}^{\bar b} = - r_{a}^{i} v_{i}^{\bar b} + r_{ai}{}^{\bar b}\beta^{i}=0 \, , \quad  \gamma_{\bar a}^{ b} = - r_{\bar a}^{i} v_{i}^{ b} + r_{\bar ai}{}^{ b}\beta^{i}=0
\ee
where the last two expressions vanish by Lorentz invariance. At the leading order in perturbation substituting the results  obtained in the previous subsections we obtain 
 \be \label{gamma1}
 \gamma_{a}^{b} = \pi^2 A_{ak}{}^{i}A_{ij}{}^{b}\lambda^{k}\lambda^{j} + {\cal O}(\lambda^3)\, , \quad  \gamma_{\bar a}^{\bar b} = \pi^2 A_{\bar ak}{}^{i}A_{ij}{}^{\bar b}\lambda^{k}\lambda^{j} + {\cal O}(\lambda^3)\, , 
  \ee
  \bea \label{gamma2}
  \gamma_{a}^{\bar b} &=&  \pi^2 A_{ak}{}^{i}A_{ij}{}^{\bar b}\lambda^{k}\lambda^{j}    + \pi^2 B_{ai}{}^{\bar b}C_{kj}^{i}\lambda^{k}\lambda^{j}=0 \, , \nonumber \\
    \gamma_{\bar a}^{ b} &=&  \pi^2 A_{\bar ak}{}^{i}A_{ij}{}^{ b}\lambda^{k}\lambda^{j}    + \pi^2 B_{\bar ai}{}^{ b}C_{kj}^{i}\lambda^{k}\lambda^{j}=0 \, .
        \eea
Formulas (\ref{gamma1}) can be obtained by an independent calculation done in \cite{FK_curv} (see formula (A.9)  in that paper). The identities in (\ref{gamma2}) follow from (\ref{eq:AABCrels}).

 Equation~\eqref{WZ1} can be also applied to   the basis of  currents 
$K_{\alpha}^{\mu}$, $K_{\tilde \alpha}^{\mu}$ defined in sections \ref{sec:red1} and \ref{sec:metricCorrection}. We have
 \begin{align} \label{gamma_g}
 \begin{split}
 \gamma_{\alpha}{}^{\beta}=-r_{\alpha}{}^{i}
 v_{i}{}^{\beta}  + \eta_{\alpha i}{}^{\beta}\beta^{i}\, .
 \end{split}
 \end{align}
and a similar expression for $\gamma_{\tilde \alpha}^{\tilde \beta}$. Since the beta functions have no values in the redundant directions, the anomalous dimensions (and mixing coefficients) of the redundant 
operators are not given by  derivatives of the beta function.  Expression (\ref{gamma_g}) shows that these mixing coefficients (which are the same as $\gamma_{\alpha}^{\beta}$)   are stored in the
coefficients in the renormalization and redundancy   anomalies.  

Using that $\eta_{\tilde \alpha i}{}^{\beta}=\eta_{\alpha i}{}^{\tilde \beta}=0$ at the leading 
order we also have 
\be
\gamma_{\alpha}^{\tilde \beta} = -r_{\alpha}{}^{i}v_{i}{}^{\tilde \beta} + {\cal O}(\lambda^3) \, , \quad 
\gamma_{\tilde \alpha}^{ \beta} = -r_{\tilde \alpha}{}^{i}v_{i}{}^{ \beta} + {\cal O}(\lambda^3)
\ee
For the models we study in section \ref{sec:models}, $v_{i}{}^{\tilde \beta}=0$ and $r_{\tilde \alpha}{}^{i}=0$  at the first two orders of perturbation 
so that there are no mixed components for the matrix $\gamma$ at least through the order ${\cal O}(\lambda^2)$.

\subsection[Perturbations by relevant operators]{Perturbations by relevant operators\footnote{The results presented in this section grew out of discussions of AK with Daniel Friedan whose contributions are gratefully acknowledged.}
}\label{relevant}
Although our main focus in this section are perturbations by marginally relevant operators, we would like to discuss briefly  perturbations by relevant 
operators that break symmetries of the fixed point. We assume that the perturbing operators $\phi_{i}$ all have anomalous dimensions 
$\epsilon_{i}= 2-\Delta_{i}> 0 $ and that there are no resonances  (for a discussion of resonances in conformal perturbation theory see e.g. \cite{Gaberdiel:2008fn}). 
The perturbation theory for correlation functions 
necessarily breaks down at some order due to the emergence of infrared divergences that signal nonperturbative effects. However for small anomalous 
dimensions this happens at high orders.
Calculations of the quantities that enter the gradient formula 
 become particularly simple as under these conditions there are no contact terms in the relevant correlators by dimensional reasons. Also by dimensional reasons 
 $J_{i}^{\mu} = 0$ and $r_{i b}{}^{a} = r_{i b}^{\bar a}=0$ to all orders in perturbation. This simplifies drastically the picture of how the redundant operators enter into the equations.
 % The situation is in some sense  boring but can nevertheless be completely analyzed as far as the perturbation theory remains applicable. 

Let us first discuss the gradient formula.
The beta functions are 
\be \label{beta_rel}
\beta^{i} = \epsilon_{i}\lambda^i \, .
\ee
In the absence of  resonances,  in the minimal subtraction scheme  formula (\ref{beta_rel}) remains exact to all orders in perturbation theory. 
The first correction to the Zamolodchikov metric is obtained by integrating the 3-point function
\be
\langle \phi_{i}(0) \phi_{j}(x)\rangle^{(1)} = \mu^{\epsilon_{i} + \epsilon_{j} + \epsilon_{k}} 
\langle \phi_{i}(0) \phi_{j}(x) \int\!\! d^{2}y \lambda^{k}\phi_{k}(y) \rangle =  \frac{P_{ijk}\mu^{\epsilon_{i} + \epsilon_{j} + \epsilon_{k}}\lambda^{k}}{|x|^{\Delta_{i} + \Delta_{j} + \Delta_{k} - 2}} 
\ee
where 
\be
P_{ijk} = \pi C_{ijk} \frac{\Gamma(\Delta_{k}-1)\Gamma\left(1 + \frac{\Delta_{i} - \Delta_{j} - \Delta_{k}}{2}\right)\Gamma\left(1 + \frac{\Delta_{j} - \Delta_{i} - \Delta_{k}}{2}\right)}
{\Gamma\left(\frac{\Delta_{i} + \Delta_{k} - \Delta_{j}}{2}\right)\Gamma\left(\frac{\Delta_{j} + \Delta_{k} - \Delta_{i}}{2}\right)\Gamma(2-\Delta_{k})} \, . 
\ee
Setting for simplicity $\Lambda=\mu$  we obtain for the Zamolodchikov metric 
\be \label{g_rel}
g_{ij} = 6\pi^2 ( \delta_{ij} + P_{ijk}\lambda^{k}) + {\cal O} (\lambda^2)\, .
\ee
For  the Osborn 1-form we have 
\bea
 w_{i} &=& 3\pi \int dx^2 x^2 \theta(1-|x|\mu ) \Bigl( \langle \phi_{i}(x) \phi_{j}(0)\rangle_{0} + \langle \phi_{i}(x)\phi_{j}(0)\rangle^{(1)} \Bigr)\epsilon_{j}\lambda^{j} +{\cal O}(\lambda^3) \nonumber \\ 
&=&3\pi^2 \lambda^{i} + 6\pi^2 \frac{P_{ijk}\epsilon_{j}}{\epsilon_{i} + \epsilon_{j} + \epsilon_{k}} \lambda^{j} \lambda^{k} + {\cal O}(\lambda^3)\, , 
\eea
so that 
\be \label{b_rel}
b_{ij} = \frac{6\pi^2}{\epsilon_{i} + \epsilon_{j} + \epsilon_{k}} [ P_{ijk}(\epsilon_{i} - \epsilon_{j}) + \epsilon_{k}(P_{kji} - P_{kij})] \lambda^k + {\cal O}(\lambda^2) \, .
\ee
For the $c$-function using (\ref{c2}) and the absence of contact terms we obtain 
\be \label{c_rel}
c= c_{0} - w_{i}\epsilon_{i}\lambda^{i} = c_{0} - 6\pi^2 \sum_{i} \epsilon_{i}(\lambda^{i})^2  - 3\pi^2 
\frac{P_{ijk}\epsilon_{j}\epsilon_{i}}{\epsilon_{i} + \epsilon_{j} + \epsilon_{k}} \lambda^{i} \lambda^{j} \lambda^{k}\, . 
\ee
It is a matter of  some elementary algebra to check that 
\be\label{grad_rel}
\partial_{i} c= -g_{ij}\beta^{j} - b_{ij}\beta^{j} 
\ee
holds through the order $\lambda^2$. It was noted in \cite{Freedman_etal} that at the  second  order in perturbation the 1-form $g_{ij}\beta^{j}$ is not closed. This is taken care of by the Osborn 
$b$-field in (\ref{grad_rel}). 

 Finally, let us discuss how the redundant operators enter into equations. The OPE of the relevant fields $\phi_{i}$ with the conserved currents has the form 
\be
J_a(z_a)\phi_i(z_i)=
\frac{i}{(z_{ai})}\oA_{ai}{}^{j}\phi_j(z_i)
+\frac{i}{(z_{ai})}\oA_{ai}{}^{\tilde{j}}\chi_{\tilde{j}}(z_i)+r.p.
\ee
and similarly for the antiholomorphic currents. We consider combinations of fundamental currents $(K_{\alpha}, \bar K_{\alpha})$ whose charges close on the perturbing fields. 
 The leading term in (\ref{ri_exp}) is universal so that we have at the leading order
 \be \label{red_rell}
\bar \partial K_{\alpha}(x) + \partial \bar K_{\alpha}(x) = \pi (Q_{\alpha})_{ j}{}^{i}\lambda^{j}\phi_{i}(x) \, , 
 \ee
 but since we have only power divergences present in the minimal subtraction scheme we do not expect any higher order corrections to (\ref{red_rell}).  The redundancy vector fields are thus 
 \be
 \hat R_{\alpha} = \pi (Q_{\alpha})_{ j}{}^{i}\lambda^{j}\partial_{i}\, . 
  \ee
  The commutator with the beta function vector field is 
  \be 
  [\hat \beta, \hat R_{\alpha}] = \pi \sum_{ij} (Q_{\alpha})_{ j}{}^{i}(\epsilon_{j}- \epsilon_{i}) \partial_{i} = 0 
  \ee
  which vanishes because $Q_{\alpha}$ at the fixed point commutes with the dilatation operator so that 
  \be
  \mbox {if }  (Q_{\alpha})_{ j}{}^{i} \ne 0 \mbox{ then } \epsilon_i = \epsilon_j \, .  
  \ee
  (This property also ensures that the tensor $P_{ijk}$ is invariant under the action of $Q_{\alpha}$'s.)
   
Since the redundancy anomaly coefficients $r_{ib}{}^{a}$ vanish, equations (\ref{red_connection}), (\ref{Gamma_a}) imply 
\be
{\cal L}_{\alpha} \langle \phi_{i_1}(x_1) \dots \phi_{i_{k}}(x_k)\rangle = 0 
\ee 
where ${\cal L}_{\alpha}$ stands for a Lie derivative with respect to  $\hat R_{\alpha}$ and the insertions are taken at finite separation. 
This implies that 
\be
{\cal L}_{\alpha} g_{ij} = {\cal L}_{\alpha} b_{ij} = 0 \, 
\ee
which together with (\ref{Rbeta_comm}) and (\ref{red_c}) means that every object in the gradient formula (\ref{grad_rel})  commutes with  the action of the redundancy vector fields. 
To perform the reduction (at a generic point in the foliation) we can locally split the coordinates into the   coordinates on the redundancy group  (the redundant directions) and 
the  coordinates invariant under the group action (nonredundant directions). This needs to be done in a special way so that the redundant directions completely drop out. 
The analysis in section \ref{sec:reduction} done for marginal perturbations in the case when the redundancy group representation is polar can be generalized to the relevant case. 
We are not going to present any details  in this paper. 

\section{Current-current perturbations of WZW models}\label{sec:models}

Let us consider a CFT  with chiral symmetry algebra $\mathcal{G}\times\overline{\mathcal{G}}$ at levels $k_L$ and $k_R$, perturbed by current-current operators 
\begin{equation}\label{cc_model}
\delta S= \int d^2x\; \lambda^i \phi_i(x)\,,\qquad \phi_i(x)=d^{a\bar{a}}_i J_{a} \bar J_{\bar a}\,.
\end{equation}
As usual, $J_a$ and $\bar{J}_{\bar{a}}$ denote the holomprhic and anti-holomorphic chiral symmetry currents of the unperturbed theory with OPE's 
\begin{gather}
\begin{aligned}
J_a(z)J_b(0)&=\frac{\eta_{ab}}{z^2}+\frac{i\fF{a}{b}{c}J_c(0)}{z}+r.p.\\
\bar{J}_{\bar{a}}(\bar z)\bar{J}_{\bar b}(0)&=\frac{\eta_{\bar{a}\bar{b}}}{\bar z_{ab}^2}
+\frac{i\fFb{a}{b}{c}\bar{J}_{\bar{c}}(0)}{\bar z}+r.p.\,,
\end{aligned}
\end{gather}%
with $\eta_{ab}=k_L \delta_{ab}$, $\eta_{\bar{a}\bar{b}}=k_R \delta_{\bar{a}\bar{b}}$.
The coefficient matrices $d^{a\bar{a}}_i$ are constrained by a number of consistency conditions. Firstly, we choose the perturbing operators to form an orthonormal set,
\begin{equation}
 \quad d_i^{a\bar{a}}d_j^{b\bar{b}}\eta_{ab}\eta_{\bar{a}\bar{b}}=\delta_{ij}\,.
\end{equation}
For convenience we also introduce operators $\chi_{\tilde{j}}$ which are orthogonal to the perturbing operators $\phi_{i}$ and which complete them to an orthonormal  basis of all current-current operators. 
 For later convenience, let $\{\Phi_I\}$ denote the full basis consisting of operators $\phi_{i}$ and $\chi_{\tilde j}$. We write
\[
 \Phi_I(x) = d^{a\bar{a}}_I J_{a}\bar J_{\bar a}\,.
\] 
Completeness implies the following relation:
\begin{equation}
d_I^{a\bar{a}}d_J^{b\bar{b}}\delta^{IJ} = \eta^{ab}\eta^{\bar{a}\bar{b}}\,.
\end{equation}
The OPE of the current-current operators $\Phi_{I}$ has the form
\begin{gather}
\begin{aligned}\label{eq:PhiPhiOPEsCC}
& \Phi_I(x)\Phi_J(y)=
\frac{\delta_{IJ}}{|x-y|^4}
+\frac{i \oA_{IJ}^cJ_c(y)}{(x-y)(\bar{x}-\bar{y})^2}
+\frac{i \bar{\oA}_{IJ}^{\bar{c}}\bar{J}_{\bar{c}}(\bar{y})}{(x-y)^2(\bar{x}-\bar{y})}
+\frac{ \oC_{IJ}{}^K\Phi_K(y)}{|x-y|^2}\\
&\hphantom{=}
+\frac{i \oD_{IJ}^{ab\bar{c}}(J_{a}J_{b})\bar J_{\bar c}  (y)}{(\bar{x}-\bar{y})}
+\frac{\oC_{IJ}{}^K(\partial \Phi_K)(y)}{2(\bar{x}-\bar{y})}
+\frac{i \bar{\oD}_{IJ}^{c\bar{a}\bar{b}}(\bar J_{\bar a}\bar J_{\bar b})J_{c}(y)}{(x-y)}
+\frac{\oC_{IJ}{}^K(\bar{\partial}\Phi_K)(y)}{2(x-y)} +\ldots 
\end{aligned}
\end{gather}%
where in the singular part we have only omitted spin 2 and spin 3 fields. 
In (\ref{eq:PhiPhiOPEsCC}) we singled out  the spin 1 
 quasiprimary fields $(J_{a}J_{b})\bar J_{\bar c}$ and $(\bar J_{\bar a}\bar J_{\bar b})J_{c}$ where 
\[
(J^a J^b)(z)    =  :\! J^{a}J^{b}\!\!:(z) - \frac{i}{2}\fF{a}{b}{c}(\partial J_c)(z)\;
\]
and similarly for the antiholomorphic currents. 

Using the orthonormality and completeness conditions  the OPE coefficients can be expressed as
\begin{alignat}{4}
   &  \oA_{IJ}^c=d^{a\bar{a}}_I d^{b\bar{b}}_J \fF{a}{b}{c}\eta_{\bar{a}\bar{b}}\,,
        &\qquad 
        \bar{\oA}_{IJ}^{\bar{c}}&=d^{a\bar{a}}_I d^{b\bar{b}}_J \eta_{ab}\fFb{a}{b}{c}\, ,\notag\\
     &\oC_{IJ}{}^K=-\delta^{KL}d^{a\bar{a}}_I d^{b\bar{b}}_Jd^{c\bar{c}}_L f_{abc}\bar{f}_{\bar{a}\bar{b}\bar{c}} \, , 
     &\qquad 
   &\quad \\
     &\oD_{IJ}^{ab\bar{c}}=d^{a\bar{a}}_I d^{b\bar{b}}_J\fFb{a}{b}{c}\,, 
     &\quad 
     \bar{\oD}_{IJ}^{c\bar{a}\bar{b}}&=d^{a\bar{a}}_I d^{b\bar{b}}_J \fF{a}{b}{c}\;\notag
  \end{alignat}%
  where 
  \be
 f_{abc}=\fF{a}{b}{d}\eta_{dc}\,,\quad \bar{f}_{\bar{a}\bar{b}\bar{c}}=\fFb{a}{b}{d}\eta_{\bar{d}\bar{c}}\, . 
  \ee
As usual the one loop renormalizability of the perturbed model requires the OPE closure of  the set of perturbing operators $\phi_i$, whence
\begin{equation}
\oC_{ij}{}^{\tilde{k}}=0\,.
\end{equation}
We will also need the OPEs of the currents $J_a$ and $\bar{J}_{\bar{a}}$ with the operators $\Phi_I$:
\begin{gather}
\begin{aligned}\label{eq:JPhiOPEsCC}
J_a(x)\Phi_I(y)&=
\frac{
\bar{\oB}_{\bar{a}I}^{b}J_b(y)
}{(\bar{x}-\bar{y})^2}
+\frac{i\bar{\oA}_{\bar{a}I}{}^R\phi_R(y)}{(\bar{x}-\bar{y})}\,,
\end{aligned}
\end{gather}%
with the OPE coefficients given by\footnote{Here again we make use of the completeness property of the coefficients $d_I^{a\bar{a}}$. }
\begin{gather}
\begin{aligned}
\oB_{aI}{}^{\bar{b}}&=d_I^{b\bar{b}}\eta_{ab}\,, &\quad&& \bar{\oB}_{\bar{a}I}{}^b&=d_I^{b\bar{b}}\eta_{\bar{a}\bar{b}}\\ 
\oA_{aB}^{C}&=d_{B}^{b\bar{b}}d_{D}^{c\bar{c}}\FF{a}{b}{c}\eta_{\bar{b}\bar{c}}\delta^{DC}\,, &\quad &&
\bar{\oA}_{\bar{a}B}^{C}&=d_{B}^{b\bar{b}}d_{D}^{c\bar{c}}\FFb{a}{b}{c}\eta_{bc}\delta^{DC}\,.
\end{aligned}
\end{gather}%
Note  the following relation between the tensors $\oA$ and $\bar{\oA}$ appearing in the OPEs~\eqref{eq:PhiPhiOPEsCC} and~\eqref{eq:JPhiOPEsCC} 
\begin{gather}
\begin{aligned}
\oA_{aI}{}^R&=\eta_{ab}\oA_{IS}^b\delta^{SR}\,, &\quad
\bar{\oA}_{\bar{a}I}{}^R&=\eta_{\bar{a}\bar{b}}\bar{\oA}_{IS}^{\bar{b}}\delta^{SR}\,.
\end{aligned}
\end{gather}
Specializing the formulae for the beta function up to two loops presented in section~\ref{sec:betaFunctionGeneral} for the general perturbation theory setup to the special case of current-current perturbations leads to 
(see appendix~\ref{CC_beta_details} for  details of the derivation)
\begin{gather}\label{GeneralCC}
\begin{aligned}
\beta^i&=\pi\oC^i_{jk}\lambda^j\lambda^k+\beta^i_{jk\ell}\lambda^j\lambda^k\lambda^{\ell}+\OL{4}\\
\beta^I_{jk\ell}&=\frac{\pi^2}{3!}\delta^{IM}d_M^{a\bar{a}}d_j^{b\bar{b}}d_k^{c\bar{c}}d_{\ell}^{d\bar{d}}\left(
E_{abcd, \bar a\bar b\bar c\bar d}+ \bar E_{abcd, \bar a\bar b\bar c\bar d}\right)\\
E_{abcd, \bar a \bar b \bar c \bar d}&=
\left(\eta_{ad}\eta_{bc}-\eta_{ac}\eta_{bd}\right)\fFb{a}{b}{r}\FFb{r}{c}{d}
+\left(\eta_{ab}\eta_{cd}-\eta_{ad}\eta_{bc}\right)\fFb{a}{c}{r}\FFb{r}{d}{b}
\\
&\qquad
+\left(\eta_{ac}\eta_{bd}-\eta_{ab}\eta_{cd}\right)\fFb{a}{d}{r}\FFb{r}{b}{c}
\\
\bar E_{abcd, \bar a \bar b \bar c \bar d}&=
\fF{a}{b}{r}\FF{r}{c}{d}\left(\eta_{\bar{a}\bar{d}}\eta_{\bar{b}\bar{c}}-\eta_{\bar{a}\bar{c}}\eta_{\bar{b}\bar{d}}\right)
+\fF{a}{c}{r}\FF{r}{d}{b}\left(\eta_{\bar{a}\bar{b}}\eta_{\bar{c}\bar{d}}-\eta_{\bar{a}\bar{d}}\eta_{\bar{b}\bar{c}}\right)
\\
&\qquad
+\fF{a}{d}{r}\FF{r}{b}{c}\left(\eta_{\bar{a}\bar{c}}\eta_{\bar{b}\bar{d}}-\eta_{\bar{a}\bar{b}}\eta_{\bar{c}\bar{d}}\right)
\,.
\end{aligned}
\end{gather}%
Upon closer inspection, we recognize the appearance of OPE coefficients of types $\oD$ and $\bar{\oD}$ described in~\eqref{eq:PhiPhiOPEsCC} and  
express the two loop beta function coefficients in a  more compact form
\begin{gather}
\begin{aligned}
\beta^I_{jk\ell}&=\frac{\pi^2}{3!}\delta^{IM}\sum_{\text{pert}(j,k,\ell)}\bigg[
\bigg(
\oD_{Mj}^{rs\bar{r}}\oD_{k\ell}^{tu\bar{s}}\bigg)\eta_{ru}\eta_{st}\eta_{\bar{r}\bar{s}}
+\bigg(
\bar{\oD}_{Mj}^{r\bar{r}\bar{s}}\bar{\oD}_{k\ell}^{s\bar{t}\bar{u}}\bigg)
\eta_{rs}\eta_{\bar{r}\bar{u}}\eta_{\bar{s}\bar{t}}
\bigg]\,.
\end{aligned}
\end{gather}%
For the group ${\rm SU}(2)$  with $k_L=k_R=k$, the special relation
\begin{equation}
\fF{a}{b}{r}\FF{r}{c}{d}=-\frac{1}{k}(\eta_{ad}\eta_{bc}-\eta_{ac}\eta_{bd})
\end{equation}
allows us to express the two loop beta function solely in terms of the OPE coefficients $C_{ink}$
\be\label{su2_beta}
\beta_{jkl}^{I}=-\frac{k\pi^2}{3}(C^{Ir}_{j}C_{rkl} + C^{Ir}_{k}C_{rlj} + C^{Ir}_{l}C_{rjk})\, .  
\ee  
Therefore, equation~\eqref{eq:cCC} for the $c$-function specializes to
\begin{equation}
C(\lambda)=c_0=2\pi^3\oC_{ijk}\lambda^i\lambda^j\lambda^k-\frac{3\pi^4}{2}\lambda^i\lambda^j\lambda^k\lambda^{\ell}\left[k+4\ln\left(\zeta\right)\right]\oC_{ij}{}^r\oC_{rk\ell}+\OL{5}\,,
\end{equation}
where $\zeta$ is an arbitrary, but fixed parameter (which is conventionally chosen as $\zeta=\Lambda/\mu$, for $\mu$ the subtraction scale and $\Lambda$ an arbitrary length scale).

As with the leading order contribution, the RG closure at two loops imposes the constraint
\begin{equation}
\beta_{(3)}^{\tilde{i}}=0\quad \forall \tilde{i}\, .
\end{equation}
In the ${\rm SU}(2)$ case any current-current perturbation which is  one loop closed is automatically two loop closed in view of formula (\ref{su2_beta}).

%% History 
A general formula for the beta function for anisotropic current-current interactions to all orders (in some scheme) was proposed in 
\cite{gerganov_beta_2001}. It was shown however in 
\cite{ludwig_4-loop_2003} that the conjectured general formula of \cite{gerganov_beta_2001} breaks down at four loops for all classical groups. 
Our two loop result agrees with all known models, such as e.g. the isotropic  Thirring  and the $U(1)$ anisotropic Thirring models,  studied in the literature.

The issues of symmetry breaking and restoration under the RG flows for current-current perturbations were studied  in \cite{sym1}, \cite{sym2}.

\subsection{Explicit examples of current-current perturbations}

In the following subsections, we will apply our formulae to a number of explicit current-current models in order to illustrate the phenomenon of redundancy. All these models will be based on a 
${\rm SU}(2)_k$ WZW model ($k_L=k_R=k$) for the unperturbed theory. For each model, we will first compute the redundancy data as described in~\eqref{rr}, i.e.\ the divergences of the chiral symmetry currents of the unperturbed theory. We will then identify those linear combinations of chiral currents   
\begin{equation} 
K_{\alpha} = \kappa_{\alpha}{}^{a}(\lambda)J_a\,,\quad \bar{K}_{\alpha} = \kappa_{\alpha}{}^{\bar{a}}(\lambda)\bar{J}_{\bar{a}}
\end{equation}
that close on the perturbing fields, $\partial_{\mu} K_{\alpha}^{\mu}=r_{\alpha}^i(\lambda)\phi_i$, and which form  the \emph{redundancy subalgebra} of the symmetry algebra 
of the fixed point theory. As described in equations~\eqref{eq:DefKappaLO}--\eqref{eq:redLO}, if we consider the set of dimension $2$ spin zero operators $\Phi_I$ as a vector space, finding the $K_{\alpha}$ at leading order in $\lambda$'s amounts to constructing a representation $Q_{\alpha}$ of the redundancy subalgebra with the block matrix form
\begin{equation}
\big(Q_{\alpha}\big)_I{}^J = \kappa^{(0)}_{\alpha}{}^a(\lambda)\big(Q_a\big)_I{}^J
+\kappa^{(0)}_{\alpha}{}^{\bar{a}}(\lambda)\big(\bar{Q}_{a}\big)_I{}^J=
\begin{pmatrix}
\big(Q_{\alpha}\big)_i{}^j & \big(0\big)_i{}^{\tilde{j}}\\
\big(0\big)_{\tilde{i}}{}^j & \big(Q_{\alpha}\big)_{\tilde{i}}{}^{\tilde{j}}
\end{pmatrix}\,,
\end{equation}
i.e.\ a fully reducible representation. The three models we will present in this section will realize at leading order fully reducible representations of the following redundancy subalgebras\footnote{Redundancy 
subalgebras should not be confused with conserved symmetry subalgebras which might be also present for the same perturbation.} of the 
unperturbed ${\rm su}(2)_L\oplus {\rm su}(2)_R$ chiral algebra:
\begin{gather}
\begin{aligned}
{\rm su}(2)_L&\subset {\rm su}(2)_L\oplus {\rm su}(2)_R&\qquad \text{(``conformal ${\rm SO}(3)$'' model)}\\
{\rm u}(1)&\subset {\rm su}(2)_L\oplus{\rm su}(2)_R&\qquad \text{(``three-coupling ${\rm U}(1)$'' model)}\\
{\rm su}(2)_{\text{diag}}&\subset {\rm su}(2)_L\oplus {\rm su}(2)_R&\qquad \text{(``six-coupling ${\rm SO}(3)$'' model)}\,.
\end{aligned}
\end{gather}%
We will refer to these models as indicated. The group present in the name is the redundancy group which up to the next-to-leading order  is generated by 
the redundancy vector fields  $\hat {\cal R}_{\alpha}$ introduced in (\ref{sp_basis}). Since the group is not changed from that identified at the leading order in practice 
we can use the leading order redundancy vector fields (\ref{eq:redLO}). 

\subsubsection{The conformal ${\rm SO}(3)$ model}

Consider a perturbation of ${\rm SU(2)}_{k}$ WZW model by three operators 
\begin{equation}
\delta S= \int d^2x\; \lambda^i \phi_{i}(x)\, , \qquad \phi_{i} =\frac{1}{k} J_{i} \bar J_{3} \qquad i\in\{1,2,3\} \, .
\end{equation}
The ${\rm SU}(2)_{\rm L}$ subgroup of the fixed point theory acts on the couplings $\lambda^i$ as on a three-vector thus forming the redundancy subgroup for this perturbation.  
Indeed our general formulas (\ref{rr})  imply 
\begin{gather}\label{chiralModcurr}
\begin{aligned}
\bar{\partial}J_1&=i\pi\left( 
 			{\lambda^{2}}\phi_{3} -{\lambda^{3}}\phi_{2}
			\right)+\OL{3}\, , \\
\bar{\partial}J_2&=i\pi \left(
 		 	{\lambda^{3}}\phi_{1}-{\lambda^{1}}\phi_{3}
			\right)+\OL{3}\, , \\
\bar{\partial}J_3&=i\pi \left(
			{\lambda^{1}}\phi_{2}-{\lambda^{2}} \phi_{1}
			\right)+\OL{3}\, , \\
\partial\bar{J}_{\bar{1}}&=-i \pi \left(
 			{\lambda^{1}}\phi_{{(12)}^{\perp}} 
			+{\lambda^{2}}\phi_{{(22)}^{\perp}} 
			+{\lambda^{3}}\phi_{{(32)}^{\perp}}\right)+\OL{3}\, , \\
\partial\bar{J}_{\bar{2}}&=i\pi \left(
			{\lambda^{1}}\phi_{{(11)}^{\perp}} 
 			+{\lambda^{2}}\phi_{{(21)}^{\perp}} 
 			+{\lambda^{3}}\phi_{{(31)}^{\perp}}\right)+\OL{3}\, , \\
\partial\bar{J}_{\bar{3}}&=0+\OL{3}\;
\end{aligned}
\end{gather}%
where   
\begin{equation}
\phi_{(ij)^{\perp}} = 
\frac{1}{k} J_{i}\bar{J}_{\bar{j}}\qquad i\in\{1,2,3\}\,,\quad j\in\{1,2\}\;
\end{equation}
are the complementary orthogonal operators.
We see from these formulae that
\begin{equation}
K_{a}=J_{a},\;\quad \bar K_{a}=0\,, \qquad (a=1,2,3)
\end{equation}
and the  redundancy vector fields are just the rotation vector fields in the 3d space of couplings
\begin{equation}
\widehat{R}_a^{(0)}=i\pi  \varepsilon_{ai}{}^j\lambda^{i}\partial_{j}+\OL{3}\,.
\end{equation}
The redundancy group is thus ${\rm SO}(3)$.
The orbits of the redundancy group are spheres centered at the origin of the coupling space, see figure~\ref{fig:chiralIllustr} below.

\begin{figure}[h!]   %[htbp]
\begin{center}
\includegraphics[width=0.5\textwidth]{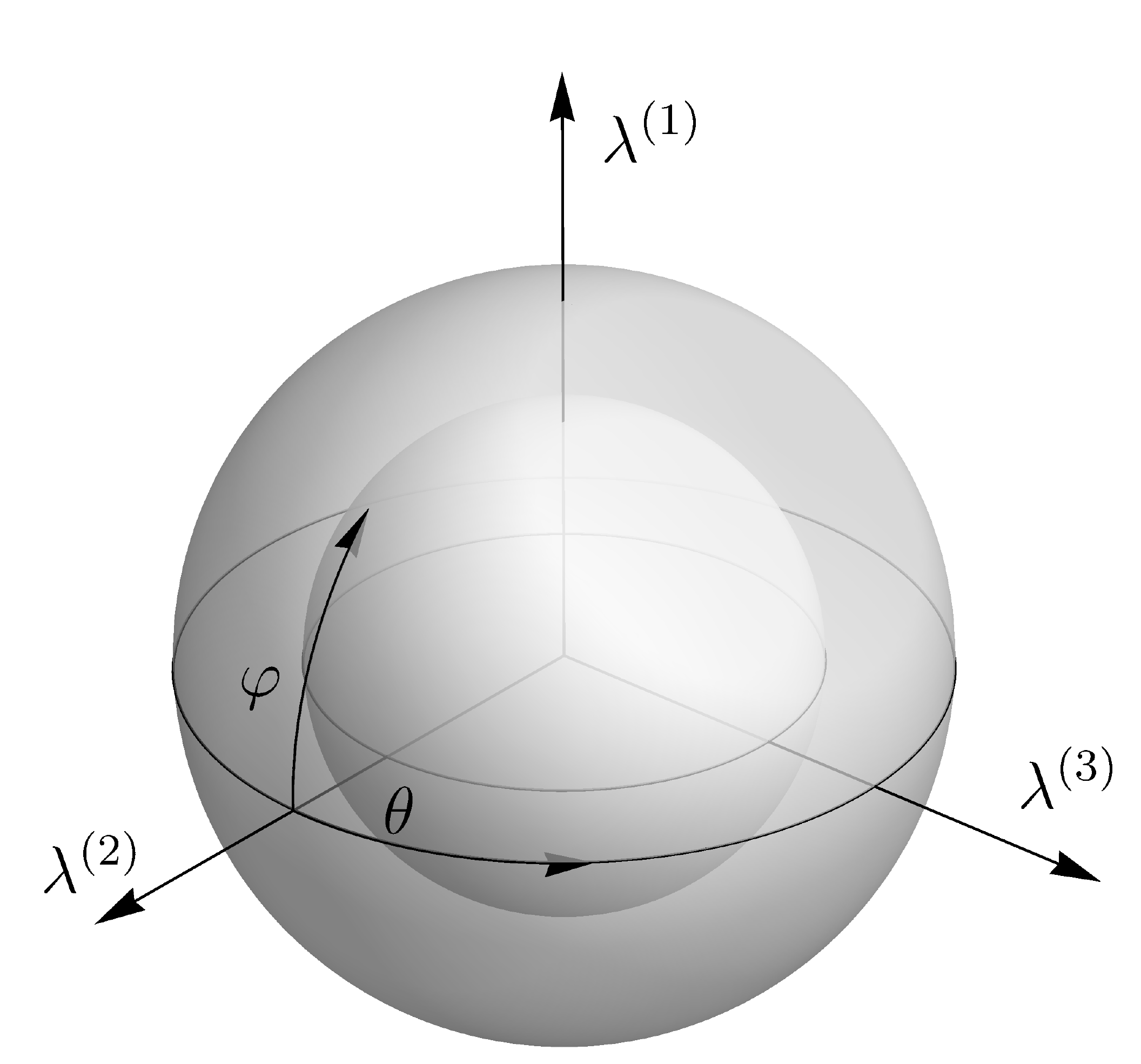}
\caption{The orbits of the action of the redundancy vector fields $\widehat{R}_a$  are spheres around the origin.}
\label{fig:chiralIllustr}
\end{center}
\end{figure}

Our formula for the beta function (\ref{GeneralCC}) implies that it vanishes at least through the two loop order. The general criterion of \cite{C-S} applies in our situation 
and says that the beta function vanishes to all orders. This is essentially due to the fact that the perturbation theory integrals are those of the free compact boson theory perturbed by a radius   
changing operator. It was shown in \cite{Moore} how to define those integrals so that the theory remains conformal. 

We also see from (\ref{chiralModcurr}) that  the perturbed theory has two conserved currents: 
\begin{equation}
J_L =\lambda^{1}J_1+\lambda^{2}J_2+\lambda^{3}J_3\, , \qquad J_{R}=\bar{J}_{\bar{3}}\, .
\end{equation}
The currents $J_{L}$, $J_{R}$   remain holomorphic  and anti-holomorphic respectively. This identifies the  ${\rm U(1)_{L}\times U(1)_{R}}$ symmetry 
currents in the deformed theory. For a particular point $\lambda^1=\lambda^2=0$ on the redundancy orbit, we are deforming by 
$J_{3}\bar J_3$. As is well known, the ${\rm SU(2)_{1}}$ theory is isomorphic to a free boson at the self dual radius. In this case, the operator $J_{3}\bar J_3$  is just the free boson radius changing operator.  For $k=1$ at the leading order we have $\lambda^3 = R - \frac{1}{R}$, where $R$ is the free boson radius (see e.g. \cite{FK_curv}, Appendix A for details). 
The all orders relationship between $R$ and $\lambda^3$ depends on the details of the subtraction scheme. In the scheme of \cite{Moore} we have 
\be
\lambda^3 = \frac{R-\frac{1}{R}}{R + \frac{1}{R}} \, . 
\ee
Evidently the T-duality transformation $R \mapsto \frac{1}{R}$ sends $\lambda^3 \mapsto - \lambda^3$. It is a well-known fact that the T-duality transformation for a free boson viewed as a deformed ${\rm SU(2)_{1}}$ theory 
can be understood as a discrete remnant of the ${\rm SU(2)_{L}\times  SU(2)_{R}}$ symmetry at the self dual radius (see e.g. \cite{Tduality}). In our case, when the redundant couplings are present, we can realize the T-duality transformation 
as a {\it  continuous} rotation in the space of couplings. In the full space of $\lambda^i$ couplings, T-duality just rotates any point on a sphere to its antipodal point.  
In fact rather than choosing  $\lambda^1=\lambda^2=0$ to specify a point on the quotient space, it is geometrically more natural to specify the nonredundant direction as a radial direction in the $\lambda^i$-space:
\be\label{rr22}
r = \sqrt{(\lambda^1)^2  + (\lambda^2)^2 + (\lambda^3)^2 }\, . 
\ee 
This variable is manifestly invariant under the redundancy group action including the T-duality. The quotient space under the redundancy group is then isomorphic to a half-line.   
While in the  $\lambda^i$ space, which includes redundant couplings, the geometry of the moduli space is smooth, in the quotient space it has a boundary singularity. 
The origin of this singularity is clear -- it came from a fixed point of the redundancy group action. This picture of the moduli space can be generalized to other exactly marginal deformations 
of WZW theories. The connection between T-duality and current-current deformations of WZW groups has been studied in \cite{Hassan_Sen}, but to the best of our knowledge the role of redundant directions 
in such deformations has not been systematically analyzed.

The RG anomaly currents are calculated using (\ref{Ji}) to be 
\be\label{Ji_cm}
J_{(i)} = \frac{i\pi}{k} \epsilon_{ij}^{k}\lambda^j J_{k} + {\cal O}(\lambda^2)\, , \qquad 
\bar J_{(i)} = {\cal O}(\lambda^2) \, .
\ee
Here we put the indices of these currents in parentheses  to distinguish them from the basis of WZW currents.
We observe that 
\be
\lambda^{1}J_{(1)}^{\mu} + \lambda^{2}J_{(2)}^{\mu} + \lambda^{3}J_{(3)}^{\mu} = 0 
\ee
(in the leading order) which means that no redundant operators admix to the invariant operator 
\be
\Phi = \frac{1}{r}\left(  \lambda^1\phi_1 + \lambda^2\phi_2 + \lambda^3\phi_3 \right)
\ee 
that couples to $r$ defined in (\ref{rr22}). 
The original perturbing operators $\phi_i$ contain some redundant operators in them. As a result they have anomalous dimensions and mix between 
themselves under the scale transformations. The anomalous dimension matrix  $\Gamma_{i}{}^j$ (cf. \ref{GammaCS}) is obtained 
by calculating  the divergences of the RG anomaly currents
\be
\Gamma_{i}{}^{j} \phi_{j}=-\partial_{\mu} J^{\mu}_{(i)}  
\ee
where 
\be
(\Gamma_{i}{}^{j}) = \frac{\pi}{k}  \left( \begin{array}{ccc} 
({\lambda^{2}})^2 + ({\lambda^{3}})^2 & -\lambda^1\lambda^2 & - \lambda^1\lambda^3 \\
-\lambda^1\lambda^2 & ({\lambda^{1}})^2 + ({\lambda^{3}})^2 & -\lambda^2\lambda^3 \\
-\lambda^1\lambda^3 & -\lambda^2\lambda^3 & ({\lambda^{1}})^2 + ({\lambda^{2}})^2
\end{array} \right) 
\ee
Evidently the invariant operator $\Phi$ does not have an anomalous dimension and does not mix with other operators. This can be made manifest by using spherical polar coordinates. 

Using (\ref{gamma_g}) we can also calculate the anomalous dimension matrices $\gamma_{a}{}^{b}$, $\bar \gamma_{\bar a}{}^{\bar b}$  for the currents $J_{a}$ and $\bar J_{\bar a}$: 
\begin{gather}
\begin{aligned}
\left(\gamma_{a}{}^{b}\right)&=\frac{\pi ^2}{k}\left(\begin{array}{ccc}
 -(\lambda^{2})^2 - (\lambda^{3})^2 &
  {\lambda^{1}} {\lambda^{2}}  &
   {\lambda^{1}} {\lambda^{3}}   \\ 
 {\lambda^{1}} {\lambda^{2}}  &
  -(\lambda^{1})^2 - (\lambda^{3})^2 &
   {\lambda^{2}} {\lambda^{3}} \\ 
 {\lambda^{1}} {\lambda^{3}} &
  {\lambda^{2}} {\lambda^{3}}  &
   -(\lambda^{1})^2 - (\lambda^{2})^2
\end{array}\right)\\
\left( \gamma_{\bar a}{}^{\bar b}\right)&=
 -\frac{\pi ^2}{k}\left(\begin{array}{ccc}
(\lambda^{1})^2 + (\lambda^{2})^2 + (\lambda^{3})^2 & 0 & 0  \\ 
 0 & (\lambda^{1})^2 + (\lambda^{2})^2 + (\lambda^{3})^2  & 0  \\ 
 0 & 0 & 0
\end{array}\right)\; . 
\end{aligned}
\end{gather}
We  see that   the currents $J_{L}$, $J_{R}$ do not develop any anomalous dimensions as expected. 

\subsubsection{The three-coupling ${\rm U}(1)$ model}
We next consider a current-current deformation of ${\rm SU(2)_{k}}$ which has a nontrivial RG flow.
The three perturbing operators are defined  as
\begin{gather} \label{3coupl}
\begin{aligned}
\phi_{(13)}& =  \frac{1}{k\sqrt{2}}\left(  J_{1}\bar J_{\bar 3}    + J_{3}\bar J_{\bar 1}\right)\\
\phi_{(22)}& = \frac{1}{k}J_{2}\bar J_{\bar 2}\\
\phi_{\widetilde{(13)}}&= \frac{1}{k\sqrt{2}}\left(J_1\bar J_{\bar 1} -J_{3}\bar J_{\bar 3}\right)
\end{aligned}
\end{gather}%
with the corresponding coupling constants $\lambda^{(13)}$, $ {\lambda^{\widetilde{(13)}}} $,  $\lambda^{(22)}$.

\begin{figure}[h!] %[htbp]
\begin{center}
\includegraphics[width=0.5\textwidth]{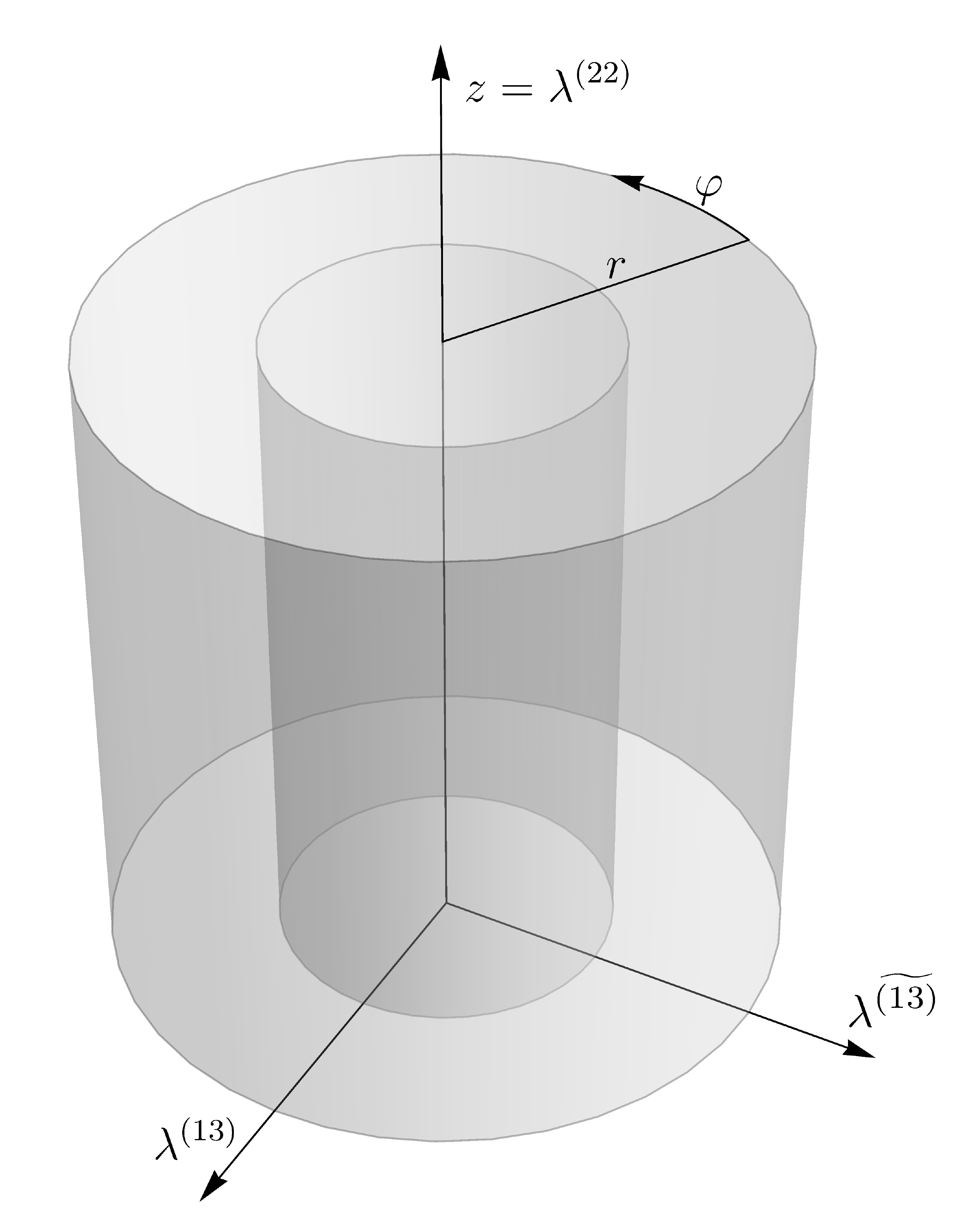}
\caption{The orbits of the action of the redundancy vector field $\widehat{R} \propto \frac{\partial}{\partial\varphi}$ in coupling space are cylinders around the $z\equiv \lambda^{(22)}$ axis.}
\label{fig:threeCouplingIllustr}
\end{center}
\end{figure}

Using (\ref{rr}) we find that the only divergences  that close on the set of perturbing operators are
\begin{gather}
\begin{aligned}
\bar{\partial}J_2 &=
i \pi \left(1 - \pi{\lambda^{(22)}}  \right)\left(
{\lambda^{(13)}}\phi_{\widetilde{(13)}}
 - {\lambda^{\widetilde{(13)}}}  \phi_{(13)}\right)+\OL{3}
\\
\partial\bar{J}_{\bar{2}} &=
 i \pi   \left(1 - \pi {\lambda^{(22)}} \right) \left(
   {\lambda^{(13)}} \phi_{\widetilde{(13)}} 
  -  {\lambda^{\widetilde{(13)}}}   \phi_{(13)}\right)+\OL{3}\,.
\end{aligned}
\end{gather}%
Hence we have 
a single redundancy vector field 
\[
\widehat{R}=2\pi i \left(1 - \pi {\lambda^{(22)}} \right) \left(
   {\lambda^{(13)}} \partial_{\widetilde{(13)}} 
  -  {\lambda^{\widetilde{(13)}}}   \partial_{(13)}\right)+\OL{3}\;
\]
generating a ${\rm U}(1)$ redundancy group. 
We also observe that the axial current  $(J_2,-\bar{J}_{\bar{2}})$ is conserved up to two loops (signaling a residual ${\rm U}(1)$ symmetry of the model).
Noticing that $\widehat{R}$ generates rotations in the $\lambda^{(13)}$--$\lambda^{\widetilde{(13)}}$ plane, we  introduce cylindrical coordinates as follows:
\begin{equation}
r=\sqrt{{\lambda^{(13)}}^2+{\lambda^{\widetilde{(13)}}}^2}\,,\quad \varphi=\arctan\left(\frac{\lambda^{\widetilde{(13)}}}{\lambda^{(13)}}\right)\,,\quad z=\lambda^{(22)}\; .
\end{equation}
Then, the redundancy vector field reads
\begin{equation}
\widehat{ {\mathcal{R}}}=2\pi i(1-\pi z)\frac{\partial}{\partial \varphi}+\OL{3}\,,
\end{equation}%
The orbits of the redundancy group are cylinders stretched 
along the $\lambda^{(22)}$ axis, as illustrated in figure~\ref{fig:threeCouplingIllustr}.

We compute  the $\beta$-function and the $c$-function in the cylindrical coordinates up to two loops
\begin{equation}\label{beta_3c}
\begin{array}{lll}
  \beta^{r}  & = &\frac{\pi r}{k}\left(2z-\pi (r^2+2z^2)\right)+\OL{4} \\
  \beta^{\varphi}  &= &0 +\OL{4} \\
  \beta^{z}  & = & \frac{\pi r^2}{k}\left(1-2\pi z\right)+\OL{4}\,,
\end{array}
\end{equation}%
 \begin{gather}\label{c_3c}
 \begin{aligned}
c(\lambda)&= {c_0} 
+ \frac{6  {\pi}^3}{k} zr^2- \frac{3 {\pi}^4}{2k^2}\left(k + 4  \ln{\zeta}\right)
 r^2 \left(r^2
 + 4z^2\right) +\mathcal{O}\left(\lambda^5\right)\;
\end{aligned}
\end{gather}%
where $c_{0}$ is the central charge of the fixed point. 
 We see that these quantities are   manifestly invariant under the action of $\widehat{R}$.\\
 The coordinates $r$ and $z$ are invariant under the redundancy action and are thus quite convenient for taking the quotient. 
 The quotient theory contains only two couplings: $r$ and $z$ with the beta functions (\ref{beta_3c}). 
 % that give an RG flow of the same universality class 
 % as the Kosterlitz-Thouless flow. 
 This two-coupling theory is a close relative of the anisotropic U(1) Thirring model and the sausage model  \cite{fateev_integrable_1993}. More precisely,  
 if we take instead of (\ref{3coupl}) the perturbing operators to be  
 \begin{gather}\label{2variant}
\begin{aligned}
\psi_{(13)}&= \frac{1}{k\sqrt{2}}\left(J_{1}\bar J_{\bar 3}-J_{3}\bar J_{\bar 1}\right)\, , \\
\psi_{(22)}&=\frac{1}{k}J_{2}\bar J_{\bar 2}\\
\psi_{\widetilde{(13)}}&= \frac{1}{k\sqrt{2}}\left(J_{1}\bar J_{\bar 1}+J_{3}\bar J_{\bar 3}\right) \,,
\end{aligned}
\end{gather}%
then  the diagonal  current  $(J_2,\bar{J}_{\bar{2}})$ is conserved, while the axial current $(J_{2}, -\bar{J}_{\bar{2}})$ generates the redundancy.
 Introducing cylindrical coordinates as before (with $\lambda^I$ coupling to $\psi_I$), the beta functions are
 \begin{gather}
 \begin{aligned}
  \beta^{r}  & = -\frac{\pi r}{k}\left(2z+\pi (r^2+2z^2)\right)+\OL{4} \\
  \beta^{\varphi}  &= 0 +\OL{4} \\
  \beta^{z}  & =  -\frac{\pi r^2}{k}\left(1+2\pi z\right)+\OL{4}\,,
\end{aligned}
\end{gather}%
and give a Kosterlitz-Thouless type flow.

The $c$-function reads
 \begin{gather}
 \begin{aligned}
c(\lambda)&= {c_0} 
- \frac{6{\pi}^3}{k} zr^2- \frac{3{\pi}^4}{2k^2}(k + 4 \ln{\zeta})
 r^2 \left(r^2
 + 4z^2\right) +\mathcal{O}\left(\lambda^5\right)\; .
\end{aligned}
\end{gather}%
Reducing this version of the model to the nonredundant directions (e.g. by keeping the $r$ and $z$ coordinates or by gauge fixing the redundancy so that $\varphi=0$ which
results in standard parameterization) we 
obtain exactly the $U(1)$ anisotropic Thirring (or sausage) model.

The RG anomaly currents for the model defined in (\ref{3coupl}) read\footnote{For the variant (\ref{2variant}) the anomaly currents are exactly the same with labels 
corresponding to \eqref{2variant}.} 
\bea
 J_{(13)}=&\frac{i\pi}{k}\lambda^{\widetilde{(13)}}J_2+\OL{2}\, , & \bar{J}_{(13)}=\frac{i\pi}{k}\lambda^{\widetilde{(13)}}\bar{J}_2+\OL{2}\, , \nonumber \\
 J_{(22)} =& 0+\OL{2}\, , & \bar{J}_{(22)}=0+\OL{2}\, , \nonumber \\
J_{\widetilde{(13)}}=&-\frac{i\pi}{k}\lambda^{(13)}J_2+\OL{2}\, , & \bar{J}_{\widetilde{(13)}}=-\frac{i\pi}{k}\lambda^{(13)}\bar{J}_2+\OL{2} \, .
\eea
 Knowing these currents allows us to calculate the complete matrix of anomalous dimensions both for the perturbing operators and for the currents. 
 For the perturbing operators the complete mixing matrix has two contributions: 
 \be
 \Gamma_{i}{}^{j} = \partial_{i}\beta^{j} - \Delta_{i}{}^{j}
 \ee
 where $\Delta_{i}^{j}$ is defined as  
 \be
 \partial_{\mu} J^{\mu}_{i} = \Delta_{i}{}^{j} \phi_{j} \, .
 \ee
 In cylindrical coordinates ordering the coordinates as $(r, \phi, z)$ we obtain 
 \be
(\partial_{i}\beta^{j}) =  \frac{2\pi}{k} \begin{pmatrix}
\left( z-\pi  (\frac{3}{2}r^2+z^2)\right) & 0 & r (1 - 2\pi z) \\
0 & 0 & 0\\
  r (1 - 2\pi z) & 0 & -\pi  r^2
\end{pmatrix}\,, 
\ee
 \be
 (\Delta_i{}^j)=\frac{2\pi^2r^2}{k} \begin{pmatrix}
0 & 0 & 0\\
0 & 1 & 0\\
0 & 0 & 0
\end{pmatrix}\,.
 \ee
The sum of these two matrices gives the mixing matrix $\Gamma_{i}{}^{j}$ up to terms of order $\lambda^3$.
 
The matrix of anomalous dimensions of the currents $\gamma_{a}^{b}$  reads
\begin{gather}
\begin{aligned}
(\gamma_{a}^{b}) & =-\frac{\pi ^2}{2k}
\left(\begin{array}{ccc}
r^2 + 2z^2& 0 & 0  \\ 
 0 & 2r^2& 0  \\ 
 0 & 0 & r^2+2z^2
\end{array}\right)\,,
%\\&=k^3 \pi ^2\left(\begin{array}{ccc}
%Q(\Lambda)-tr(\lambda)& 0 & 0  \\ 
% 0 & 2Q(\Lambda)& 0  \\ 
% 0 & 0 & Q(\Lambda)-tr(\lambda)
%\end{array}\right)\,,
\end{aligned}
\end{gather}%
and the same expression also gives the matrix elements of  $\bar \gamma_{\bar a}^{\bar b}$. 

 The metric with one loop correction in cylindrical coordinates ordered as $(r, \varphi, z)$ is
\begin{equation}
(g^{(0)}_{ij})+ (g^{(1)}_{ij})=6\pi^2\left( \begin{array}{ccc}
1&0&0\\
0&r^2&0\\
0&0&1
\end{array} \right) 
-\frac{24\pi^3}{ k} \ln \zeta  \begin{pmatrix}
z & 0 & r\\
0 & r^2 z & 0\\
r & 0 & 0
\end{pmatrix}\,.
\end{equation}
We observe that in the corrected metric the redundant coordinate $\varphi$ remains orthogonal to the nonredundant ones. Moreover the metric for the 
nonredundant coordinates is independent of $\varphi$. This, together with the form of the $\beta$ and $c$-function (\ref{beta_3c}), (\ref{c_3c})  makes the reduction of the gradient formula straightforward.

\subsubsection{The six-coupling ${\rm SO(3)}$ model}

We will finally present a current-current perturbation model which is nonconformal and exhibits a non-Abelian redundancy symmetry. To this end, again starting from the ${\rm SU}(2)_k$ WZW model, 
define six perturbing operators
\begin{gather}
\begin{aligned}
\phi_{(ij)} & = \left\lbrace
\begin{array}{ccl}
\frac{1}{k}J_{i}\bar J_{\bar i}&\quad& i=1,2,3\\
\frac{1}{k\sqrt{2}}\left(J_{i}\bar J_{\bar j}+ J_{j}\bar J_{\bar i}\right)&\quad& i<j,\;i,j\in\{1,2,3\}
\end{array}\right.
\end{aligned}
\end{gather}%
and the orthonormal operators
\begin{equation}
\phi_{(ij)^{\perp}} = \frac{1}{k\sqrt{2}}\left(J_{i}\bar J_{\bar j}-J_{j}\bar J_{\bar i}\right)\quad i<j,\;i,j\in\{1,2,3\}\,.
\end{equation}
It is convenient to consider the six couplings at hand as entries of  a symmetric matrix  $\Lambda$, defined as
\begin{equation}
\Lambda = ( d^{a\bar a}_{A}\lambda^A)=\frac{1}{k}\begin{pmatrix}
\lambda^{(11)} & \frac{1}{\sqrt{2}}\lambda^{(12)} & \frac{1}{\sqrt{2}}\lambda^{(13)}\\
\frac{1}{\sqrt{2}}\lambda^{(12)} & \lambda^{(22)} & \frac{1}{\sqrt{2}}\lambda^{(23)}\\
\frac{1}{\sqrt{2}}\lambda^{(13)} & \frac{1}{\sqrt{2}}\lambda^{(23)} & \lambda^{(33)}
\end{pmatrix}\; .
\end{equation}
This matrix possesses three invariants, which may be computed as the coefficients of powers of the variable $\mu$ in the characteristic polynomial of $\Lambda$:
\be \label{3invariants}
p(\Lambda)=\det\left(\Lambda-\mu\; {\mathbb 1}_{3\times 3}\right)
=-\mu^3+\mu^2{\rm tr}(\Lambda)-\mu\mathcal{Q}(\Lambda)+\det(\Lambda)\,,
\ee
where ${\rm tr}(\Lambda)$ and ${\rm det}(\Lambda)$ are the trace and determinant, 
while 
\be
Q(\Lambda)=\frac{1}{2}\left(
{\rm tr} (\Lambda)^2-{\rm tr}(\Lambda^2)
\right)\, . 
\ee
The matrix $\Lambda$ can be diagonalized by an orthogonal transformation $O$
\begin{equation}
O\Lambda O^{T}=\begin{pmatrix}
\tau^{1} & 0 & 0\\
0 & \tau^{2} & 0\\
0 & 0 & \tau^{3}
\end{pmatrix} \,,
\end{equation}
where the entries $\tau^{i}$ are the eigenvalues of $\Lambda$. Since the matrices $O$ form the Lie group $O(3)$, which has rank $3$, we may consider a reparameterization of our coupling matrix $\Lambda$ in terms of the 
eigenvalues $\tau^{i}$ and three parameters $\theta^{j}$ for the matrix $O$. The invariants of the matrix $\Lambda$  depend on the variables $\tau^{i}$ only,
\be
{\rm tr}(\Lambda) = \tau^{1} + \tau^2 + \tau^3\, , \quad 
Q(\Lambda) = \tau^1\tau^2 + \tau^1\tau^3 + \tau^2\tau^3 \, , \quad 
{\rm det}(\Lambda) = \tau^1\tau^2\tau^3 \, . 
\ee

In the first two orders of perturbation the redundancy currents are just the diagonal currents $K_{a}=(J_{a}, \bar J_{a})$, $a=1,2,3$. 
We omit the explicit expressions as they are quite long and confine ourselves to spelling out the net result. 
We have checked that the corresponding redundancy vector fields generate infinitesimal orthogonal transformations on the coupling matrix $\Lambda$
\be
\Lambda \mapsto \Lambda + \epsilon [\Lambda, X] \, , \quad X^{T} = -X \, .
\ee
The redundancy transformations form a group isomorphic to ${\rm SO}(3)$ that acts on the coupling space by similarity transformations 
\be \label{ort_act}
\Lambda \mapsto O\Lambda O^{T} \, , \quad O\in {\rm SO}(3) \, .
\ee
We note that the  representation of ${\rm SO}(3)$ given by (\ref{ort_act}) is a well known example of a polar representation (see e.g. \cite{Palais}). 
The subset of diagonal matrices forms what is called a section of the foliation - a submanifold that meets 
all leaves orthogonally. We will discuss polar representations further in the next subsection.

We have also checked that the two loop beta functions commute with this action and found that 
the two-loop $c$-function can be compactly expressed in terms of the 3 invariants  as
\begin{equation}
c(\lambda)=c_0
-\frac{12\pi^3}{k}\det(\Lambda)  
- \frac{6\pi^4}{k^2} (k + 4\ln\zeta) (Q(\Lambda)^2 - 2 \det(\Lambda) {\rm tr}(\Lambda)) +\OL{5}\,.
\end{equation}
It is convenient to introduce instead of $\lambda^{(ij)}$ a different set of local coordinates  taking the eigenvalues $\tau^1, \tau^2, \tau^3$ and any local coordinates $\theta^1, \theta^2, \theta^3$ on the group   
${\rm SO}(3)$. In this splitting the redundancy group only acts on the $\theta^{i}$ coordinates.
% except for the permutations of $\tau^i$ which are a global remnant of the redundancy group.   
 The beta functions only have components in the $\tau^i$ directions given by 
\begin{gather}
\begin{aligned}
\beta^{1} &= -\frac{2\pi}{k} {\tau^{2}} {\tau^{3}} 
 - \frac{2\pi^2}{k} {\tau^{1}} \left(({\tau^{2}})^2 + ({\tau^{3}})^2\right)   +\mathcal{O}(\tau^3)\\ 
\beta^{2} &= -\frac{2\pi}{k}{\tau^{1}} {\tau^{3}}  
- \frac{2\pi^2}{k}{\tau^{2}} \left(({\tau^{1}})^2 + ({\tau^{3}})^2\right)   +\mathcal{O}(\tau^3)\\ 
\beta^{3} &= -\frac{2 \pi }{k}  \tau^{1} {\tau^{2}} 
 - \frac{2\pi^2}{k}{\tau^{3}}\left(({\tau^{1}})^2 + ({\tau^{2}})^2\right)+\mathcal{O}(\tau^3)\, .
\end{aligned}
\end{gather}%
We can thus reduce the theory to the one in which only the three invariant coordinates $\tau^i$ are present. The reduced theory is isomorphic to the following perturbation of the ${\rm SU(2)}$ WZW theory 
\be
\delta S = \int\!\! d^{2} z\, (\tau^{1}J_{1}\bar J_{\bar 1} + \tau^{2} J_{2}\bar J_{\bar 2} + \tau^{3}J_{3}\bar J_{\bar 3}) \, . 
\ee

We have worked out the RG anomaly currents $J_{i}^{\mu}$ and checked that they vanish in the nonredundant directions $\tau^{i}$.
In the next subsection we discuss in more detail how the gradient formula can be reduced to the nonredundant directions for this and other models in which the redundancy group 
representation is polar. The reduced metric for the 6-coupling model can be obtained by reducing the perturbative Zamolodchikov metric to diagonal matrices $\Lambda$. 
We omit the corresponding formulas.

\subsubsection{Reducing the gradient formula}\label{sec:reduction}
For the ${\rm U(1)}$ and the  6-coupling ${\rm SO}(3)$ models the two loop gradient formula has the form 
\be \label{grad_again}
\partial_{i}c = -g_{ij}\beta^{j} \, .
\ee
As we have seen above for each model, both the $c$-function and the beta functions reduce naturally to the submanifold parameterized by 
the invariant coordinates. We now would like  to discuss the reduction of the metric and the reduced gradient formula in greater generality. 
We will consider perturbed CFTs with redundancy group $G$  up to two loops as analyzed in section \ref{general_cp}.
To reduce the gradient formula (\ref{grad_again})  we pick new local coordinates 
\begin{equation}
\{\tilde \lambda^{\alpha}\}=\{\tau^a\}\cup\{\theta^r\}
\end{equation}
 such that $\theta^{a}$ are the coordinates in the  redundant directions and $\tau^{a}$ are the nonredundant ones. In this subsection we will use the indices $a,b,c$ in tensors for the $\tau$-directions and 
 $r,s,t$ for the $\theta$-directions. To distinguish all quantities calculated in the $\tilde \lambda^{\alpha}$ coordinates we will put a twiddle above them.  For the nonredundant directions we get 
 \be
 \partial_{a} \tilde c = - \tilde g_{ab}\tilde \beta^{b} -\tilde  g_{ar}\tilde \beta^{r} \, . 
 \ee
To reduce this formula consistently we need to pick coordinates in which 
\be \label{conditions}
\tilde g_{ar}\tilde \beta^{r}=0\, , \quad \tilde g_{ab} = \tilde g_{ab}(\tau) \, , \quad  \tilde \beta^{a}= \tilde \beta^{a}(\tau)\, , 
\ee
that is the last two quantities  are functions of the coordinates $\tau^{a}$  only. 

We showed in section \ref{general_cp} that in the original coordinates $\lambda^{i}$ the leading order metric is up to a constant factor the standard Euclidean metric: 
\be
g^{(0)}_{ij} = 6\pi^2 \delta_{ij} \,  
\ee
while the metric correction can be written as 
\be
g^{(1)}_{ij} = -\partial_i\partial_jG_{(1)}+\OL{2}\, , 
\qquad G_{(1)} = 4\pi^3\ln{\zeta}\left(\oC_{ijk}\lambda^i\lambda^j\lambda^k\right)\,.
\ee
The beta functions up to two loops can be written as 
\bea
&\beta^i=\partial_i \left(\beta_{(2)}+\beta_{(3)}\right)+\OL{4}\,, \\ 
 &  \beta_{(2)} = \frac{\pi}{3} \sum_{ijk} C_{ijk}\lambda^{i}\lambda^{j}\lambda^{k} \, , \qquad 
  \beta_{(3)} = \frac{1}{4}\sum_{ijkl} \beta^{l}_{ijk}\lambda^{i}\lambda^{j}\lambda^{k}\lambda^{l}\, . 
\eea
(The apparent noncovariant look of this equation is due to having particular coordinates with  a flat metric $\delta_{ij}$.)
Since the tensors $C_{ijk}$, $\beta^{l}_{ijkl}$ are invariant under the action of $G$,  so are the potential functions $G_{(1)}$, $\beta_{(2)}$, $\beta_{(3)}$. 
We further choose  the coordinates $\tau^{a}$ to be invariant under the action of $G$, that is 
\be
{(Q_{\alpha})_{ i}}^{j} \lambda^{i} \partial_{j} \tau^{a} = 0 \, . 
\ee
The above potential functions are thus functions of $\tau^{a}$ only. Using this we obtain for the leading order beta function and metric in the $\tilde \lambda^{\alpha}$ coordinates 
\be
 \tilde \beta^{(2)a} = \eta^{ab} \partial_{b}\tilde \beta_{(2)}  \, , \quad  \tilde g^{(0)}_{\alpha \beta} = 6\pi^2 \eta_{\alpha \beta} 
\, , 
\ee
\be
\eta^{\alpha \beta} = \sum_{i} \frac{\partial \tilde \lambda^{\alpha}}{\partial \lambda^{i}} 
  \frac{\partial \tilde \lambda^{\beta}}{\partial \lambda^{i}} 
  \ee
where the matrix $\eta_{\alpha \beta}$ is the  inverse to $\eta^{\alpha \beta}$ and in the first two equations one should only 
retain in $\eta_{\alpha \beta}$ and $\eta^{\alpha \beta}$ the leading order terms  in the $\tilde \lambda^{a}$ expansion.    
The one-loop gradient formula then is 
\be
\partial_{a} \tilde c^{(3)} = -\tilde g_{ab}^{(0)}\tilde \beta^{(2)b} - 6\pi^2 \tilde g^{(0)}_{ar}\tilde g^{(0)rc}\partial_{c}\tilde \beta^{(2)} \, . 
\ee
We see that the conditions (\ref{conditions}) at this order imply that the metric $\tilde g^{(0)}_{\alpha \beta}$ must be of the form
\begin{equation}\label{polar_metric} 
(\tilde g^{(0)}_{\alpha\beta})=\begin{pmatrix}
\tilde g^{(0)}_{ab}(\tau^i) & 0\\
0 & \tilde g^{(0)}_{rs}(\tau^i,\theta^r)
\end{pmatrix}\, .
\end{equation}
As we will see shortly, the following stronger condition is more natural and will also ensure a consistent reduction at two loops, namely we 
will require that the tensor $\eta_{\alpha \beta}$ has the block form 
\be
\label{polar_metric2} 
(\eta_{\alpha\beta})=\begin{pmatrix}
\eta_{ab}(\tau^i) & 0\\
0 & \eta_{rs}(\tau^i,\theta^r)
\end{pmatrix}\, .
\ee
This means 
that  the coordinates $\theta^{r}$ are orthogonal to the coordinates $\tau^{a}$ with respect to the standard flat space metric and the invariant coordinates block depends only on $\tau^a$.
Such coordinates can be considered as an analogue of spherical coordinates associated with the standard ${\rm SO}(n)$ action in ${\mathbb R}^n$. It was shown in 
\cite{Dadok} that such coordinates can be constructed when the representation ${(Q_{\alpha})_{i}}^{j}$ of $G$ is {\it polar}. An orthogonal  representation 
 is called polar if there exists a complete connected submanifold 
that meets all  orbits orthogonally.  Such a submanifold is called a section and in physics language it is a special gauge slice. 
In the three examples   considered before the representation of $G$ was polar and thus the gradient formula (at least at one loop) can be 
consistently reduced as our explicit calculations indeed showed. Assuming the metric is of the form (\ref{polar_metric}) we further obtain that $\tilde \beta^{r}$ vanishes at one loop. 
 
 At two loops we obtain for the metric correction 
 \begin{equation} \label{g_11}
\tilde g_{\alpha\beta}^{(1)}=
\partial_{\alpha}\partial_{\beta}\tilde{G}_{(1)}-\tilde{\Gamma}^{\gamma}_{\alpha\beta}\partial_{\gamma}\tilde{G}_{(1)}\,,
\end{equation}
where 
$$
\tilde{\Gamma}_{\alpha\beta}^{\gamma}
= \frac{1}{2}\eta^{\gamma\delta}\left(\partial_{\alpha}\eta_{\delta\beta}+\partial_{\beta}\eta_{\alpha\delta}-\partial_{\delta}\eta_{\alpha\beta}\right)
$$
 are the Christoffel symbols for the flat metric $\delta_{ij}$ in the $\tilde \lambda^{\alpha}$ coordinates that in (\ref{g_11}) we assume to be truncated at the leading order. 
Using (\ref{polar_metric2}) we find that $\tilde \Gamma_{ab}^{c}$ is a function of $\tau^d$ only and hence so is $\tilde g_{ab}^{(1)}$. Moreover 
since $\tilde \Gamma_{ar}^b=0$ we have $\tilde g_{ar}^{(1)}=0$. This means that the metric correction $\tilde g^{(1)}_{\alpha \beta}$ is of the same form 
as (\ref{polar_metric}). The two loop beta function for the nonredundant coordinates has the form 
\be
\tilde \beta^{(3)a} = \eta^{(0)ab}\partial_{b} \tilde \beta_{(3)} + \eta^{(1)ab}\partial_{b} \tilde \beta_{(2)} 
\ee
where the upper bracketed index of $\eta^{(i)ab}$ labels the corresponding order of expansion in $\tilde \lambda^{\alpha}$. Formula 
(\ref{polar_metric2}) implies that the two loop beta function is independent of $\theta^r$ and $\tilde \beta^r = 0$. 
The two loop gradient formula then reduces to the $\tau^a$-directions: 
\be
\partial_{a} \tilde c^{(4)} = - \tilde g^{(0)}_{ab}\tilde \beta^{(3)b} - \tilde g^{(1)}_{ab}\tilde \beta^{(2)b} \, . 
\ee
It is tempting to conjecture that  the Zamolodchikov metric $g_{\alpha \beta}$ will remain polar to all orders as long as 
all perturbative corrections will be expressed in terms of $G$-invariant tensors.

%%%%%%%%%%%%%%%%%%%%%%%%%%%%%%%%%%%%%%%%%%%%%%%%%%%%%%%% CONCLUSIONS
\section{Concluding remarks}\label{discussion}

In this section we will try to summarize what we have  learned and will talk about   the open questions and future directions. 

What we have seen in the conformal perturbation theory analysis is that in the vicinity of fixed points with symmetry we can construct theories in which redundant operators originate from the
broken symmetries. At the two loop level we observed that the redundancy vector fields  close under the Lie bracket and the corresponding integral surfaces give a {\it foliation} in the coupling space. 
Theories on the same leaf of this foliation differ only by parameterization of observables  and are physically equivalent. 

Moreover, in conformal perturbation theory the leafs are generated by an action of a certain group -- the {\it redundancy group}.
The appearance of this group has a simple origin. At the fixed point we can construct this group as a subgroup of the symmetry group that preserves the form of the perturbation, i.e.\ its action 
on the perturbing operators can be undone by reparameterizing the couplings. In the perturbed theory one can imagine a subtraction scheme that will preserve this action to all orders. 
For example for the current-current perturbations,  correlators of operators constructed using currents only are rational functions multiplied by tensors invariant under the action of the above specified subgroup.
Thus any subtraction scheme that modifies the rational functions only and leaving the tensors intact will do.  In particular, point splitting plus minimal subtraction will preserve the redundancy group. 

Although this picture of a foliation associated with a certain group action, which we observe in conformal perturbation analysis, is very suggestive, it is not clear that this is the case in general. 
One can show however that a collection of vector fields closed under the Lie bracket and the associated foliation  do arise at least perturbatively to all orders. 
This is a consequence of the Wess-Zumino consistency conditions applied to the
redundancy anomalies:
\be\label{RaRb}
[{\cal R}_{a}(x), {\cal R}_{b}(y)] = 0 \, . 
\ee 
This result will be presented elsewhere \cite{FK_redundancy}. 

Another salient feature that was present in  our examples is that {\it the foliation associated with redundancy is preserved by the RG flow}. In fact this is a general  consequence of the Wess-Zumino consistency 
condition (\ref{Rbeta_comm}). The RG flow moves any two physically equivalent theories on the same leaf to a pair of physically equivalent theories. In particular this implies that one 
can reduce the beta function to a transverse section of the foliation. One can think of such a transverse section as a gauge choice, i.e. a choice of nonredundant directions. 
   We have also shown in section \ref{sec:red} that for a fairly general class of perturbations\footnote{The situation is more complicated for nonlinear sigma models, see footnote \ref{ft}. }
   the $c$-function is invariant under the shifts in redundant directions (\ref{red_c}). To reduce the gradient formula to a transverse section (a nonredundant gauge slice) we also 
   need to reduce the metric and the antisymmetric form.   This has to be done in such a way that the reduced tensors are independent of the choice of the section (up to the change of 
   coordinates in the reduced theory). We have seen in the particular models studied in section \ref{sec:models} that this is possible to do by choosing coordinates invariant under the redundancy group action. 
   Moreover in section \ref{sec:reduction} we showed that at two loops in conformal perturbation there is a consistent reduction for any model in which the (fixed point) {\it representation of 
   the redundancy group is polar}. 
   One important  property of the analysis in section \ref{sec:reduction} was the invariance of the metric tensor under the redundant vector fields 
   $$
   {\cal L}_{\alpha} g_{ij} = 0 \, 
   $$ 
   that holds up to two loops in conformal perturbation in certain coordinates. 
   In general the Lie derivative of the Zamolodchikov metric can be written as (see (\ref{red_connection}), (\ref{Gamma_a})) 
\be \label{metric_Lie}
{\cal L}_{a} g_{ij} = - r_{ai}^{c}r_{c}^{k} g_{kj}  - r_{aj}^{c}r_{c}^{k} g_{ik} \, . 
\ee
The Wess-Zumino conditions (\ref{RaRb}) imply that the connection coefficients $r_{ai}^{b}$ satisfy a zero curvature condition \cite{FK_redundancy}. One may hope to use this fact to 
bring the right hand side of (\ref{metric_Lie}) under control. It is plausible then that an analogue of the coordinate split associated with the redundancy group action which we 
have exploited in conformal perturbation does exist more generally.  One also needs to analyze the action of the redundancy vector fields on the antisymmetric form $b_{ij}$ and the 
metric correction $\Delta g_{ij}$. 
Moreover, having shown that one can consistently reduce  all the geometric objects to a transverse section, one still needs to work out how the reduced objects transform 
under a change of scale (cf.\ (\ref{CS2})). In the examples analyzed in sections  \ref{relevant} and \ref{sec:models}, we showed that the RG anomaly currents for the invariant (transverse) coordinates are 
absent and thus the reduced objects transform geometrically (by the ${\cal L}_{\beta}$ Lie derivative) under the change of scale. More generally the transformation will be geometric if the nonredundant directions are 
orthogonal to the redundant ones.  As we showed in section \ref{sec:reduction}, one can choose such coordinates (up to two loops) for any model in which the redundancy group representation 
is polar.   
It remains to see whether these results can be generalized. We leave these questions to future work.  

There are other more technical questions which would be interesting to pursue further. 
At the level of analyzing specific perturbations, two interesting closure questions have arisen. It may be the case that new redundant operators, which are not combinations of the original perturbing operators, 
 emerge in the commutator of the original redundancy vector fields with the beta function. Another point where we may need to enlarge the space of couplings to include extra redundant 
operators is when expressing the total derivatives of the RG anomaly currents $J_{i}^{\mu}$.   Although we have not succeeded in constructing interesting examples exhibiting such situations, 
as far as we can see there is no general principle that would forbid them. % It would be interesting to construct models exhibiting such features. 

While the discussion in this paper focused on the redundancy aspect, it was interesting to see the models discussed as examples of the geometric objects present in the gradient formula. 
We saw that the antisymmetric form $b_{ij}$ at the two loop order appeared only for relevant perturbations (see section \ref{relevant}). For marginal perturbations one could detect the 
appearance of $b_{ij}$ by checking whether the 1-form $g_{ij}\beta^{j}$ is closed (given that we showed that $\Delta g_{ij} $ may appear only at very high orders).
In the perturbative corrections to the Zamolodchikov metric however at the next-to-leading order we may  see a nontrivial  curvature tensor. In Riemann 
normal coordinates we have 
$$g_{ij} = 6\pi^2(\delta_{ij} + \frac{1}{3}R_{ikjl}\lambda^{k}\lambda^{l}+ \dots ) $$ 
where $R_{ikjl}$ is the Riemann curvature tensor for the Zamolodchikov 
metric (see \cite{FK_curv} for a recent discussion). There is no reason to expect that the 1-form 
$$
f_{i} \equiv R_{ikjl}C^{j}_{mn}\lambda^{k}\lambda^{l}\lambda^{m}\lambda^{n}\, , 
$$  
which we obtain contracting the metric correction with the leading order beta function,  is closed so one may expect a nontrivial 2-form $b_{ij}$ to appear at the order 
${\cal O}(\lambda^2)$\footnote{One also needs to analyze along with $f_{i}$ the 3-loop beta function calculated in Riemann normal coordinates}.
As the role (and possible use) of $b_{ij}$ is not understood, it would be interesting to do more calculations exhibiting its appearance. The same goes for the tensor $\Delta g_{ij}$ 
which so far only has been detected for nonlinear sigma models. 

\begin{center}
{\bf Acknowledgements}
\end{center}
We are grateful to Daniel Friedan and Hugh Osborn for useful discussions.
This work was  supported  by the Leverhulme grant   RPG-184.

\appendix
\renewcommand{\theequation}{\Alph{section}.\arabic{equation}}
\setcounter{equation}{0}

\section{Details of the beta function computations}\label{app:DerBeta}
\subsection{The method}
We first remind the reader of the  method for computing the beta functions presented in~\cite{Gaberdiel:2008fn}, specializing to the case of perturbations by dimension $2$  operators $\phi_i$. 
 Consider  the series expansion formula for the partition function of the perturbed theory in orders of $\lambda$'s:
\begin{align}
\begin{split}
\left\langle \mathbb{1}\right\rangle_{\lambda}
&=\left\langle  e^{\delta S}\right\rangle_0\\
&=\left\langle \mathbb{1} \right\rangle_0
+\sum_i\lambda^i \int d^2z\; \left\langle\phi_i(z,\bar{z})\right\rangle_0\\
&+ \frac{1}{2!}\sum_{j,k}\lambda^j\lambda^k \int d^2z_j\int d^2 z_k\; \Theta_{jk} 
\left\langle\phi_j(z_j,\bar{z}_j)\phi_k(z_k,\bar{z}_k)\right\rangle_0\\
&+ \frac{1}{3!}\sum_{j,k,\ell} \lambda^j\lambda^k\lambda^{\ell} \int d^2z_j\int d^2 z_k\int d^2 z_{\ell}\; \Theta_{jk}\Theta_{jl}\Theta_{kl} 
\left\langle\phi_j(z_j,\bar{z}_j)\phi_k(z_k,\bar{z}_k)\phi_{\ell}(z_{\ell},\bar{z}_{\ell})\right\rangle_0\\
&+ \mathcal{O}(\lambda^4)\:
\end{split}
\end{align}%
where 
\begin{equation}\label{eq:defThetaDiff}
\Theta_{jk}= H(\vert z_j-  z_k\vert -\epsilon)H(L-\vert z_j -z_k \vert)\;
\end{equation}
are the cutoff functions that ensure $\varepsilon < \vert z_j-z_k\vert< L$ for any pair of variables $z_j$ and $z_k$. The RG invariance implies 
\begin{equation}\label{eq:basicBF}
\lim_{\epsilon\to 0} \epsilon\frac{d}{d\epsilon} e^{\delta S}=0
\end{equation}
provided that $\epsilon \partial_{\epsilon}\lambda^i = \beta^{i}(\lambda)$.
We treat (\ref{eq:basicBF}) as an  operator equation, i.e. inside   correlation functions.  
To pick a particular operator content we   insert  \eqref{eq:basicBF} into a correlator with an asymptotic state $\phi_m(\infty)$ (of dimension $2$ and spin $0$), where as usual
\begin{equation}
\left\langle\phi_m(\infty) \ldots\right\rangle\equiv \lim\limits_{z\to\infty}\vert z\vert^4\left\langle \phi_m(z,\bar{z})\ldots\right\rangle\,.
\end{equation}
We obtain 
\begin{align}
\begin{split}
&=\epsilon\partial_{\epsilon}\bigg\lbrace
\sum_i\lambda^i \int d^2z\; \left\langle \phi_m(\infty)\phi_i(z,\bar{z})\right\rangle_0\\
&+ \frac{1}{2!}\sum_{j,k}\lambda^j\lambda^k  \int d^2z_1\int d^2 z_2\; \Theta_{12} 
\left\langle \phi_m(\infty)\phi_j(z_1,\bar{z}_1)\phi_k(z_2,\bar{z}_2)\right\rangle_0\\
&+ \frac{1}{3!}\sum_{j,k,\ell}\lambda^j\lambda^k\lambda^{\ell} \int d^2z_j\int d^2 z_k\int d^2 z_{\ell}\; \Theta_{jk}\Theta_{jl}\Theta_{kl} 
\left\langle \phi_m(\infty)\phi_j(z_j,\bar{z}_j)\phi_k(z_k,\bar{z}_k)\phi_{\ell}(z_{\ell},\bar{z}_{\ell})\right\rangle_0 + \dots \bigg\rbrace\\
&= {\cal O}(\epsilon) \, .  \label{eq:betaFunctionAnsatz}
\end{split}
\end{align}%
Using translation invariance to factor out the volume element and introducing the quantities 
\begin{align}
\begin{split}\label{eq:TwoPointlthOrder}
&\langle \phi_m(\infty)\phi_j(0)\rangle_{(\ell)}=\frac{1}{\ell !}\lambda^{i_1}\ldots\lambda^{i_{\ell}}\times\\
&\quad\times \int d^2 z_1\ldots\int d^2z_{\ell}\; \left(\Theta_1\ldots \Theta_{\ell}\prod_{r<s}\Theta_{rs}\right)\langle \phi_m(\infty)\phi_j(0)\phi_{i_1}(z_1)\ldots\phi_{i_{\ell}}(z_{\ell})\rangle_{0;c}\;
\end{split}
\end{align}%
where 
\begin{equation}\label{eq:defTheta}
\Theta_{j}= H(\vert z_j\vert -\epsilon)H(L-\vert z_j \vert)\,,
\end{equation}
we recast (\ref{eq:betaFunctionAnsatz}) into the form 
\begin{gather}\label{eq:betaV2}
\begin{aligned}
\sum_{\ell=0}^{\infty}\bigg\lbrace
\beta^n\langle \phi_m(\infty)\phi_n(0)\rangle_{(\ell)}
+\frac{\epsilon}{(\ell+1)!}\lambda^n\lambda^{i_1}\ldots\lambda^{i_{\ell}}
\frac{\partial}{\partial \epsilon}\partial_{i_1}\ldots \partial_{i_{\ell}}
\langle \phi_m(\infty)\phi_n(0)\rangle_{(\ell)}
\bigg\rbrace  ={\cal O}(\epsilon)\,.
\end{aligned}
\end{gather}
Substituting into (\ref{eq:betaV2}) the expansion  
\[
\beta^{i}=\sum_{\ell>0}\beta^{i}_{(\ell)}\equiv\sum_{\ell>0}\beta^{i}_{r_1\ldots r_{\ell}}\lambda^{r_1}\ldots\lambda^{r_{\ell}}\, , 
\] 
we obtain the following recursion relations
\begin{equation}\label{eq:betaRecA}
\beta^i(\lambda)=-\lim_{\epsilon\to 0}\delta^{ij}\sum_{\ell>0}\bigg\{
\beta^k(\lambda)+
\frac{\epsilon}{(\ell+1)!}\lambda^k\lambda^{r_1}\ldots\lambda^{r_{\ell}}\frac{\partial}{\partial\epsilon}\partial_{r_1}\ldots\partial_{r_{\ell}}\bigg\}\langle \phi_j(\infty)\phi_k(0)\rangle_{(\ell)}\,.
\end{equation}
The explicit formulae for the $\beta$-function coefficients up to $\OL{3}$ read
\begin{align}
\begin{split}\label{eq:betaToNLO}
\beta^i_{(2)}&=-\lim_{\epsilon\to 0}\delta^{ij}\frac{\epsilon}{2!}\lambda^k\lambda^r\frac{\partial}{\partial\epsilon}\partial_r\langle\phi_j(\infty)\phi_k(0)\rangle_{(1)}
\\
\beta^i_{(3)}&=-\lim_{\epsilon\to 0}\delta^{ij}\bigg\{
\beta^k_{(2)}\langle \phi_j(\infty)\phi_k(0)\rangle_{(1)}
+
\frac{\epsilon}{3!}\lambda^k\lambda^r\lambda^s\frac{\partial}{\partial \epsilon}
\partial_{r}\partial_s\langle\phi_j(\infty)\phi_k(0)\rangle_{(2)}\bigg\rbrace\,.
\end{split}
\end{align}%
%The recursion relation may also be written more compactly for the coefficients $\beta_{(\ell)}^i$ in the following form:
%\begin{align}
%\beta_{(\ell)}^i&=\delta^{ij}\bigg\{
%\frac{1}{\ell!}G^{(\ell-1)}_{jr_1;r_2\ldots r_{\ell}}\lambda^{r_1}\ldots\lambda^{r_{\ell}}
%-\sum_{m=1}^{\ell-2}\beta_{(\ell-m)}^k\langle\phi_j(\infty)\phi_k(0)\rangle_{(m)}
%\bigg\}\,.
%\end{align}

\subsection{The two loop beta function}\label{app:betaNLOcomputations}
\subsubsection{Derivation of the general formula}
In this appendix we  show how to derive formulas (\ref{NLOgen1}) and (\ref{beta2}) for  the two loop beta functions $\beta^i_{(3)}$.
We calculate
\begin{align}
\begin{split}
\langle\phi_i(\infty)\phi_j(0)\rangle_{(1)}=2\pi\ln(L/\epsilon)\oC_{ijk}\lambda^k\; ,
\end{split}
\end{align}%
which upon insertion into (\ref{eq:betaToNLO}) reproduces  the well-known one loop result
\begin{equation}
\beta^i_{(2)}(\lambda)=\pi\oC_{ijk}\lambda^j\lambda^k\,.
\end{equation}
At the two loop order the counterterm part is
\begin{align} \label{2lc}
-\delta^{ij}\beta^k_{(2)}\langle \phi_j(\infty)\phi_k(0)\rangle_{(1)}
&=-2\pi^2\delta^{ij}\ln(L/\epsilon)\oC_{rs}{}^k\oC_{jkt}\lambda^r\lambda^s\lambda^t\nonumber \\
&=-\frac{2\pi^2}{3!}\ln(L/\epsilon)\sum_{\text{perm}(r,s,t)}
\oC^i_{rm}\oC^m_{st}\lambda^r\lambda^s\lambda^t\,.
\end{align}%
For the remaining term we calculate 
\begin{align*}
&-\epsilon\frac{\partial}{\partial\epsilon}\partial_r\partial_s\langle \phi_i(\infty)\phi_j(0)\rangle_{(2)}\\
%&\quad=-\epsilon\int d^2z_1\int d^2z_2\bigg(
%\frac{\partial}{\partial\epsilon}\bigg[\Theta_1\Theta_{12}\Theta_2\bigg]
%\bigg)\langle \phi_i(\infty)\phi_j(0)\phi_r(z_1)\phi_s(z_2)\rangle_{0;c}\\
&\quad=\epsilon\int d^2z_1\int d^2z_2\bigg(
\delta^{\epsilon}_1\Theta_{12}\Theta_2
+\Theta_1\delta^{\epsilon}_{12}\Theta_2
+\Theta_1\Theta_{12}\delta^{\epsilon}_2
\bigg)\langle \phi_i(\infty)\phi_j(0)\phi_r(z_1)\phi_s(z_2)\rangle_{0;c}\,,
\end{align*}%
where we introduced the notations
\begin{equation}
\delta^{\epsilon}_a=\delta(|z_a|-\epsilon)H(L-|z_a|)\,,\quad\delta^{\epsilon}_{ab}=\delta(|z_a-z_b|-\epsilon)H(L-|z_a-z_b|)\,.
\end{equation}
Focusing for the moment  on the term involving $\delta^{\epsilon}_1\Theta_{12}\Theta_2$, we perform a global conformal transformation $f(z)=z/z_1$ on the $4$-point function, followed by a coordinate redefinition $g:z_2\mapsto\eta = \frac{z_2}{z_1}$. 
Taking further the $z_1$-integral we obtain  
\begin{align*}
&\epsilon\int d^2z_1\int d^2z_2\;
\delta^{\epsilon}_1\Theta_{12}\Theta_2\langle \phi_i(\infty)\phi_j(0)\phi_r(z_1)\phi_s(z_2)\rangle_{0;c}\\
&=
2\pi\int d^2\eta\;
\Theta(\epsilon|1-\eta|)\Theta(\epsilon|\eta|)\langle \phi_i(\infty)\phi_j(0)\phi_r(1)\phi_s(\eta)\rangle_{0;c}\\
&= 2\pi\int_{U_I} d^2\eta\;
\langle \phi_i(\infty)\phi_j(0)\phi_r(1)\phi_s(\eta)\rangle_{0;c}\,.
\end{align*}%
Here, the integration region $U_I$, illustrated in figures~\ref{fig:integrationRegionsAndDetails} (blue colored region) and~\ref{fig:contIllustr}, is defined via the product of cutoff  functions
\begin{equation}
\begin{split}
\Theta(\epsilon|x|)&= H(\epsilon(1-|x|))H(L-\epsilon|x|)\,.
\end{split}
\end{equation}
Analogously using the transformation  $f(z)=z/z_1$ and the change of integration variable $g:z_2\mapsto \eta = \frac{z_2}{z_2+\epsilon e^{i\phi_1}}$ we get
{\allowdisplaybreaks
\begin{align*}
&\epsilon\int d^2z_1\int d^2z_2\;
\Theta_1\delta^{\epsilon}_{12}\Theta_2\langle \phi_i(\infty)\phi_j(0)\phi_r(z_1)\phi_s(z_2)\rangle_{0;c}\\
&=
2\pi\int d^2\eta\;
\Theta\left(\epsilon\left|\frac{1}{\eta-1}\right|\right)
\Theta\left(\epsilon\left|1+\frac{1}{\eta-1}\right|\right)\langle \phi_i(\infty)\phi_j(0)\phi_r(1)\phi_s(\eta)\rangle_{0;c}\\
&= 2\pi\int_{U_{II}} d^2\eta\;
\langle \phi_i(\infty)\phi_j(0)\phi_r(1)\phi_s(\eta)\rangle_{0;c}\,.
\end{align*}}%
Finally, choosing $f(z)=z/z_1$ and $g:z_2\mapsto \eta = \frac{\epsilon e^{i\phi_2}}{z_1}$,  
{\allowdisplaybreaks
\begin{align*}
&\epsilon\int d^2z_1\int d^2z_2\;
\Theta_1\Theta_{12}\delta^{\epsilon}_2\langle \phi_i(\infty)\phi_j(0)\phi_r(z_1)\phi_s(z_2)\rangle_{0;c}\\
&=
2\pi\int d^2\eta\;
\Theta\left(\epsilon\left|\frac{1}{\eta}\right|\right)
\Theta\left(\epsilon\left|1-\frac{1}{\eta}\right|\right)\langle \phi_i(\infty)\phi_j(0)\phi_r(1)\phi_s(\eta)\rangle_{0;c}\\
&= 2\pi\int_{U_{III}} d^2\eta\;
\langle \phi_i(\infty)\phi_j(0)\phi_r(1)\phi_s(\eta)\rangle_{0;c}\,.
\end{align*}}%
The regions $U_{II}$, $U_{III}$ are described by the corresponding Heaviside functions.
Thus we obtain
\begin{align}\label{w3regs}
-\frac{\epsilon}{3!}\lambda^r\lambda^s\frac{\partial}{\partial\epsilon}\partial_r\partial_s\langle \phi_i(\infty)\phi_j(0)\rangle_{(2)}
&=\frac{\pi}{3}\lambda^r\lambda^s\int_{U_{I}\cup U_{II} \cup U_{III}}d^2\eta\;
\langle \phi_i(\infty)\phi_j(0)\phi_r(1)\phi_s(\eta)\rangle_{0;c}\,.
\end{align}%
Substituting the last expression along with (\ref{2lc}) into  (\ref{eq:betaToNLO})  and sending the cutoff parameter $\epsilon/L$ to zero we obtain formula (\ref{NLOgen1}).

Now, as detailed in appendix~\ref{app:permRules}, there exist combinations of conformal transformations and coordinate redefinitions such that the integrals over the regions $U_{II}$ and $U_{III}$ may be expressed as integrals over $U_I$ with permuted insertion points of the $4$-point function in the integrands. Moreover, the anti-cyclic permutations of insertion points may also be obtained by the aforementioned combined operations, which leads to
\begin{align*}
&%-\frac{\epsilon}{3!}\lambda^j\lambda^r\lambda^s\frac{\partial}{\partial\epsilon}\partial_r\partial_s\langle \phi_i(\infty)\phi_j(0)\rangle_{(2)}\\
\lambda^r\lambda^s\lambda^j\int_{U_{I}\cup U_{II} \cup U_{III}}d^2\eta\;
\langle \phi_i(\infty)\phi_j(0)\phi_r(1)\phi_s(\eta)\rangle_{0;c}\\
&=\frac{1}{2}\lambda^r\lambda^s\lambda^t\sum_{\text{perm}(r,s,t)}\int_{U_{I}}d^2\eta\;
\langle \phi_i(\infty)\phi_r(0)\phi_s(1)\phi_t(\eta)\rangle_{0;c}\;
\end{align*}%
that proves formula (\ref{beta2}).

\subsubsection{Description of the three integration regions}\label{app:integrationRegionDescr}

The three integration regions $U_{I}$, $U_{II}$, $U_{III}$ are explicitly described as follows  
\begin{equation}
U_I:\quad\left\{\begin{array}{rcccl}
1&\leq& x^2+y^2 &\leq&\frac{L^2}{\epsilon^2}\\
1&\leq& (x-1)^2+y^2&\leq &\frac{L^2}{\epsilon^2}
\end{array}\right.\,,
\end{equation}\\
\begin{equation}\label{eq:regionUII}
U_{II}:\quad 
\left\{ \begin{array}{rcccl}
x&\geq & \frac{1}{2} &&\\
\frac{\epsilon^2}{L^2}&\leq & (x-1)^2+y^2&\leq &1\\
 \frac{\epsilon^2}{L^2}(1+\delta)^2&\leq &(x-(1+\delta))^2+y^2&&
 \end{array}\right.\,,\qquad \delta = \frac{1}{\frac{L^2}{\epsilon^2}-1}\, , 
\end{equation}
\begin{equation}\label{eq:regionUIII}
U_{III}:\quad \left\{ \begin{array}{rcccl}
x&\leq &\frac{1}{2} &&\\
\frac{\epsilon^2}{L^2}&\leq&  x^2+y^2 &\leq&1\\
 \frac{\epsilon^2}{L^2}(1+\delta)^2&\leq& (x+\delta)^2+y^2&&
 \end{array}\right.\,,\qquad \delta = \frac{1}{\frac{L^2}{\epsilon^2}-1}\,.
\end{equation}
Since $\delta\to0$ in the limit $\tfrac{L}{\epsilon}\to\infty$, we observe that the union of the three regions $U_i$ converges to the entire $\eta$ plane, 
with the approximate integration region being bounded by a very large circle of radius $\frac{L}{\epsilon}$ around $\eta=\frac{1}{2}$ and with two discs of vanishing radius cut out around $\eta=0$ and $\eta=1$.

\subsubsection{List of  transformations generating permutations of insertion points and integration regions}\label{app:permRules}

The combined operation of first performing a conformal transformation $f_{\sigma}$ on the $4$-point function followed by a coordinate transformation $g_{\sigma}:\; \tilde{\eta}= g(\eta)$ results in a permutation of insertion points and integration regions in the integrals  that appear in (\ref{w3regs}). The transformation results in the following identity:
\begin{align*}
&\int_{U_i}d^2\eta\; \langle \phi_m(\infty)\phi_j(\eta)\phi_k(1)\phi_{\ell}(0)\rangle\\
&\qquad \qquad \overset{f_{\sigma}}{=}\int_{U_i}d^2\eta\;\left\vert \frac{\partial f_{\sigma}}{\partial z}\right\vert^4 \left\langle
\phi_m(\infty)\phi_{\sigma(j)}\left(f_{\sigma}(\eta)\right)
\phi_{\sigma(k)}(f_{\sigma}(1))\phi_{\sigma(\ell)}(f_{\sigma}(0))\right\rangle\\
&\qquad \qquad \overset{g_{\sigma}}{=}\int_{U_{\Sigma(i)}}d^2\tilde{\eta}\;\left\langle
\phi_m(\infty)\phi_{\sigma(j)}\left(\tilde{\eta}\right)
\phi_{\sigma(k)}(1)\phi_{\sigma(\ell)}(0)\right\rangle\; .
\end{align*}
The full list of combined permutations $(\sigma,\Sigma)$ generated by operations $(f_{\sigma},g_{\sigma})$ is given below
{\allowdisplaybreaks
\begin{align}
\begin{split}\label{eq:permA1}
\left\lbrace \begin{matrix}
f_{\mbox{\scriptsize$\begin{pmatrix}
j&k&\ell\\
k&\ell&j
\end{pmatrix}$}}(z)&\equiv& \frac{z-1}{\eta-1}\\[2em]
g_{\mbox{\scriptsize$\begin{pmatrix}
j&k&\ell\\
k&\ell&j
\end{pmatrix}$}}(\eta)&\equiv& \frac{1}{1-\eta}
\end{matrix}\right.
\quad \Rightarrow\quad\sigma\equiv\begin{pmatrix}
j&k&\ell\\
k&\ell&j
\end{pmatrix}\,,\quad \Sigma\equiv\begin{pmatrix}
I&II&III\\
II&III&I
\end{pmatrix}\,.
\end{split}\\
%%%%%%%%%%%%%%%%%%%%
\begin{split}\label{eq:permA2}
\left\lbrace \begin{matrix}
f_{\mbox{\scriptsize$\begin{pmatrix}
j&k&\ell\\
\ell&j&k
\end{pmatrix}$}}(z)&\equiv& 
1-\frac{z}{\eta}\\[2em]
g_{\mbox{\scriptsize$\begin{pmatrix}
j&k&\ell\\
\ell&j&k
\end{pmatrix}$}}(\eta)&\equiv& 
1-\frac{1}{\eta}
\end{matrix}\right.
\quad \Rightarrow\quad\sigma\equiv\begin{pmatrix}
j&k&\ell\\
\ell&j&k
\end{pmatrix}\,,\quad \Sigma\equiv\begin{pmatrix}
I&II&III\\
III&I&II
\end{pmatrix}\,.
\end{split}\\
%%%%%%%%%%%%%%%%%%%%
\begin{split}\label{eq:permA3}
\left\lbrace \begin{matrix}
f_{\mbox{\scriptsize$\begin{pmatrix}
j&k&\ell\\
k&j&\ell
\end{pmatrix}$}}(z)&\equiv& 
\frac{z}{\eta}\\[2em]
g_{\mbox{\scriptsize$\begin{pmatrix}
j&k&\ell\\
k&j&\ell
\end{pmatrix}$}}(\eta)&\equiv& 
\frac{1}{\eta}
\end{matrix}\right.
\quad \Rightarrow\quad\sigma\equiv\begin{pmatrix}
j&k&\ell\\
k&j&\ell
\end{pmatrix}\,,\quad \Sigma\equiv\begin{pmatrix}
I&II&III\\
III&II&I
\end{pmatrix}\,.
\end{split}\\
%%%%%%%%%%%%%%%%%%%%%%%
\begin{split}\label{eq:permA4}
\left\lbrace \begin{matrix}
f_{\mbox{\scriptsize$\begin{pmatrix}
j&k&\ell\\
j&\ell&k
\end{pmatrix}$}}(z)&\equiv& 
1-z\\[2em]
g_{\mbox{\scriptsize$\begin{pmatrix}
j&k&\ell\\
j&\ell&k
\end{pmatrix}$}}(\eta)&\equiv&
1-\eta
\end{matrix}\right.
\quad \Rightarrow\quad\sigma\equiv\begin{pmatrix}
j&k&\ell\\
j&\ell&k
\end{pmatrix}\,,\quad \Sigma\equiv\begin{pmatrix}
I&II&III\\
I&III&II
\end{pmatrix}\,.
\end{split}\\
%%%%%%%%%%%%%%%%%%%%%%%
\begin{split}\label{eq:permA5}
\left\lbrace \begin{matrix}
f_{\mbox{\scriptsize$\begin{pmatrix}
j&k&\ell\\
\ell&k&j
\end{pmatrix}$}}(z)&\equiv& 
\frac{z-\eta}{1-\eta}\\[2em]
g_{\mbox{\scriptsize$\begin{pmatrix}
j&k&\ell\\
\ell&k&j
\end{pmatrix}$}}(\eta)&\equiv&
1+\frac{1}{\eta-1}
\end{matrix}\right.
\quad \Rightarrow\quad\sigma\equiv\begin{pmatrix}
j&k&\ell\\
\ell&k&j
\end{pmatrix}\,,\quad \Sigma\equiv\begin{pmatrix}
I&II&III\\
II&I&III
\end{pmatrix}\,.
\end{split}
\end{align}}%
We see that  for each occurrence of the integration regions $\cU_{II}$ and $\cU_{III}$ in the two loop  $\beta$-function formula, there exists a combined operation that transforms it into an integral over the region $\cU_{I}$.

In addition to these pairs of global conformal transformations and coordinate transformations, which realize all permutations of the three insertion points $0$, $1$ and $\eta$, we will now 
introduce an additional operation that permutes the insertion points $0$ and $\infty$. 
Consider the transformation
\begin{equation}
f(w) = \frac{1}{w}\,,\quad \left\vert \frac{\partial f}{\partial w}\right\vert^2=\frac{1}{|w|^4}\,.
\end{equation}
In order to apply it to the $4$-point function, we need to regularize the transformation as follows:
\begin{align*}
&\int_{U_i}d^2\eta\; \langle \phi_m(\infty)\phi_j(\eta)\phi_k(1)\phi_{\ell}(0)\rangle
=\int_{U_i}d^2\eta\; \lim\limits_{\delta\to0}\lim\limits_{|x|\to\infty}|x|^4
\langle \phi_m(x)\phi_j(\eta)\phi_k(1)\phi_{\ell}(\delta)\rangle\\
& \qquad \overset{f(w)}{=}
\int_{U_i}d^2\eta\;
\lim\limits_{\delta\to0}\lim\limits_{|x|\to\infty}|x|^4\frac{1}{|x|^4}\frac{1}{|\eta|^4}\frac{1}{\delta^4}
\left\langle
\phi_m(\tfrac{1}{x})\phi_{j}\left(\frac{1}{\eta}\right)
\phi_{k}(1)\phi_{\ell}(\tfrac{1}{\delta})\right\rangle\\
&\quad\overset{R = 1/\delta}{=}
\int_{U_i}d^2\eta\;
\frac{1}{|\eta|^4}\lim\limits_{R\to\infty}R^4
\left\langle
\phi_m(0)\phi_{j}\left(\frac{1}{\eta}\right)
\phi_{k}(1)\phi_{\ell}(R)\right\rangle\\
&\qquad=\int_{U_i}d^2\eta\;
\frac{1}{|\eta|^4}
\left\langle
\phi_m(0)\phi_{j}\left(\frac{1}{\eta}\right)
\phi_{k}(1)\phi_{\ell}(\infty)\right\rangle\,.
\end{align*}%
Finally, applying the coordinate transformation
\begin{equation}
\tilde{g}(\eta) = \frac{1}{\eta}\,,
\end{equation}
we know from~\eqref{eq:permA3} that this transformation will permute the integration regions as
\begin{equation}
\tilde{\Sigma}=\begin{pmatrix}
I & II & III\\
III & II & I
\end{pmatrix}\,,
\end{equation}
hence we finally obtain
\begin{align*}
\int_{U_i}d^2\eta\; \langle \phi_m(\infty)\phi_j(\eta)\phi_k(1)\phi_{\ell}(0)\rangle
&\overset{\tilde{g}\circ f(w)}{=}\int_{U_{\tilde{\Sigma}(i)}}d^2\tilde{\eta}\;
\left\langle
\phi_m(0)\phi_{j}\left(\eta\right)
\phi_{k}(1)\phi_{\ell}(\infty)\right\rangle\,.
\end{align*}%
Combining this transformation  with the previously introduced ones shows that we may realize all possible permutations of insertion points in the formula defining the two loop coefficients, which proves
 that the tensor $\beta^{i}_{jk\ell}$ is invariant  under permutations of all four indices. (Obviously this is not a coordinate independent statement but rather the special property of the renormalization scheme employed.)

\subsubsection{Explicit parametrization of $\partial U_{I}$}

\begin{figure}[b!]%[htbp]
\begin{center}
\includegraphics[width=0.9\textwidth]{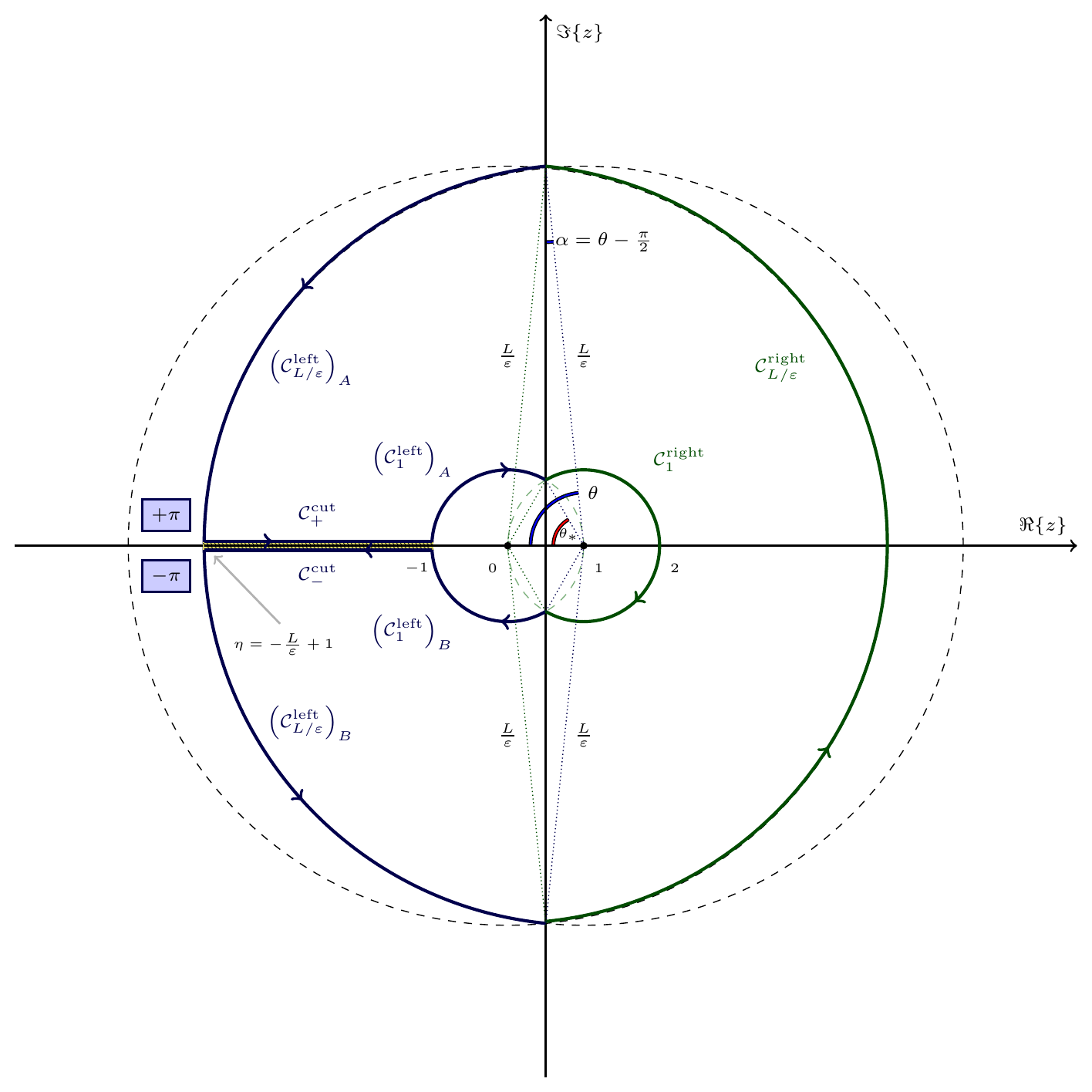}
\caption{Illustration of $\partial U_I$ and the choice of branch cut. The imaginary axis is centered at $\eta=\tfrac{1}{2}$ in order to illustrate the reflection symmetry of the integration contour (up to the branch cut pieces).}
\label{fig:contIllustr}
\end{center}
\end{figure}

The boundary of the integration region $U_I$ has to be augmented by a branch cut whenever contour integrals over logarithms are involved upon applying the complex Stokes theorem. 
A particularly convenient choice for this branch cut as well as the different segments of $\partial U_I$ is presented in figure~\ref{fig:contIllustr}. %
The different segments of the integration contour $\partial U_I$ may be parametrized as follows:
{\allowdisplaybreaks
\begin{subequations}
\begin{alignat}{6}
&\left(\mathcal{C}^{\mbox{\tiny{left}}}_{L/\varepsilon}\right)_A&: &\qquad& 
	\eta&=1+\frac{L}{\varepsilon}e^{i\varphi}\,, &\qquad& 
	\pi-\theta&{}\leq{}&&\varphi&{}\leq \pi\\
&\left(\mathcal{C}^{\mbox{\tiny{left}}}_{L/\varepsilon}\right)_B&: &\qquad& 
	\eta&=1+\frac{L}{\varepsilon}e^{i\varphi}\,, &\qquad& 
	-\pi&{}\leq{}&&\varphi&{}\leq -\pi+\theta\\
&{}\;\mathcal{C}^{\mbox{\tiny{right}}}_{L/\varepsilon}&: &\qquad& 
	\eta&=\frac{L}{\varepsilon}e^{i\varphi}\,, &\qquad& 
	-\theta&{}\leq{}&&\varphi&{}\leq \theta\\
	\nonumber\\
&\left(\mathcal{C}^{\mbox{\tiny{left}}}_1\right)_A&: &\qquad& 
	\eta&=e^{i\varphi}\,, &\qquad& 
	\pi&{}\geq{}&&\varphi&{}\geq \theta_{*}\\
&\left(\mathcal{C}^{\mbox{\tiny{left}}}_1\right)_B&: &\qquad& 
	\eta&=e^{i\varphi}\,, &\qquad& 
	-\theta_{*}&{}\geq{}&&\varphi&{}\geq -\pi\\
&{}\;\mathcal{C}^{\mbox{\tiny{right}}}_1&: &\qquad& 
	\eta&=1+e^{i\varphi}\,, &\qquad& 
	\pi-\theta_{*}&{}\geq{}&&\varphi&{}\geq -\pi+\theta_{*}\\
	\nonumber\\
&{}\;\mathcal{C}^{\mbox{\tiny{cut}}}_{+}&: &\qquad& 
	\eta&=xe^{i\pi}\,, &\qquad& 
	-\frac{L}{\varepsilon}+1&{}\leq{}&&x&{}\leq -1\\
&{}\;\mathcal{C}^{\mbox{\tiny{cut}}}_{-}&: &\qquad& 
	\eta&=xe^{-i\pi}\,, &\qquad& 
	-1&{}\geq{}&&x&{}\geq -\frac{L}{\varepsilon}+1\\
\end{alignat}
\end{subequations}}%
In the above equations, we introduced the notations 
\begin{align}
\begin{split}
\theta&= \frac{\pi}{2}-\alpha\,, \qquad \alpha=\arcsin\left(\frac{\varepsilon}{2L}\right)\, , \\
\theta_{*}&= \arccos\left(\frac{1}{2}\right)=\frac{\pi}{3}\,.
\end{split}
\end{align}%

\subsection{Derivation of the two loop beta function for current-current perturbations}\label{CC_beta_details}
 
The connected 4-point function of the current-current operators (\ref{cc_model}) is 
 \begin{align*}
&\langle \Phi_I(0)\Phi_J(1)\Phi_K(\eta)\Phi_L(\infty)\rangle_{0;c}=d_I^{a\bar{a}}d_J^{b\bar{b}}d_K^{c\bar{c}}d_L^{d\bar{d}}\bigg\lbrace\\
&\qquad
\eta_{ab}\eta_{cd}
\left(
\frac{
\eta_{\bar{a}\bar{c}}\eta_{\bar{b}\bar{d}}
}{\bar{\eta}^2}+\frac{
\eta_{\bar{a}\bar{d}}\eta_{\bar{b}\bar{c}}
}{(\bar{\eta}-1)^2}
\right)
+\frac{
\eta_{ac}\eta_{bd}
}{\eta^2}
\left(
\eta_{\bar{a}\bar{b}}\eta_{\bar{c}\bar{d}}
+\frac{
\eta_{\bar{a}\bar{d}}\eta_{\bar{b}\bar{c}}
}{(\bar{\eta}-1)^2}
\right)
+\frac{
\eta_{ad}\eta_{bc}
}{(\eta-1)^2}
\left(
\eta_{\bar{a}\bar{b}}\eta_{\bar{c}\bar{d}}+
\frac{
\eta_{\bar{a}\bar{c}}\eta_{\bar{b}\bar{d}}
}{\bar{\eta}^2}
\right)\\
&\qquad-\bigg[
\left(
\eta_{ab}\eta_{cd}+\frac{\eta_{ac}\eta_{bd}}{\eta^2}+\frac{\eta_{ad}\eta_{bc}}{(\eta-1)^2}
\right)\left(\fFb{a}{b}{r}\FFb{r}{c}{d}+\frac{
\fFb{a}{c}{r}\FFb{r}{d}{b}
}{\bar{\eta}}\right)\frac{1}{(\bar{\eta}-1)}
\bigg]-\overline{\bigg[\ldots\bigg]}\\
&\qquad +\frac{1}{|\eta-1|^2}\bigg(
\fF{a}{b}{r}\FF{r}{c}{d}\fFb{a}{b}{r}\FFb{r}{c}{d}
+\frac{
\fF{a}{c}{r}\FF{r}{d}{b}\fFb{a}{c}{r}\FFb{r}{d}{b}
}{|\eta|^2}
+\frac{
\fF{a}{c}{r}\FF{r}{d}{b}\fFb{a}{b}{r}\FFb{r}{c}{d}
}{\eta}
++\frac{
\fF{a}{b}{r}\FF{r}{c}{d}\fFb{a}{c}{r}\FFb{r}{d}{b}
}{\bar{\eta}}
\bigg)
\bigg\rbrace
\end{align*}%
 The symmetrized connected $4$-point function is
  \begin{align} \label{symm_4pt}
&\frac{1}{4!}\sum_{\text{perm}(I,J,K,L)}\langle \Phi_I(0)\Phi_J(1)\Phi_K(\eta)\Phi_L(\infty)\rangle_{0;c}
=\frac{1}{3!}d_I^{a\bar{a}}d_J^{b\bar{b}}d_K^{c\bar{c}}d_L^{d\bar{d}}\bigg[
E_{\eta\eta\bar{\eta}\bar{\eta}}F_{\eta\eta\bar{\eta}\bar{\eta}}(0,1,\eta)\nonumber \\
&\;+
E_{\eta\eta\bar{f}\bar{f}}F_{\eta\eta\bar{f}\bar{f}}(0,1,\eta)+
E_{ff\bar{\eta}\bar{\eta}}F_{ff\bar{\eta}\bar{\eta}}(0,1,\eta)+E_{f\bar{f}\bar{f}}F_{ff\bar{f}\bar{f}}(0,1,\eta)\bigg] %\nonumber \\
%&\;
\end{align}%
where we use the following shorthand notations   %$E_{\ldots}$ and $F_{\ldots}$ listed below:
{\allowdisplaybreaks
\begin{align*}
E_{\eta\eta\bar{\eta}\bar{\eta}}&=
\eta_{ab}\eta_{cd}\left(
\eta_{\bar{a}\bar{c}}\eta_{\bar{b}\bar{d}}
+\eta_{\bar{a}\bar{d}}\eta_{\bar{b}\bar{c}}
\right)
+\eta_{ac}\eta_{bd}\left(
\eta_{\bar{a}\bar{b}}\eta_{\bar{c}\bar{d}}
+\eta_{\bar{a}\bar{d}}\eta_{\bar{b}\bar{c}}
\right)
+\eta_{ad}\eta_{bc}\left(
\eta_{\bar{a}\bar{b}}\eta_{\bar{c}\bar{d}}
+\eta_{\bar{a}\bar{c}}\eta_{\bar{b}\bar{d}}
\right)\\
F_{\eta\eta\bar{\eta}\bar{\eta}}(0,1,\eta)&=
\frac{1}{\eta^2}+\frac{1}{(\eta-1)^2}
+\frac{1}{\bar{\eta}^2}+\frac{1}{(\bar{\eta}-1)^2}
+\frac{1}{\eta^2(\bar{\eta}-1)^2}
+\frac{1}{(\eta-1)^2\bar{\eta}^2}\\
\\
E_{\eta\eta\bar{f}\bar{f}}&=
\eta_{ab}\eta_{cd}\left(\fFb{a}{c}{r}\FFb{r}{d}{b} -\fFb{a}{d}{r}\FFb{r}{b}{c}\right)
+\eta_{ac}\eta_{bd}\left(\fFb{a}{d}{r}\FFb{r}{b}{c}-\fFb{a}{b}{r}\FFb{r}{c}{d}\right)\\
&\qquad
+\eta_{ad}\eta_{bc}\left(\fFb{a}{b}{r}\FFb{r}{c}{d}-\fFb{a}{c}{r}\FFb{r}{d}{b}\right)\\
F_{\eta\eta\bar{f}\bar{f}}(0,1,\eta)&=
-\frac{1}{(\bar{\eta}-1)}
+\frac{1}{\bar{\eta}}
-\frac{1}{(\eta-1)^2\bar{\eta}}
+\frac{1}{\eta^2(\bar{\eta}-1)}\\
E_{ff\bar{f}\bar{f}}&=
\fF{a}{b}{r}\FF{r}{c}{d}\fFb{a}{b}{r}\FFb{r}{c}{d}
+\fF{a}{c}{r}\FF{r}{d}{b}\fFb{a}{c}{r}\FFb{r}{d}{b}
+\fF{a}{d}{r}\FF{r}{b}{c}\fFb{a}{d}{r}\FFb{r}{b}{c}\\
F_{ff\bar{f}\bar{f}}(0,1,\eta)&=
\frac{2}{|\eta-1|^2}+\frac{2}{|\eta|^2}-\frac{1}{(\eta-1)\bar{\eta}}-\frac{1}{\eta(\bar{\eta}-1)}
\end{align*}}%
We also have the relations 
$$
E_{ff\bar{\eta}\bar{\eta}}=\overline{E_{\eta\eta\bar{f}\bar{f}}}\,, \quad 
F_{ff\bar{\eta}\bar{\eta}}(0,1,\eta)=\overline{F_{\eta\eta\bar{f}\bar{f}}(0,1,\eta)}\,,
$$
where the notation $\overline{\vphantom{X}\ldots}$ amounts to replacing all holomorphic quantities by anti-holomorphic quantities and vice versa. The coefficient function $E_{ff\bar{f}\bar{f}}$ contracted with the tensors $d$ yield contractions of $3$-point function coefficient tensors $\oC$:
\begin{align}
\begin{split}
d_I^{a\bar{a}}d_J^{b\bar{b}}d_K^{c\bar{c}}d_L^{d\bar{d}}E_{ff\bar{f}\bar{f}}
&=\oC_{IJ}{}^R\oC_{RKL}+\oC_{IK}{}^R\oC_{RJL}+\oC_{IL}{}^R\oC_{RJK}\\
&=\frac{1}{2}\sum_{perm(J,K,L)}\oC_{IJ}{}^R\oC_{RKL}\,.
\end{split}
\end{align}%

The two loop beta function coefficients are computed from the general formula (\ref{beta2}) using (\ref{symm_4pt}).
 We need the following integrals over the integration region $U_I$ described in appendix~\ref{app:integrationRegionDescr}, which are computed using Stokes theorem:
{\allowdisplaybreaks
\begin{align}
\begin{split}
\int_{U_I} d^2\eta\; \frac{1}{(\eta+ \tfrac{1}{2})^2}&=\int_{U_I} d^2\eta\; \frac{1}{(\bar\eta+ \tfrac{1}{2})^2}
\xrightarrow{
\tfrac{L}{\varepsilon}\to\infty}-\frac{\pi}{3}+\frac{\sqrt{3}}{4}\\
\int_{U_I} d^2\eta\; \frac{1}{(\eta- \tfrac{1}{2})^2}&=\int_{U_I} d^2\eta\; \frac{1}{(\bar\eta- \tfrac{1}{2})^2}
\xrightarrow{
\tfrac{L}{\varepsilon}\to\infty}\frac{2\pi}{3}+\frac{\sqrt{3}}{4}\\
\int_{U_I} d^2\eta\; \frac{1}{(\eta- \tfrac{1}{2})^2(\bar\eta+ \tfrac{1}{2})^2}&=
\int_{U_I} d^2\eta\; \frac{1}{(\eta+\tfrac{1}{2})^2(\bar\eta- \tfrac{1}{2})^2}
\xrightarrow{
\tfrac{L}{\varepsilon}\to\infty}-\frac{\pi}{3}-\frac{\sqrt{3}}{2}
\end{split}\\
\begin{split}
\int_{U_I} d^2\eta\; \frac{1}{(\eta+ \tfrac{1}{2})}&=\int_{U_I} d^2\eta\; \frac{1}{(\bar\eta+ \tfrac{1}{2})}
\xrightarrow{
\tfrac{L}{\varepsilon}\to\infty}\frac{\pi}{3}+\frac{\sqrt{3}}{2}\\
\int_{U_I} d^2\eta\; \frac{1}{(\eta- \tfrac{1}{2})}&=\int_{U_I} d^2\eta\; \frac{1}{(\bar\eta- \tfrac{1}{2})} \xrightarrow{
\tfrac{L}{\varepsilon}\to\infty}\frac{2\pi}{3}-\frac{\sqrt{3}}{2}\\
\int_{U_I} d^2\eta\; \frac{1}{(\eta\pm \tfrac{1}{2})(\bar\eta\pm \tfrac{1}{2})^2}&=
\int_{U_I} d^2\eta\; \frac{1}{(\eta\pm \tfrac{1}{2})^2(\bar\eta\pm \tfrac{1}{2})}
\xrightarrow{
\tfrac{L}{\varepsilon}\to\infty}\pm\left(\frac{\pi}{3}-\sqrt{3}\right)\\
\int_{U_I} d^2\eta\; \frac{1}{(\eta\pm \tfrac{1}{2})(\bar\eta\mp \tfrac{1}{2})^2}&=
\int_{U_I} d^2\eta\; \frac{1}{(\eta\mp \tfrac{1}{2})^2(\bar\eta\pm \tfrac{1}{2})}
\xrightarrow{
\tfrac{L}{\varepsilon}\to\infty}\pm\left(\frac{\sqrt{3}}{2}-\frac{2\pi}{3}\right)
\end{split}\\
\begin{split}
\int_{U_I} d^2\eta\; \frac{1}{(\eta\pm \tfrac{1}{2})(\bar\eta\pm \tfrac{1}{2})}&\xrightarrow{
\tfrac{L}{\varepsilon}\to\infty}2\pi\ln(L/\varepsilon)-\Delta\\
\int_{U_I} d^2\eta\; \frac{1}{(\eta\pm \tfrac{1}{2})(\bar\eta\mp \tfrac{1}{2})}&\xrightarrow{
\tfrac{L}{\varepsilon}\to\infty}2\pi\ln(L/\varepsilon)-2\Delta
\end{split}
\end{align}
}
The symbol $\Delta$ stands for the contribution
\begin{align}
\Delta&\equiv \frac{i}{2}\left( Li_2\left(e^{\frac{i\pi}{3}}\right)- Li_2\left(e^{\frac{-i\pi}{3}}\right)\right)\,.
\end{align}%
Collecting all contributions and using 
\begin{gather}
\begin{aligned}
\int_{U_I}d^2\eta\;F_{\eta\eta\bar{\eta}\bar{\eta}}(0,1,\eta)&\xrightarrow{
\tfrac{L}{\varepsilon}\to\infty}0\\
\int_{U_I}d^2\eta\;F_{\eta\eta\bar{f}\bar{f}}(0,1,\eta)&\xrightarrow{
\tfrac{L}{\varepsilon}\to\infty}\pi\\
\int_{U_I}d^2\eta\;F_{ff\bar{\eta}\bar{\eta}}(0,1,\eta)&\xrightarrow{
\tfrac{L}{\varepsilon}\to\infty}\pi\\
\int_{U_I}d^2\eta\;F_{ff\bar{f}\bar{f}}(0,1,\eta)&\xrightarrow{
\tfrac{L}{\varepsilon}\to\infty}4\pi\ln(L/\epsilon)\,,
\end{aligned}
\end{gather}%
we obtain formula (\ref{GeneralCC}).

\section{Details on the computation of the redundancy coefficients}\label{app:redundancyLOandNLOv2}

\subsection{Leading order calculation}

The leading order coefficients $r^{(1)}_{ai}{}^I$ may be computed via\footnote{To avoid potential complications arising from permuting the differential $\partial_{\bar{x}}$ with the integration necessary to obtain $\langle J_a(x)\Phi_{I}(y)\rangle_{(1)}$, we will first compute the integral and take the derivative on the result, rather than first taking the derivative on the correlator $\langle J_a(x)\Phi_{I}(y)\phi_i(v)\rangle_{0}$ (which would result in $\delta$ functionals).} 
\begin{align*}
\begin{split}
r^{(1)}_{ai}{}^{I}&=\delta^{IJ}|x-y|^4\partial_{i}\partial_{\bar{x}}\langle J_a(x)\Phi_{J}(y)\rangle_{(1)}\,.
\end{split}
\end{align*}%
With the $3$-point functions
\begin{align}
\begin{split}
\left\langle J_a(x)\Phi_{J}(y)\phi_i(v)\right\rangle_{0}
&=\frac{i\oA_{aJ i}}{(x-y)(x-v)(y-v)(\bar{y}-\bar{v})^2}\\
&=-\frac{i\oA_{aJ i}}{(x-y)^2}\left(\frac{1}{(x-v)}-\frac{1}{(y-v)}\right)\frac{1}{(\bar{y}-\bar{v})^2}\\
&=-\frac{i\oA_{aJ i}}{(x-y)^2}\partial_{\bar{v}}\bigg\lbrace
\left(\frac{1}{(x-v)}-\frac{1}{(y-v)}\right)\frac{1}{(\bar{y}-\bar{v})}\bigg\rbrace
\,,
\end{split}
\end{align}%
and implementing the explicit normalization convention and point-splitting for the integrals in the perturbed correlators $\langle J_a(x)\Phi_{J}(y)\rangle_{(1)}$, a straightforward computation by means of the complex Stokes theorem yields:
\begin{align*}
r^{(1)}_{ai}{}^{I}&=\delta^{IJ}|x-y|^4\partial_{i}\partial_{\bar{x}}\langle J_a(x)\Phi_{J}(y)\rangle_{(1)}\\
&=-i\delta^{IJ}(\bar{x}-\bar{y})^2\partial_{\bar{x}}
\oA_{aJ i}\lim\limits_{\tfrac{\epsilon}{L}\to0}\int d^2v\;\Theta_{xv}\Theta_{yv}
\partial_{\bar{v}}\bigg\lbrace
\left(\frac{1}{(x-v)}-\frac{1}{(y-v)}\right)\frac{1}{(\bar{y}-\bar{v})}\bigg\rbrace\\
&=i\pi \delta^{IJ}\oA_{aJ i}(\bar{x}-\bar{y})^2\partial_{\bar{x}}\bigg\lbrace\frac{1}{(\bar{x}-\bar{y})}\bigg\rbrace\\
&=i\pi\oA_{ai}{}^I\,.
\end{align*}%
Analogously, we may compute
\begin{align*}
\bar{r}^{(1)}_{\bar{a}i}{}^{I}&=i\pi\bar{\oA}_{\bar{a}i}{}^{I}\,.
\end{align*}

\subsection{Next-to-leading order calculation}

Let us focus on the computation of the next-to-leading order coefficients $r^{(2)}_{aij}{}^{I}$ for concreteness 
(since the coefficients $\bar{r}^{(2)}_{\bar{a}ij}{}^I$ may be computed in an entirely analogous fashion). Inspecting the defining equation
\begin{align}
\begin{split}
r^{(2)}_{aij}{}^{I}&=\delta^{IJ}|x-y|^4\partial_{i}\partial_{j}\bigg\lbrace
\partial_{\bar{x}}\langle J_a(x)\Phi_{J}(y)\rangle_{(2)}
-r^{(1)}_{ak}{}^{K}\lambda^k\langle \Phi_{K}(x)\Phi_{J}(y)\rangle_{(1)}
\bigg\rbrace\,,
\end{split}
\end{align}%
we need the formulae for the two perturbed correlator contributions $\langle \Phi_{K}(x)\Phi_{J}(y)\rangle_{(1)}$ and $\langle J_a(x)\Phi_{J}(y)\rangle_{(2)}$. First of all, we may compute 
\begin{align}
\langle \Phi_{K}(x)\Phi_{J}(y)\phi_{\ell}(v)\rangle_0
&=\frac{\oC_{JK\ell}}{|x-y|^2|x-v|^2|y-v|^2}\label{eq:appPPP}
\end{align}%
and
\begin{align}
&\left\langle J_a(x)\Phi_{J}(y)\phi_r(v)\phi_s(w)\right\rangle_{0}=\nonumber\\
&\qquad\frac{i}{(\bar{y}-\bar{v})(\bar{y}-\bar{w})(\bar{v}-\bar{w})}\bigg\lbrace
\frac{\oB_{aJ}{}^{\bar{b}}\bar{\oA}_{\bar{b}rs}}{(x-y)^2(v-w)^2}
+\frac{\oB_{ar}{}^{\bar{b}}\bar{\oA}_{\bar{b}sJ}}{(x-v)^2(y-w)^2}
+\frac{\oB_{as}{}^{\bar{b}}\bar{\oA}_{\bar{b}J r}}{(x-w)^2(y-v)^2}\bigg\rbrace\nonumber\\
&\qquad+\frac{i}{|y-v|^2|y-w|^2|v-w|^2}\bigg\lbrace
\frac{\oA_{aJ}{}^t\oC_{trs}}{(x-y)}
+\frac{\oA_{ar}{}^t\oC_{tsJ}}{(x-v)}
+\frac{\oA_{as}{}^t\oC_{tJ r}}{(x-w)}\bigg\rbrace\,.\label{eq:appJPPP}
\end{align} %
To obtain $\langle  \Phi_{K}(x)\Phi_{J}(y)\rangle_{(1)}$, we may use the formula
\begin{equation}
\frac{1}{|x-y|^2|x-v|^2|y-v|^2}=\frac{1}{|x-y|^4}\left\vert \frac{1}{(x-v)}-\frac{1}{(y-v)}\right\vert^2\,,
\end{equation}
and compute the integral over $d^2v$ by means of the complex Stokes theorem as
\begin{equation}
\int d^2v\; \Theta_{xv}\Theta_{yv}\left\vert \frac{1}{(x-v)}-\frac{1}{(y-v)}\right\vert^2
\xrightarrow{\tfrac{\epsilon}{L}\to0}4\pi\ln\left(\frac{|x-y|}{\epsilon}\right)\,.
\end{equation}
The double integral necessary to obtain $\langle J_a(x)\Phi_{J}(y)\rangle_{(2)}$ is rather complicated due to the fact that we have to ``disentangle'' the product of cut-off functions $\Theta$ implementing the point-splitting in order to apply the complex Stokes theorem. This procedure results in a contour integral
\[
\int d^2v\; \Theta_{xv}\Theta_{yv}\int d^2w\; \Theta_{xw}\Theta_{yw}\Theta_{vw}\ldots =\sum_{\alpha=I}^{III}\int_{\mathcal{U}^{\alpha}_v}d^2v\; \int_{\mathcal{U}^{\alpha}_{w;v}}d^2w\;\ldots\,,
\]
where the three combinations of integration regions $(\mathcal{U}^{\alpha}_v,\mathcal{U}^{\alpha}_{w;v})$ are those obtained from decomposing the product of $\Theta$ cut-off functions in such a fashion that we can perform the integral over $d^2w$ first. After a tedious computation, the double integral may be evaluated as
\begin{align}
\begin{split}
&\partial_{\bar{x}}\bigg\lbrace
\sum_{\alpha=I}^{III}\int_{\mathcal{U}^{\alpha}_v}d^2v\; \int_{\mathcal{U}^{\alpha}_{w;v}}d^2w\;  
\left\langle J_a(x)\Phi_{J}(y)\phi_{(r}(v)\phi_{s)}(w)\right\rangle_0\bigg\rbrace\\
&\qquad\xrightarrow{\tfrac{\epsilon}{L}\to0} 
-\frac{4i\pi^2\ln\left(\frac{|x-y|}{\varepsilon}\right)}{|x-y|^4}\oA_{aJ}{}^t\oC_{trs}+\frac{i\pi^2}{|x-y|^4}\left(
\oB_{ar}{}^{\bar{b}}\bar{\oA}_{\bar{b}J s}
+\oB^{as}{}^{\bar{b}}\bar{\oA}_{\bar{b}J r}\right)\,.
\end{split}
\end{align}%
Thus, the divergent parts of the two contributions to the next-to-leading order coefficient tensors cancel each other, and we finally obtain:
\begin{equation}
r^{(2)}_{ars}{}^{I}=-i\pi^2\left(
\oB_{ar}{}^{\bar{b}}\bar{\oA}_{\bar{b}s}{}^{I}
+\oB^{as}{}^{\bar{b}}\bar{\oA}_{\bar{b} r}{}^{I}\right)\,.
\end{equation}
In an entirely analogous procedure, the next-to-leading order coefficients $\bar{r}^{(2)}_{\bar{a}rs}{}^{I}$ may be computed as
\begin{equation}
\bar{r}^{(2)}_{\bar{a}rs}{}^{I}=-i\pi^2\left(
\bar{\oB}_{\bar{a}r}{}^{b}\oA_{bs}{}^{I}
+\bar{\oB}_{\bar{a}s}{}^{b}\oA_{b r}{}^{I}\right)\,.
\end{equation}

\section{Some contour integrals}
By virtue of the complex Stokes theorem 
\begin{align}
\begin{split}
\int_M d^2x\; \partial_{\mu}F^{\mu}&=
\int_M d^2z\; \left(\bar{\partial}F^{z}+\partial F^{\bar{z}}\right)=\int_{\partial M}\bigg\lbrace dz \varepsilon_{\bar{z}z}F^{\bar{z}}+d\bar{z}\varepsilon_{z\bar{z}}F^{z}\bigg\rbrace\\
&=\frac{i}{2}\int_{\partial M}\bigg\lbrace
-dz F^{\bar{z}}+d\bar{z}F^z
\bigg\rbrace \,,
\end{split}
\end{align}
we may evaluate the complex single and double integrals of interest in this paper once we find an explicit description of the integration contour $\partial M$. The contours are computed from the combinations of cutoff functions $\Theta_{xy}$,
\[
\Theta_{xy} = H(|x-y|-\epsilon)H(L-|x-y|)\,,
\]
which are used to implement a point-splitting regularization scheme for the integrals. For brevity, the presence of the cutoff functions is indicated by the notation $[\ldots]$. Amongst the list of integrals of interest in this paper, there are two divergent integrals:
\begin{gather}
\begin{aligned}
\int \left[\frac{d^2x}{(y-x)(\bar{z}-\bar{x})}\right]&=2\pi\ln\left(\frac{\Lambda}{\epsilon+|y-z|}\right)\,, 
&\quad & &\int \left[\frac{d^2x}{(y-x)(\bar{z}-\bar{x})}\right]&=2\pi\ln(\Lambda/\epsilon)\,.
\end{aligned}
\end{gather}
We also encounter a number of convergent integrals:
\begin{gather}
\begin{aligned}
\int d^2x\;\left[\frac{1}{(y-x)(\bar{z}-\bar{x})^2}\right]&=-\frac{\pi}{(\bar{y}-\bar{z})}\,,
&\quad& &\int d^2x\;\left[\frac{1}{(y-x)(\bar{y}-\bar{x})^2}\right]&=0\, , \\
\int d^2x\;\left[\frac{1}{(y-x)^2(\bar{z}-\bar{x})}\right]&=\frac{\pi}{(y-z)}\,,
&\quad& &\int d^2x\;\left[\frac{1}{(y-x)^2(\bar{y}-\bar{x})}\right]&=0\,.
\end{aligned}
\end{gather}%

%% commands necessary to implement the bibliography style amsplain:

\providecommand{\bysame}{\leavevmode\hbox to3em{\hrulefill}\thinspace}
\providecommand{\MR}{\relax\ifhmode\unskip\space\fi MR }
\providecommand{\MRhref}[2]{%
  \href{http://www.ams.org/mathscinet-getitem?mr=#1}{#2}
}
\providecommand{\href}[2]{#2}

\end{document}